\documentclass[10pt]{article}

\usepackage{amssymb,amsmath,eurosym}

\usepackage{graphicx}
\usepackage{subcaption}
\usepackage{dcolumn}
\usepackage{authblk,hyperref}

\usepackage{tikz}
\usetikzlibrary{positioning,shapes,math,arrows.meta,decorations}
\usepackage{caption}
\usepackage{mwe}
\usepackage[percent]{overpic}

\usepackage{multirow}

\newcommand{\qv}{Q}

\newcommand{\cecs}{Centro de Estudios Cient\'ificos, Arturo Prat 514, Valdivia, Chile}
\newcommand{\uss}{Universidad San Sebasti\'an, General Lagos 1163, Valdivia, Chile}

\begin{document}

\title{Exact partition function of the Potts model on the Sierpinski gasket and the Hanoi lattice}

\author[1,2]{P. D. Alvarez \thanks{E-mail: \href{mailto:pedro.alvarez@uss.cl}{\nolinkurl{pedro.alvarez@uss.cl}}}}
\affil[1]{\cecs}
\affil[2]{\uss}

\maketitle

\begin{abstract}
We present an analytic study of the Potts model partition function on the Sierpinski and Hanoi lattices, which are self-similar lattices of triangular shape with non integer Hausdorff dimension.
Both lattices are examples of non-trivial thermodynamics in less than two dimensions, where mean field theory does not apply.
We used and explain a method based on ideas of graph theory and renormalization group theory to derive exact equations for appropriate variables that are similar to the restricted partition functions.
We benchmark our method with Metropolis Monte Carlo simulations.
The analysis of fixed points reveals information of location of the Fisher zeros and we provide a conjecture about the location of zeros in terms of the boundary of the basins of attraction.
\end{abstract}


\section*{Introduction}

In the analysis of critical phenomena and phase transitions the dimensionality of the lattice plays a key role, for a detailed review, see the classic book \cite{Zinn-Justin:2002ecy}. In the case of two-dimensional lattice models many interesting phase transitions do indeed occur and many exact results are known \cite{Baxter:1982zz}. However, the case of fractals, whose Hausdorff dimension is generically a non-integer number and in cases of interest is also tipically smaller than four, is much less understood.

We focus on the Potts model which is a natural generalization of the Ising model. This model is able to describe both first and second order phase transitions and has connections to many models of physics and mathematics \cite{Wu:1982ra}, that makes the developing of new exact methods a very important task.  For detailed reviews on its connections with other areas in physics and mathematics see \cite{Baxter:1982zz} and \cite{Wu:1982ra,Yang:1994qs,Cardy:2001vr,Yaffe:1982qf}. It is much harder to solve the Potts model for generic spin states number $q$ rather than the Ising model: only the Ising model on a square lattice has been solved in the thermodynamical limit by Onsager \cite{O44}, while the analytic solution of the Potts model in two dimensions on infinite square lattice is still to be found. Very few exact results of the Potts model are known in dimensions higher than one, see \cite{Baxter:1982zz}. One family of lattices that is solvable are the strip lattices, i.e. periodic lattices whose length is much larger than their width, can provide interesting qualitative information about the thermodynamics of the Potts model \cite{Sokal,Shrock1,Shrock2}. On the other hand, strip lattices are effectively one-dimensional systems: in particular, their Hausdorff dimension is one.

The main goal of this paper is to show a technique based on graph theory that allows us to obtain exact equations for the partition function. This permits to apply the tools of dynamical systems theory to study the Potts model on self similar lattices of fractal dimensions higher than one. We will focus on two special cases, the Sierpinski gasket and the Hanoi graph but the same ideas could, in principle, applied to more general cases.

The analysis of the Potts model on the Sierpinski gasket began with the pioneering papers \cite{Dhar:1977cu, 10.1063/1.523515}. The first studies of the Potts model on hierarchical lattices employing real-space renormalization group methods was performed in \cite{ANBerker_1979} and widely explored in a series of following papers \cite{Andelman:1981yt, GK81, GK82, McKay82, McKay84, GK84, GASM83, GASM, DDI} which have been an important source of inspiration as far as the present paper is concerned. Let us remark that other schemes, such as mean field theory, do not provide very useful results when the dimensionality is low \cite{Andelman:1981yt}.

Lattice models on hierarchical lattices has been studied in the literature for many reasons. First of all, the thermodynamic limit is well defined \cite{GASM}. Second, given that a fractal does have an inherent scale symmetry then critical phenomena could appear and the scale invariance also allows to associate exact or approximate RG-equations. We will illustrate our method on the Sierpinski gasket, which has Hausdorff dimension is $d_{H}=\log 3/ \log 2\approx 1.6$\footnote{The\ Hausdorff dimension coincides with the usual definition of dimension in the case of regular lattices: for any regular two dimensional lattices in the thermodynamical limit $d_{H}=2$, for regular three dimensional lattices $d_{H}=3$ and so on.} See for instance \cite{2017PhRvB..96q4407P}, where a path counting approach is used to study the Ising model in fractal lattices.

A dimensionality higher than one also makes the Sierpinski gasket more interesting arena than the tree-type lattices, which are effectively one dimensional. Another reason to study exact methods on the Sierpinski lattice is the fact that the size of the sampling required is comparatively bigger than in higher dimensional lattices ought to the fact that a higher lacunarity number makes very easy to induce spontaneos magnetization of a local patch of the lattice. Because of this estimates on the critical exponents, mostly based on Monte Carlo simulations, fluctuate in the literature, see \cite{Genzor2023PhRvE.107c4131G} and references therein. 

Moreover, the lack of translation invariance and the lower dimensionality imlies that fractal lattices may belong to a different universality classes. The hyperscaling hypothesis involves the system dimension $d$. An open question is the validity of the hyperscaling hypothesis in the case of non-integer dimensional, where critical phenomena is not fully understood \cite{Genzor2023PhRvE.107c4131G}.

In particular, in \cite{McKay82}, it was shown that hierarchical lattices develop, under frustration, chaotic mapping between different generations, thus proving a first connection between self-similar lattices, frustration and dynamical systems. Similar analysis were also performed on other specific types fractals lattices such as the Hanoi graph (see, for example 
\cite{Boe08} \cite{OS87, BKM, MM, CC, MST98, MS99, MVS02, BL}). Fractals often appear in the analysis of dynamical systems \cite{strogatz:2000,wainwright_ellis_1997}. Here it will be shown that, in a sense, the converse is also true: namely, the computation of the partition function on a large class of fractal lattices can be reduced to the analysis of a dynamical system.

The key idea to analyse the partition function of the Potts model on self-similar lattices is to exploit the recursive symmetry deriving a closed set of dynamical equations describing the possible phases of the system. The great importance of recursive symmetry in statistical systems can be recognized also in the cases of more usual lattices as in \cite{Canfora:2006hv,Canfora:2007di,Astorino:2008sm,Astorino:2008wh,Astorino:2009ha}. In these works, recursive symmetry has been used to get a phenomenological description of the Ising model in three dimensions in a quite good agreement with the available numerical data. In \cite{ACRR}, following the results presented in \cite{BDC,BERAHA19791,BERAHA198052}, it has been proposed to analyze strip lattice using the formalism of the dichromatic polynomial (a very well known tool in graph and knot theory: see, for instance \cite{kau1,Kauffman:1991ds,Wu2}): this allows to derive a linear recursive system of equations which determines the partition function. Here we will extend these results to the cases of self-similar lattices. 

The paper is organized as follows: in section \ref{sectionsierpinski}, the Sierpinski gasket is introduced along with its recursive definition and its fractal properties. In section \ref{sectiondelcont}, we present the basics of the Potts model and the Fortuin-Kasteleyn formalism. In section \ref{sec-coeff}, we define suitable geometric coefficients related to the connectivity pattern of the vertices of the lattice that are essential for our methods. In section \ref{sec-patt}, a recursive procedure to compute the partition function of the Potts model using the dichromatic polynomial is described. An algorithm is outlined aimed to achieve a closed set of recursive equations for the connectivity coefficients. A decomposition of the partition function for the Potts model at generic recursive step $n+1$ in terms of the partition function at step $n$ is described. The procedure leads to a discrete dynamical system which completely characterize the relevant thermodynamical properties of the Potts model. In section \ref{sectiondynhanoi}, we apply the technique to the Hanoi lattice. In section \ref{sectionthermo}, we run benchmarks and obtain exact values for the peak of the specific heat and zero temperature entropy. In section \ref{sectionfixed}, we study the fixed points of the dynamical systems. New sets of initial conditions that produce new lattices for same dynamical system are considered in section \ref{sec-change}. In section \ref{zerosandbasins}, we provide an study of the loci of Fisher zeros and the map of basins of the dynamical system. Lastly, in section \ref{sectionconclusions}, we provide a summary and outlook of our study.

\section{The Sierpinski gasket}\label{sectionsierpinski}

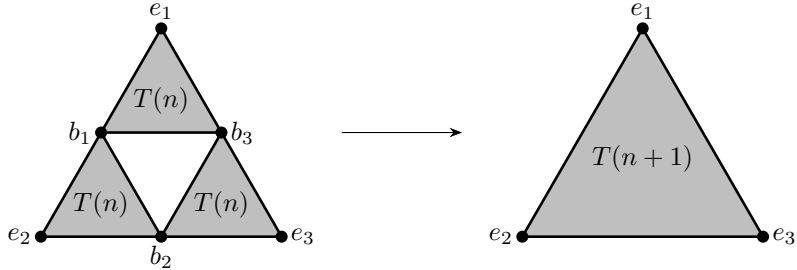
\begin{figure}
\centering
\begin{tikzpicture}[scale=0.8]
\tikzmath{\shift = 8;}
\draw [line width=1pt, fill=gray!50] (\shift,0) -- (2+\shift,3.464) -- (4+\shift,0) -- cycle;

\coordinate[label=above:$e_1$] (e1) at (2+\shift,3.464);
\coordinate[label=left:$e_2$]  (e2) at (0+\shift,0);
\coordinate[label=right:$e_3$] (e3) at (4+\shift,0);

\coordinate[label=above:$T(n+1)$] (nplus1) at (2+\shift,3.464/2-0.8);
\fill   (0+\shift,0) circle (1mm) 
        (4+\shift,0) circle (1mm)
        (2+\shift,3.464) circle (1mm);
\draw[-{Stealth[black]}] (5,3.464/2)   -- (7,3.464/2);
\draw [line width=1pt, fill=gray!50] (1,3.464/2) -- (2,3.464) -- (3,3.464/2) -- cycle;
\draw [line width=1pt, fill=gray!50] (0,0) -- (1,3.464/2) -- (2,0) -- cycle;
\draw [line width=1pt, fill=gray!50] (2,0) -- (3,3.464/2) -- (4,0) -- cycle;

\coordinate[label=above:$e_1$] (e1) at (2,3.464);
\coordinate[label=left:$e_2$]  (e2) at (0,0);
\coordinate[label=right:$e_3$] (e3) at (4,0);

\coordinate[label=above:$T(n)$] (n) at (2,3.464*3/4-0.7);
\coordinate[label=above:$T(n)$] (n) at (1,3.464/4-0.7);
\coordinate[label=above:$T(n)$] (n) at (3,3.464/4-0.7);
\fill   (0,0) circle (1mm) 
        (4,0) circle (1mm)
        (2,3.464) circle (1mm);
        
\fill   (1,3.464/2) circle (1mm) 
        (3,3.464/2) circle (1mm)
        (2,0) circle (1mm);
        
\coordinate[label=left:$b_1$] (b1) at (1,3.464/2);
\coordinate[label=below:$b_2$]  (b2) at (2,0);
\coordinate[label=right:$b_3$] (b3) at (3,3.464/2);

\end{tikzpicture}
\caption{Sierpinski iteration procedure. $T(1)$ is a lattice with $V=3$ and $E=3$, $T(2)$ is a lattice with $V=6$ and $E=9$, $T(3)$ is a lattice with $V=15$ and $E=27$ and so on. In general, the number of vertices at step $n$ is $V_n=(3^n+3)/2$ while the number of edges is $E_n=3^n$. The number of internal (bulk) sites of $T(n)$ is $(3^n-3)/2$ and the number of external sites is always $3$.}
\label{figsierpinski}
\end{figure}

In this paper we will focus on two types of hierarchical lattices with triangular shape, the Sierpinski triangle and the Hanoi triangle. For the Sierpinski triangle the recursive procedure of building is as follow. Take a lattice with external vertices $\{e_1,e_2,e_3\}$, make three copies and relabel vertices in the following way $\{e_1,b_1,b_2\}$, $\{b_1,e_2,b_3\}$ and $\{b_3,e_3,b_2\}$, and join them together at the vertices $b_1$, $b_2$ and $b_3$. Now we have new external vertices $\{e_1,e_2,e_3\}$ that respect the original triangular structure so that it is ready to repeat the multiplication and gluing procedure, see fig. \ref{figsierpinski}). For the sake of simplicity we considered triangular-shape structures with invariance under $(2 \pi/3)$ rotations, but this condition is not mandatory.

\begin{figure}
\centering
\begin{tikzpicture}[scale=0.8]
\tikzmath{\shift = 8; \shifty =-1;}
\draw [line width=1pt, fill=gray!50] (\shift,\shifty) -- (2+\shift,3.464+\shifty) -- (4+\shift,\shifty) -- cycle;

\coordinate[label=above:$e_1$] (e1) at (2+\shift,3.464+\shifty);
\coordinate[label=left:$e_2$]  (e2) at (0+\shift,0+\shifty);
\coordinate[label=right:$e_3$] (e3) at (4+\shift,0+\shifty);

\coordinate[label=above:$T(n+1)$] (nplus1) at (2+\shift,3.464/2-0.8+\shifty);
\fill   (0+\shift,0+\shifty) circle (1mm) 
        (4+\shift,0+\shifty) circle (1mm)
        (2+\shift,3.464+\shifty) circle (1mm);
\draw[-{Stealth[black]}] (5,3.464/2+\shifty)   -- (7,3.464/2+\shifty);
\draw [line width=1pt, fill=gray!50] (1,3.464/2) -- (2,3.464) -- (3,3.464/2) -- cycle;
\draw [line width=1pt, fill=gray!50] (-1,-3.464/2) -- (0,0) -- (1,-3.464/2) -- cycle;
\draw [line width=1pt, fill=gray!50] (3,-3.464/2) -- (4,0) -- (5,-3.464/2) -- cycle;
\draw [line width=1pt] (0,0) -- (1,3.464/2);
\draw [line width=1pt] (4,0) -- (3,3.464/2);
\draw [line width=1pt] (1,-3.464/2) -- (3,-3.464/2);

\coordinate[label=above:$e_1$] (e1) at (2,3.464);
\coordinate[label=left:$e_2$]  (e2) at (-1,-3.464/2);
\coordinate[label=right:$e_3$] (e3) at (5,-3.464/2);

\coordinate[label=above:$T(n)$] (n) at (2,3.464*3/4-0.7);
\coordinate[label=above:$T(n)$] (n) at (0,-3.464/4-0.7);
\coordinate[label=above:$T(n)$] (n) at (4,-3.464/4-0.7);
\fill   (0,0) circle (1mm) 
        (4,0) circle (1mm)
        (2,3.464) circle (1mm)
        (1,-3.464/2) circle (1mm)
        (3,-3.464/2) circle (1mm)
        (-1,-3.464/2) circle (1mm)
        (5,-3.464/2) circle (1mm);
        
\fill   (1,3.464/2) circle (1mm) 
        (3,3.464/2) circle (1mm);
        
\coordinate[label=left:$b_1$] (b1) at (1,3.464/2);
\coordinate[label=left:$b_2$] (b2) at (0,0);
\coordinate[label=below:$b_3$] (b3) at (1,-3.464/2);
\coordinate[label=below:$b_4$] (b4) at (3,-3.464/2);
\coordinate[label=right:$b_5$] (b5) at (4,0);
\coordinate[label=right:$b_6$] (b6) at (3,3.464/2);
\end{tikzpicture}
\caption{Hanoi iteration procedure. $T(1)$ is a lattice with $V=3$ and $E=3$, $T(2)$ is a lattice with $V=9$ and $E=12$, $T(3)$ is a lattice with $V=27$ and $E=39$ and so on. In general, the number of vertices at step $n$ is $V_n=3^n$ while the number of edges is $E_n=(3/2) (3^n - 1)$.}
\label{fighanoi}
\end{figure}
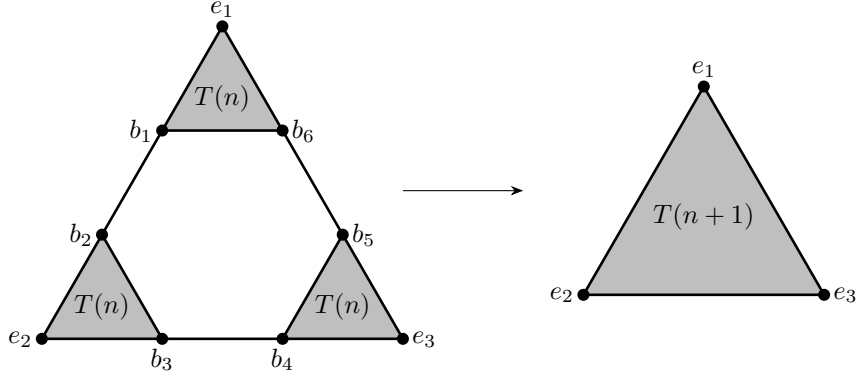

The recursive relation to produce the ``Hanoi'' lattice is similar to that of Sierpinski, except an auxiliary structure is used in very step to join the three triangular-shape structures. In particular that auxiliary structure can be a simple edge, see fig.\ref{fighanoi}. The reason behind considering this recursive relation is two fold, on the one hand it is a very simple example of a recursive procedure that will imply that the corresponding dynamical equations depend explicitly on the temperature variable, this will be clear below. On the other hand, the resulting lattice has neighbourg connectivities three instead of four for all internal vertices. 

\section{Dichromatic Polynomial and the Partition function}\label{sectiondelcont}

In this section we briefly review the basics of the dichromatic polynomial which will be essential for the derivation of the RG equations in the present formalism. 

The dichromatic polynomial $K\left[ X\right] $ of a lattice $X$ coincides with the partition function of the Potts model on $X$ \cite{ForKas}, see also the reviews \cite{ForKas,Wu:1982ra,Kauffman:1991ds}. $K\left[ X\right]$ depends on the variable $q$, the number of spin states \textit{per site}, and on the variable $v=\exp \left( \beta J\right) -1$ where $\beta=1/k_{B}T$, $k_{B}$ is the Boltzmann's constant, $T$ is the temperature and $J$ is the coupling, $J=+1$ corresponds to ``ferromagnetic'' regime and $J=-1$ to ``antiferromagnetic regime'', in the sense that neighbouring sites prefer to stay at the same state or at distinct states. The ranges $-1\leq v\leq 0$ and $\infty\geq v\geq 0$ correspond to the range $0 \leq T \leq \infty$ in the antiferromagnetic regime and the ferromagnetic regime respectively.

The dichromatic polynomial $K\left[X\right] $ can be computed by a successive application of the following rules:
\begin{enumerate}
\item \textit{deletion-contraction algorithm} the dichromatic polynomial of a graph $X$ in which two points are connected by an edge $K\left[ \bullet - \bullet \right] $ is equal to the sum of the dichromatic polynomial $K\left[ \bullet \ \bullet \right] $ of a graph obtained from $X$ by removing the edge plus $v$ times the dichromatic polynomial $K\left[ \bullet =\bullet \right] $ of a graph obtained from $X$ by collapsing the edge \footnote{we will also use the terms \textquotedblleft contracting" and \textquotedblleft identifying".} so that the external vertices of the edge are identified, 
\begin{equation}
K\left[ \bullet -\bullet \right] =K\left[ \bullet \ \bullet \right] +v\ K \left[ \bullet =\bullet \right] ,  \label{dp1}
\end{equation}

\item the dichromatic polynomial of a graph constructed by the disjoint union of two graphs is equal to the product of the dichromatic polynomials of each graph,
\begin{equation}
K\left[ X_1 \cup X_2\right] = K\left[ X_1\right] K\left[ X_2\right],  \label{dp2}
\end{equation}

\item the dichromatic polynomial of a single vertex is equal to $q$, 
\begin{equation}
K\left[ \bullet \right] =q.  \label{dp3}
\end{equation}
\end{enumerate}

As far as the goals of the present work are concerned, the most important feature of the definitions in eqs. (\ref{dp1}), (\ref{dp2}) and (\ref{dp3}) is that they are \emph{local} rules: namely, such rules act on an edge or a point and they do not depend on the structure of the graph far from the edge or the point on which they are acting.

The geometry of self-similar lattices suggests to search for an inductive procedure which should allow one to construct the partition function at step $n+1$ from the information of the step $n$. Thus, the question is, \textit{which is the minimal amount of physical information coming from the step} $n$ \textit{needed in order to construct} $K(n+1)$?

\subsection{Connectivity coefficients}\label{sec-coeff}

In this section we will introduce the concept of \emph{connectivity coefficients}, as a natural set of variables that describes the state of the system in the space of graphs. A useful analogy comes from real space RG methods: the connectivity coefficients are similar to coefficients multiplying the \textit{restricted partition functions}.. The connectivity coefficients can be seen as geometric coefficients characterizing the connectivity patterns of the external vertices of the Sierpinski gasket.

\begin{figure}
\centering
\begin{tikzpicture}[scale=0.3]
\tikzmath{\shift = 12; \shifty =0; \rad=0.3;}

\coordinate[label=above:$e_1$] (e1) at (2+\shift,3.464+\shifty);
\coordinate[label=left:$e_2$]  (e2) at (0+\shift,0+\shifty);
\coordinate[label=right:$e_3$] (e3) at (4+\shift,0+\shifty);

\fill   (0+\shift,0+\shifty) circle (\rad) 
        (4+\shift,0+\shifty) circle (\rad)
        (2+\shift,3.464+\shifty) circle (\rad);

\coordinate[label=above:$e_1 e_2 e_3$] (base1) at (2,3.464/2+\shifty-0.7);
\draw[{Stealth[black]}-{Stealth[black]}] (6,3.464/2+\shifty)   -- (9,3.464/2+\shifty);
\tikzmath{\shifty =-6;}
\draw [line width=1pt,double] (\shift,\shifty) -- (2+\shift,3.464+\shifty);

\coordinate[label=above:$e_1$] (e12) at (2+\shift,3.464+\shifty);
\coordinate[label=left:$e_2$]  (e22) at (0+\shift,0+\shifty);
\coordinate[label=right:$e_3$] (e32) at (4+\shift,0+\shifty);

\fill   (0+\shift,0+\shifty) circle (\rad) 
        (4+\shift,0+\shifty) circle (\rad)
        (2+\shift,3.464+\shifty) circle (\rad);

\coordinate[label=above:${e_3 e_1 = e_2}$] (base2) at (2,3.464/2+\shifty-0.7);
\draw[{Stealth[black]}-{Stealth[black]}] (6,3.464/2+\shifty)   -- (9,3.464/2+\shifty);
\tikzmath{\shifty =-12;}
\draw [line width=1pt,double] (\shift,\shifty) -- (4+\shift,0+\shifty);

\coordinate[label=above:$e_1$] (e13) at (2+\shift,3.464+\shifty);
\coordinate[label=left:$e_2$]  (e23) at (0+\shift,0+\shifty);
\coordinate[label=right:$e_3$] (e33) at (4+\shift,0+\shifty);

\fill   (0+\shift,0+\shifty) circle (\rad) 
        (4+\shift,0+\shifty) circle (\rad)
        (2+\shift,3.464+\shifty) circle (\rad);

\coordinate[label=above:${e_1 e_2 = e_3}$] (base3) at (2,3.464/2+\shifty-0.7);
\draw[{Stealth[black]}-{Stealth[black]}] (6,3.464/2+\shifty)   -- (9,3.464/2+\shifty);
\tikzmath{\shifty =-18;}
\draw [line width=1pt,double] (4+\shift,\shifty) -- (2+\shift,3.464+\shifty);

\coordinate[label=above:$e_1$] (e14) at (2+\shift,3.464+\shifty);
\coordinate[label=left:$e_2$]  (e24) at (0+\shift,0+\shifty);
\coordinate[label=right:$e_3$] (e34) at (4+\shift,0+\shifty);

\fill   (0+\shift,0+\shifty) circle (\rad) 
        (4+\shift,0+\shifty) circle (\rad)
        (2+\shift,3.464+\shifty) circle (\rad);

\coordinate[label=above:${e_2 e_3 = e_1}$] (base4) at (2,3.464/2+\shifty-0.7);
\draw[{Stealth[black]}-{Stealth[black]}] (6,3.464/2+\shifty)   -- (9,3.464/2+\shifty);
\tikzmath{\shifty =-24;}
\draw [line width=1pt,double] (\shift,\shifty) -- (2+\shift,3.464+\shifty)-- (4+\shift,0+\shifty) -- cycle;

\coordinate[label=above:$e_1$] (e14) at (2+\shift,3.464+\shifty);
\coordinate[label=left:$e_2$]  (e24) at (0+\shift,0+\shifty);
\coordinate[label=right:$e_3$] (e34) at (4+\shift,0+\shifty);

\fill   (0+\shift,0+\shifty) circle (\rad) 
        (4+\shift,0+\shifty) circle (\rad)
        (2+\shift,3.464+\shifty) circle (\rad);

\coordinate[label=above:${e_1 = e_2 = e_3}$] (base4) at (2,3.464/2+\shifty-0.7);
\draw[{Stealth[black]}-{Stealth[black]}] (6,3.464/2+\shifty)   -- (9,3.464/2+\shifty);
\end{tikzpicture}
\caption{Diagrammatic illustration of the base used in eq. (\ref{connectiondefinition1nosymm}). The double line means that such pair of vertex are idenfied, according to the second term in (\ref{dp1}).}
\label{base}
\end{figure}
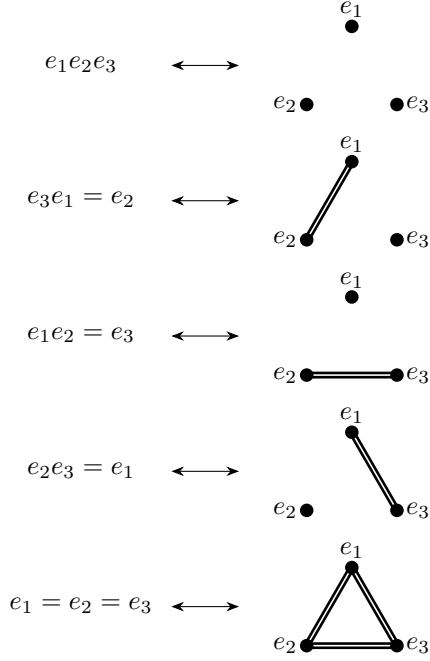

We will denote the lattices we are interested in at step $n$ by $T(n)$. The recursive relation used to build the lattice $T(n+1)$ from three lattices $T(n)$ is such that the three external vertices $e_1$, $e_2$ and $e_3$ remain unvarying in those privileged positions so they will play a special role in the decomposition, see figs. \ref{figsierpinski} and \ref{fighanoi}. Let us suppose that we apply the rules of the dichromatic polynomial to every edge and to every internal vertex of the graph $T(n)$ \textit{but without evaluating the three external vertices} $e_1$, $e_2$ and $e_3$ of $T(n)$\footnote{Namely, when the dichromatic polynomial $K[ \ \bullet \ ]$ acts on some of the three vertices $e_1$, $e_2$ and $e_3$ we will hold the usage of rule in Eq. (\ref{dp3}).}. Five different types of terms arise: in first place there are the terms where none of the marked points $e_1$, $e_2$ and $e_3$ are identified. Secondly, there are the terms in which just two of the three marked points are actually identified (there are three possible pairs: $e_1=e_2$, or $e_1=e_3$, or $e_2=e_3$). The fifth possibility is that all the marked points $e_1$, $e_2$ and $e_3$ are identified. Thus, one can express $K\left[ T(n)\right] $ in the following way
\begin{eqnarray}
K\left[ T(n)\right] &=& x(n) K [e_1e_2e_3] \notag\\
&&+y_1(n) K[ e_1=e_2 e_3] +y_2(n) K[ e_1 e_2= e_3] + y_3(n) K[ e_2 e_3 = e_3]\notag\\
&& +z(n)K[ e_1=e_2=e_3]\,, \label{connectiondefinition1nosymm}
\end{eqnarray}
where the connectivity coefficients are $x(n)$, $y_1(n)$, $y_2(n)$, $y_3(n)$ and $z(n)$. These coefficients multiply the elements of the base of triangular graphs
\begin{align}
 R \ =& \ [e_1 e_2 e_3]\,,\\
 T_1 \ =& \ [e_1  e_2 = e_3]\,,\\
 T_2 \ =& \ [e_2 e_3 = e_1]\,,\\
 T_3 \ =& \ [e_3 e_1 = e_2]\,,\\
 S \ =& \ [e_1 = e_2 = e_3]\,,
\end{align}
see Fig. \ref{base} for the corresponding diagrams.

Next, for the sake of simplicity, we will assume $(2\pi/3)$-rotational symmetry which implies $y_1 = y_2 = y_3 \equiv y$, leading to
\begin{eqnarray}  
K\left[ T(n)\right] &=& x(n) K [e_1e_2e_3] \notag\\
&&+y(n) ( K[ e_1 e_2 = e_3] +K[e_2 e_3 = e_1] +K [ e_3 e_1 = e_2] )  \notag \\
&& +z(n)K[ e_1=e_2=e_3]\,.  \label{connectiondefinition1}
\end{eqnarray}
The relevant connectivity coefficients are $x(n)$, $y(n)$ and $z(n)$. So far, the expression (\ref{connectiondefinition1}) is a generic ansatz for any lattice with the same structure of external points $e_1$, $e_2$ and $e_3$ and symmetry under $(2\pi /3)$-rotations. Of course, the partition function $K(n)\equiv K(T(n))$ and the connectivity coefficients satisfy the following
relation 
\begin{equation}  \label{norm3}
K\left( n\right) =q^{3}x\left( n\right) +3q^{2}y\left( n\right) +q z(n).
\end{equation}
which is just (\ref{connectiondefinition1}) after evaluating the dichromatic polynomial over the external vertices, 
\begin{eqnarray}
&K\left[ e_1\ e_2\ e_3\right] =q^{3}\,,& \\
&K[e_1 e_2 = e_3] =K[ e_2 e_3 = e_1] =K[ e_3 e_1 = e_2] = q^2\,,& \\
&K\left[ e_1=e_2=e_3\right]=q\,.&
\end{eqnarray}

To clarify the above definitions, let us consider the simplest example: the partition function of $T(1)$, a simple triangle, which is a 3-vertex cycle. Using the deletion-contraction algorithm, it is easy to check that the connectivity coefficients of the triangle $T(1)$ are 
\begin{equation}  \label{firsttriangl2}
x(1)=1,\qquad y(1)=v,\qquad z(1)=v^{2}\left( 3+v\right) ,
\end{equation}
thus replacing (\ref{firsttriangl2}) in (\ref{norm3}) one recover the usual expression for the partition function for the periodic chain of three sites $K(1)$.

An heuristic insight can be given about the physical meaning of the connectivity coefficients. In the thermodynamic limit ($n \rightarrow \infty$) one can classify the phases of the system as follows: the range of temperatures in which 
\begin{equation}
\left\vert \frac{y\left( n\right) }{x\left( n\right) }\right\vert \underset{%
n\rightarrow \infty }{\ll }1,\qquad \left\vert \frac{z\left( n\right) }{%
x\left( n\right) }\right\vert \underset{n\rightarrow \infty }{\ll }1,
\label{hight}
\end{equation}%
is a sort of a high temperature phase since, when the above inequalities hold, long range correlations are suppressed. This fact can be inferred from (\ref{connectiondefinition1}), since $x(n)$ is the coefficient which corresponds to the terms of $K(n)$ where the vertices $e_{1}$, $e_{2}$ and $e_{3}$ of $T(n)$ are not identified.

On the other hand, if one has 
\begin{equation}
\left\vert \frac{y\left( n\right) }{x\left( n\right) }\right\vert \underset{%
n\rightarrow \infty }{\gtrapprox }1,\qquad \left\vert \frac{z\left( n\right) 
}{x\left( n\right) }\right\vert \underset{n\rightarrow \infty }{\gtrapprox }%
1,  \label{modt}
\end{equation}%
then (at least) one pair of the external vertices are identified. For definiteness, let us say that $e_{1}$ and $e_{2}$ are identified. Then, by definition of\ $y\left(n\right) $\ and $z\left( n\right) $, this implies that there is (at least)
one path joining $e_{1}$ and $e_{2}$ such that \textit{all} the spins belonging to this path are in the same state (not only the external spins as in the usual restricted partition functions). In particular, the present variables are well suited to detect the arising of frustration where in such case we will have $x(n) \ll y(n)$ and $x(n) \ll z(n)$ \footnote{Actually in order to have frustration is necessary that the number of adjacent spins in the same state grows proportional to the total number of sites $y(n), z(n) \sim V_n$.}.

The next step in the derivation of the dynamical equations is to exploit the self similarity of the Sierpinski gasket in order to derive a set of equations that relate the connectivity coefficients of the step $n+1$ to the ones of the step $n$. Indeed, one could compute directly by ``brute force'' the three geometric coefficients $x(n)$, $y(n)$ and $z(n)$: one should consider the Sierpinski triangle after the $n-$th step and use the basic definitions in Eqs. (\ref{dp1}), (\ref{dp2}
) and (\ref{dp3}) without evaluating the three special points of $T(n)$. However, it is easy to see that in this way the number of terms to compute grows faster than exponentially with the number of edges. The connectivity coefficients in eq.(\ref{connectiondefinition1}) are defined in order to avoid a brute force computation as it will be shown in the next section.

\section{Dynamical equations}\label{sec-patt}

In this section we will show how to express the partition function $K(n+1)$ of $T(n+1)$ and the coefficients $x(n+1)$, $y(n+1)$ and $z(n+1)$ at step $n+1$ in terms of the coefficients $x(n)$, $y(n)$ and $z(n)$ at step $n$. The external vertices of $T(n+1)$ (which will be called $e_1$, $e_2$ and $e_3$) will be the marked points. Each of the three $T(n)-$blocks composing $T(n+1)$ can be further decomposed into five terms according to Eq. (\ref{connectiondefinition1}). Then, one has to collect the terms which multiply: $K\left[ e_1\ e_2\ e_3\right] $ (they will form the coefficient $x(n+1)$); all the terms which multiply $K\left[ e_1=e_2\ e_3\right]$, $K\left[ e_1 \ e_2 = e_3\right]$ or $K\left[ e_2 \ e_3 = e_1\right]$ (they will form the coefficient $y(n+1)$) and all the terms which multiply $K\left[ e_1=e_2=e_3\right]$ (they will form the coefficient $z(n+1)$). After this, one simply needs to compare to the coefficients that multiply the connectivity patterns of the vertices $e_1$, $e_2$ and $e_3$ of $T(n+1)$ (see Fig.\ref{figsierpinski}- Fig.\ref{fighanoi}). The result of this operations is the following discrete dynamical system: 
\begin{eqnarray}
x(n+1)&=&f(x(n),y(n),z(n);q), \label{dyn-sier1}\\
y(n+1)&=&g(x(n),y(n),z(n);q), \label{dyn-sier2}\\
z(n+1)&=&h(x(n),y(n),z(n);q), \label{dyn-sier3}
\end{eqnarray}
with the initial data given by Eqs. (\ref{firsttriangl2}). The functions $f$%
, $g$ and $h$ are homogeneous polynomials of degree $3$ in the dynamical
variables:
\begin{eqnarray}
f(x,y,z;q)&=&q^{3}x^{3}+9q^{2}x^{2}y+3qx^{2}z+24qxy^{2}+12xyz+(14+q)y^{3}+3y^{2}z,\label{f1} \\
g(x,y,z;q) &=&4qy^{3}+q^{2}xy^{2}+7y^{2}z+2qxyz+yz^{2}+xz^{2},  \label{f2} \\
h(x,y,z;q) &=&z^{3}+6yz^{2}+3qy^{2}z.\label{f3}
\end{eqnarray}

\begin{table}[h!]
\begin{center}
\begin{tabular}{cc}
$R\wedge A\wedge B$ & $T_3\wedge A\wedge B$ \\ 
\begin{tabular}{|l|c|c|c|c|c|}
\hline
& $R$ & $T_3$ & $T_1$ & $T_2$ & $S$ \\ \hline
$R$ & $q^3 R$ & $q^2 R$ & $q^2 R$ & $q^2 R$ & $q R$ \\
$T_3$ & $q^2 R$ & $q R$ & $q R$ & $q R$ & $R$ \\
$T_1$ & $q^2 R$ & $q R$ & $q^2 R$ & $q R$ & $q T_1$ \\
$T_2$ & $q^2 R$ & $q R$ & $q R$ & $q R$ & $R$ \\
$S$ & $q R$ & $R$ & $q T_1$ & $R$ & $T_1$ \\ \hline
\end{tabular}
& 
\begin{tabular}{|l|c|c|c|c|c|}
\hline
& $R$ & $T_3$ & $T_1$ & $T_2$ & $S$ \\ \hline
$R$ & $q^2 R$ & $q R$ & $q R$ & $q R$ & $R$ \\
$T_3$ & $q^2 T_3$ & $q T_3$ & $q T_3$ & $q T_3$ & $T_3$ \\
$T_1$ & $q R$ & $R$ & $q T_1$ & $R$ & $T_1$ \\
$T_2$ & $q R$ & $R$ & $q T_2$ & $R$ & $T_2$ \\
$S$ & $R$ & $T_3$ & $q S$ & $T_3$ & $S$ \\ \hline
\end{tabular}%
\end{tabular}
\begin{tabular}{cc}
$T_1\wedge A\wedge B$ & $T_2\wedge A\wedge B$ \\ 
\begin{tabular}{|l|c|c|c|c|c|}
\hline
& $R$ & $T_3$ & $T_1$ & $T_2$ & $S$ \\ \hline
$R$ & $q^2 R$ & $q R$ & $q R$ & $q R$ & $R$ \\
$T_3$ & $q R$ & $R$ & $R$ & $q T_1$ & $T_1$ \\
$T_1$ & $q R$ & $R$ & $q T_1$ & $R$ & $T_1$ \\
$T_2$ & $q R$ & $q R$ & $R$ & $R$ & $R$ \\
$S$ & $R$ & $R$ & $T_1$ & $T_1$ & $T_1$ \\ \hline
\end{tabular}
& 
\begin{tabular}{|l|c|c|c|c|c|}
\hline
& $R$ & $T_3$ & $T_1$ & $T_2$ & $S$ \\ \hline
$R$ & $q^2 R$ & $q R$ & $q R$ & $q^2 T_2$ & $q T_2$ \\
$T_3$ & $q R$ & $R$ & $R$ & $q T_2$ & $T_2$ \\
$T_1$ & $q R$ & $q T_3$ & $q T_1$ & $q T_2$ & $q S$ \\
$T_2$ & $q R$ & $R$ & $R$ & $q T_2$ & $T_2$ \\
$S$ & $R$ & $T_3$ & $T_1$ & $T_2$ & $S$ \\ \hline
\end{tabular}%
\end{tabular}
\begin{tabular}{c}
$S\wedge A\wedge B$ \\ 
\begin{tabular}{|l|c|c|c|c|c|}
\hline
& $R$ & $T_3$ & $T_1$ & $T_2$ & $S$ \\ \hline
$R$ & $q R$ & $R$ & $R$ & $q T_2$ & $T_2$ \\
$T_3$ & $q T_3$ & $T_3$ & $T_3$ & $q S$ & $S$ \\
$T_1$ & $R$ & $T_3$ & $T_1$ & $T_2$ & $S$ \\
$T_2$ & $R$ & $R$ & $R$ & $T_2$ & $T_2$ \\
$S$ & $T_3$ & $T_3$ & $S$ & $S$ & $S$ \\ \hline
\end{tabular}%
\end{tabular}%
\end{center}
\caption{Sierpinski composition of basis elements. The connectivity patterns of the external vertices can be coded in an algebraic structure. The tables should be read as $A\in1^{\mathrm{st}}$column and $B\in1^{\mathrm{st}}$row. See Fig. \ref{composition} for a diagram of the graph $A \wedge B \wedge C$. This table correspond to the relevant compositions appearing in the Sierpinski recursive relation of Figure \ref{figsierpinski}. When computing this table we need to evaluate $K[ A \wedge B \wedge C ]$ applying rules (\ref{dp1}), (\ref{dp2}) and (\ref{dp3}) to every edge and bulk vertices ($b_i$), but holding the rules when acting on external vertices ($e_i$).}\label{connpattable}
\end{table}

\begin{figure}
\centering
\begin{tikzpicture}[scale=0.7]
\tikzmath{\shift = 8; \rad=0.15;}

\draw [line width=1pt, fill=gray!50] (1,3.464/2) -- (2,3.464) -- (3,3.464/2) -- cycle;
\draw [line width=1pt, fill=gray!50] (0,0) -- (1,3.464/2) -- (2,0) -- cycle;
\draw [line width=1pt, fill=gray!50] (2,0) -- (3,3.464/2) -- (4,0) -- cycle;

\coordinate[label=above:$e_1$] (e1) at (2,3.464);
\coordinate[label=left:$e_2$]  (e2) at (0,0);
\coordinate[label=right:$e_3$] (e3) at (4,0);

\coordinate[label=left:$b_1$] (b1) at (1,3.464/2);
\coordinate[label=below:$b_2$]  (b2) at (2,0);
\coordinate[label=right:$b_3$] (b3) at (3,3.464/2);

\coordinate[label=above:$A$] (A) at (2,3.464*3/4-0.6);
\coordinate[label=above:$B$] (B) at (1,3.464*1/4-0.6);
\coordinate[label=above:$C$] (C) at (3,3.464*1/4-0.6);

\fill   (0,0) circle (\rad) 
        (4,0) circle (\rad)
        (2,3.464) circle (\rad)
        (1,3.464/2) circle (\rad) 
        (2,0) circle (\rad)
        (3,3.464/2) circle (\rad);

\end{tikzpicture}
\caption{Sierpinski composition of three graphs $A \wedge B \wedge C$. The products summarized in Table \ref{connpattable}are understood according to this diagram. }\label{composition}
\end{figure}
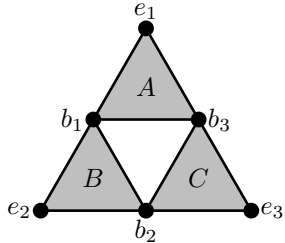

It is, perhaps, more satisfactory to express the recursive composition of graphs in an algebraic manner, by defining
\begin{equation}
 T(n+1) = T(n) \wedge T(n) \wedge T(n)\,,
\end{equation}
where the meaning of $A \wedge B \wedge C$ is defined in Fig. \ref{composition}. Now, expanding the graphs as follows,
\begin{equation}
 T(n) = x(n) R + y_1(n) T_1+ y_2(n) T_2+ y_3(n) T_3 + z(n) S\,,
\end{equation}
we can deduce the equations for $x(n+1)$, $y(n+1)$ and $z(n+1)$ from
\begin{align}
 T(n+1) \dot{=}& (x(n) R + y_1(n) T_1+ y_2(n) T_2+ y_3(n) T_3 + z(n) S)  \notag\\
 &\wedge(x(n) R + y_1(n) T_1+ y_2(n) T_2+ y_3(n) T_3 + z(n) S)\notag\\
 &\wedge (x(n) R + y_1(n) T_1+ y_2(n) T_2+ y_3(n) T_3 + z(n) S) \label{recursiverelation}\,.
\end{align}
where the notation $\dot{=}$ means that the equality is valid at the level of the polynomials, that is after applying $K[ \cdot ]$ to both sides of the equation. When computing $K[ A \wedge B \wedge C ]$ we need to hold the application of rules (\ref{dp1}), (\ref{dp2}) and (\ref{dp3}) to the external vertices ($e_i$) but release the application of the rules when acting on bulk vertices ($b_i$). See Fig. \ref{composition} for a diagram clarifying the meaning of of the graph $A \wedge B \wedge C$. All we need to know in order to deduce the dynamical equations is the table of composition rules $A \wedge B \wedge C$. This composition is associative and all the possible ways that the elements of the basis combine for the Sierpinski iteration (see Fig. \ref{figsierpinski}) are summarized in Table \ref{connpattable}.

In this framework, the discrete evolution parameter $n$ of the dynamical system (which is the recursive label of the Sierpinski triangles $T(n)$) corresponds to the hierarchical size of the Sierpinski gasket $T(n)$. Hence, the thermodynamical limit corresponds to the asymptotic analysis of the dynamical system in eqs. (\ref{dyn-sier1})-(\ref{dyn-sier3}) together with the initial data in eq. (\ref{firsttriangl2}) when $n\rightarrow \infty $.

It is also worth noting that the temperature does not enter directly the dynamical system for the Sierpinski composition: in the present scheme, the dependence on the temperature only appears in the initial data through $v$. On the other hand, the parameter $q$ of the Potts model enters directly the equations of the dynamical system i.e. eqs. (\ref{f1}), (\ref{f2}) and (\ref{f3}). However, the value of $q$ is arbitrary and does not even has to be an integer, meaning that in the subsequent analysis $q$ can be promoted to take real or complex values. This is great advantage of the present method in comparison to RG in real space. Another advantage with respect to real space RG techiniques is that the composition rules do not depend on the value of $q$, that is we do not need to worry about the combinatorics that appear when combining restricted partition functions. Let us comment that non-integer values of $q$ have also been considered before \cite{Wagneretal2001}

From the computational point of view, if one would compute by ``brute force'' the partition function of the Potts model on the Sierpinski gasket after $n$ steps, one should compute $2^{E_n}$ terms using the definitions in Eqs. (\ref{dp1}), (\ref{dp2}) and (\ref{dp3}), where $E_n$ is the number of edges of the $n$-step Sierpinski triangle. Therefore, roughly speaking, one should compute $\sim 2^{3^n}$ terms after $n$ steps. On the other hand, the derivation of the dynamical system can be achieved from the recursive relation (\ref{recursiverelation}) that requires to perform just $B^3$ algebraic operations where $B$ is the size of the basis of connectivity patterns. For the triangle-shaped lattices $B=5$ and all the possible compositions are summarized in the 125 elements of Table \ref{connpattable}.

\subsection{Dynamical equations on Hanoi graph}\label{sectiondynhanoi}

Now we will apply the method to another interesting class of triangular self-similar lattices known in the literature as Hanoi graph. Hanoi graph is defined recursively, see figure \ref{fighanoi}. This class of lattices will be denoted as $T_h (n)$ while, as in the Sierpinski case, we also have three external vertices that will be denoted as $e_1$, $e_2$ and $e_3$. The geometrical structure of the recursive transformation from the step $n$ to the step $n+1$ is quite similar to the one of the Sierpinski lattice but there is an important difference: the lattice $T_h(n+1)$ can be see as three lattices $T_h(n)$ glued in such a way that the three triangles $T_h(n)$ constituting $T_h(n+1)$ are pairwise joined to each other trough an edge.

\begin{table}[h!]
\begin{center}
\resizebox{\columnwidth}{!}{%
\begin{tabular}{c}
$R\wedge A\wedge B$ \\ 
\begin{tabular}{|l|c|c|c|c|c|}
\hline
& $R$ & $T_3$ & $T_1$ & $T_2$ & $S$ \\ \hline
$R$ & $q^3 \qv^3 R$ & $q^2 \qv^3 R$ & $q^2 \qv^3 R$ & $q^2 \qv^3 R$ & $q \qv^3 R$ \\
$T_3$ & $q^2 \qv^3 R$ & $q \qv^3 R$ & $q \qv^3 R$ & $q \qv^3 R$ & $\qv^3 R$ \\ 
$T_1$ & $q^2 \qv^3 R$ & $q \qv^3  R$ & $q^2\qv^2( R+v T_1)$ & $q \qv^3  R$ & $q\qv^2( R+v T_1)$ \\
$T_2$ & $q^2 \qv^3 R$ & $q \qv^3 R$ & $q \qv^3 R$ & $q \qv^3 R$ & $\qv^3 R$ \\
$S$ & $q \qv^3 R$ & $\qv^3 R$ & $q\qv^2 (R+v T_1)$ & $\qv^3 R$ & $\qv^2(R+v T_1)$ \\ \hline
\end{tabular}\\
\\
$T_3\wedge A\wedge B$ \\
\begin{tabular}{|l|c|c|c|c|c|}
\hline
& $R$ & $T_3$ & $T_1$ & $T_2$ & $S$ \\ \hline
$R$  & $q^2 \qv^3 R$ & $q \qv^3 R$ & $q \qv^3 R$ & $q \qv^3 R$ & $\qv^3 R$ \\ 
$T_3$ & $q^2 \qv^2 (R+vT_3)$ & $q \qv^2 (R+v T_3)$ & $q \qv^2 (R+v T_3)$ & $q \qv^2 (R+v T_3)$ & $\qv^2 (R +v T_3)$ \\
$T_1$ & $q\qv^3 R$ & $q\qv^2 R$ & $q \qv^2 (R+v T_1)$ & $\qv^3 R$ & $q \qv^2 (R+v T_1)$ \\ 
$T_2$ & $q\qv^3 R$ & $\qv^3 R$ & $q\qv((1+2v)R+v^2 T_2)$ & $\qv^3 R$ & $\qv ((1+2v)R+v^2T_2)$ \\ 
$S$ & $q \qv^2 (R+v T_3)$ & $\qv^2 (R+vT_3)$ & $\qv (R+vT_2+vT_3+v^2S)$ & $\qv^2 (R+vT_3)$ & $\qv (R+vT_1+vT_3+v^2S)$ \\ \hline
\end{tabular}\\
\\
$T_1\wedge A\wedge B$\\ 
\begin{tabular}{|l|c|c|c|c|c|}
\hline
& $R$ & $T_3$ & $T_1$ & $T_2$ & $S$ \\ \hline
$R$ & $q^2\qv^3 R$ & $q\qv^3 R$ & $q\qv^3 R$ & $q\qv^3 R$ & $\qv^3 R$ \\
$T_3$ & $q\qv^3 R$ & $\qv^3R$ & $\qv^3R$ & $q\qv((1+2v)R+v^2T_1)$ & $\qv((1+2v)R+v^2T_1)$ \\
$T_1$ & $q\qv^3 R$ & $\qv^3R$ & $q \qv^2(R+v T_1)$ & $\qv^3R$ & $\qv^2(R+vT_1)$ \\
$T_2$ & $q\qv^3 R$ & $U R$ & $\qv^3 R$ & $\qv^3 R$ & $U R$ \\
$S$ & $\qv^3 R$ & $U R$ & $\qv^2(R+vT_1)$ & $\qv ((1+2v)R+v^2T_1)$ & $(\qv+v)R+(\qv v+v^2(2+v))T_1$ \\ \hline
\end{tabular}\\
\\
$T_2\wedge A\wedge B$ \\
\begin{tabular}{|l|c|c|c|c|c|}
\hline
& $R$ & $T_3$ & $T_1$ & $T_2$ & $S$ \\ \hline
$R$ & $q^2\qv^3 R$ & $q\qv^3 R$ & $q\qv^3 R$ & $q^2\qv^2(R+v T_2)$ & $q\qv^2(R+v T_2)$ \\
$T_3$ & $q\qv^3 R$ & $\qv^3R$ & $\qv^3R$ & $q\qv^2(R+v T_2)$ & $\qv^2(R+vT_2)$ \\
$T_1$ & $q\qv^3 R$ & $q\qv ((1+2v)R+v^2 T_3)$ & $q\qv^2 ((1+2v)R+v T_1)$ & $q\qv^2 (R+v T_2)$ & $q\qv (R+v T_1+v T_2+v^2S)$ \\
$T_2$ & $q\qv^3 R$ & $\qv^3R$ & $\qv^3R$ & $q\qv^2(R+v T_2)$ & $\qv^2(R+vT_2)$ \\
$S$ & $\qv^3R$ & $\qv ((1+2v)R+v^2 T_1)$ & $\qv^2(R+vT_1)$ & $\qv^2(R+vT_2)$ & $\qv (R+v T_1+v T_2+v^2S)$ \\ \hline
\end{tabular}\\
\\
$S\wedge A\wedge B$ \\ 
\begin{tabular}{|l|c|c|c|c|c|}
\hline
& $R$ & $T_3$ & $T_1$ & $T_2$ & $S$ \\ \hline
$R$ & $q\qv^3 R$ & $\qv^3R$ & $\qv^3R$ & $q\qv^2(R+v T_2)$ & $\qv^2(R+vT_2)$ \\
$T_3$ & $q\qv^2(R+v T_3)$ & $\qv^2(R+vT_3)$ & $\qv^2(R+vT_3)$ & $q\qv(R+vT_2+vT_3+v^2 S)$ & $\qv(R+vT_2+vT_3+v^2 S)$ \\
$T_1$ & $\qv^3R$ & $\qv((1+2v)R+v^2T_3)$ & $\qv^2(R+vT_1)$ & $\qv^2(R+vT_2)$ & $\qv(R+vT_1+vT_2+v^2S)$ \\
$T_2$ & $\qv^3R$ & $U R$ & $\qv^3 R$ & $\qv^2(R+vT_2)$ & $(\qv+v)R+(\qv v+v^2(2+v))T_2$ \\
$S$ & $\qv^2(R+vT_3)$ & $(\qv+v)R+(\qv v+v^2(2+v))T_3$ & $\qv (R+vT_1+vT_3+v^2S)$ & $\qv (R+vT_2+vT_3+v^2S)$ & $R+vT_1+vT_2+vT_3+v^2(3+v)S$ \\ \hline
\end{tabular}
\end{tabular}
}
\end{center}
\caption{``Hanoi'' composition of basis elements. The tables should be read as $A\in1^{\mathrm{st}}$column and $B\in1^{\mathrm{st}}$row. See Fig. \ref{compositionhanoi} for a diagram of the graph $A \wedge B \wedge C$. This table correspond to the relevant compositions appearing in the Hanoi recursive relation of Figure \ref{compositionhanoi}. When computing this table we need to evaluate $K[ A \wedge B \wedge C ]$ applying rules (\ref{dp1}), (\ref{dp2}) and (\ref{dp3}) to every edge and bulk vertices ($b_i$), but holding the rules when acting on external vertices ($e_i$). Here we defined the shorthand notations $\qv=q+v$ and $U=q\qv^2+qv\qv+qv^3(1+v)$. $U$ is the dichromatic polynomial of the 3-vertex cycle.}\label{connpattablehanoi}
\end{table}

\begin{figure}
\centering
\begin{tikzpicture}[scale=0.7]
\tikzmath{\shiftx = 1;\shifty = 3.464/2; \rad=0.15;}

\draw [line width=1pt, fill=gray!50] (1,3.464/2) -- (2,3.464) -- (3,3.464/2) -- cycle;
\draw [line width=1pt, fill=gray!50] (0-\shiftx,0-\shifty) -- (1-\shiftx,3.464/2-\shifty) -- (2-\shiftx,0-\shifty) -- cycle;
\draw [line width=1pt, fill=gray!50] (2+\shiftx,0-\shifty) -- (3+\shiftx,3.464/2-\shifty) -- (4+\shiftx,0-\shifty) -- cycle;

\draw [line width=1pt, fill=gray!50] (1,3.464/2) -- (0,0);
\draw [line width=1pt, fill=gray!50] (1,-3.464/2) -- (3,-3.464/2);
\draw [line width=1pt, fill=gray!50] (3,3.464/2) -- (4,0);

\coordinate[label=above:$e_1$] (e1) at (2,3.464);
\coordinate[label=left:$e_2$]  (e2) at (0-\shiftx,0-\shifty);
\coordinate[label=right:$e_3$] (e3) at (4+\shiftx,0-\shifty);

\coordinate[label=left:$b_1$] (b1) at (1,3.464/2);
\coordinate[label=left:$b_2$]  (b2) at (0,0);
\coordinate[label=below:$b_3$]  (b3) at (2-\shiftx,0-\shifty);
\coordinate[label=below:$b_4$]  (b4) at (2+\shiftx,0-\shifty);
\coordinate[label=right:$b_5$] (b5) at (4,0);
\coordinate[label=right:$b_6$] (b6) at (3,3.464/2);

\coordinate[label=above:$A$] (A) at (2,3.464*3/4-0.6);
\coordinate[label=above:$B$] (B) at (1-\shiftx,3.464*1/4-0.6-\shifty);
\coordinate[label=above:$C$] (C) at (3+\shiftx,3.464*1/4-0.6-\shifty);

\fill   (2,3.464) circle (\rad)
        (0-\shiftx,0-\shifty) circle (\rad)
        (4+\shiftx,0-\shifty) circle (\rad)
        
        (1,3.464/2) circle (\rad)
        (0,0) circle (\rad)
        (2-\shiftx,0-\shifty) circle (\rad)
        (2+\shiftx,0-\shifty) circle (\rad)
        (4,0) circle (\rad)
        (3,3.464/2) circle (\rad);

\end{tikzpicture}
\caption{``Hanoi'' composition of three graphs $A \wedge B \wedge C$. The products summarized in Table \ref{connpattablehanoi} are understood according to this diagram. }\label{compositionhanoi}
\end{figure}
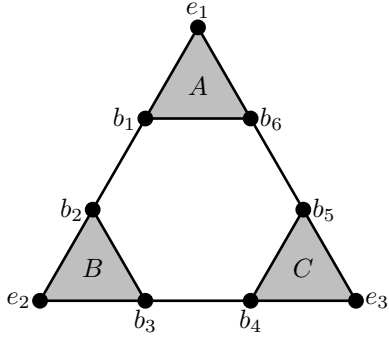

Following the general procedure explained in sections \ref{sec-coeff} and \ref{sec-patt}, we expand the partition function of $K_h(n) \equiv K_h\left[ T(n)\right]$ in terms of the connectivity coefficients as follows, 
\begin{eqnarray}  
K_h (n) &=& x(n) K_h [e_1e_2e_3] \notag\\
&&+y(n) ( K_h[ e_1 e_2 = e_3] +K_h[e_2 e_3 = e_1] +K_h [ e_3 e_1 = e_2] )  \notag \\
&& +z(n)K_h[ e_1=e_2=e_3]\,.  \label{connectiondefinition2}
\end{eqnarray}

This definition allows one to express the connection coefficients $x(n+1)$, $y(n+1)$ and $z(n+1)$ of $T_{h}\left( n+1\right) $ in terms of the coefficients $x(n)$, $y(n)$ and $z(n)$ of the step $n$ along the same line used to analyze the Sierpinski gasket. The resulting dynamical system is the following
\begin{align}
x(n+1) =& f_h(x(n),y(n),z(n);q,v)\,, \label{dyn-hanoi1}\\
y(n+1) =& g_h(x(n),y(n),z(n);q,v)\,, \label{dyn-hanoi2}\\
z(n+1) =& h_h(x(n),y(n),z(n);q,v)\,, \label{dyn-hanoi3}
\end{align}
the corresponding initial data could be chosen again as a simple 3-vertex cycle, given in (\ref{firsttriangl2}). The functions $f_h$, $g_h$ and $h_h$ are homogeneous polynomials of degree $3$ and, in contrast with the ones for the Sierpinski lattice, they include explicit dependence on the temperature via the $v$ parameter, their explicit expressions are lengthy and can be written in the following way 
\begin{align}
f_h(x,y,z;q,v) =& \left( \frac{k}{q}\right) ^{3}+3v\frac{k w_1}{q}\left( \frac{k}{q}+w_2\right)\notag\\
&+ 3v^2\left( x w_2^2+ w_1 ^2\left( \frac{k}{q}+2 w_2\right) \right)+v^3f(x,y,z;q)\,,  \label{hanoifirst}\\
g_h(x,y,z;q,v) =& v\frac{k w_2^2}{q}+v^2 w_2^2\left( 3y+2w_1\right) +v^3g(x,y,z;q)\,,\label{hanoisecond} \\
h_{h}(x,y,z;q,v) = & 3v^2 z w_2 ^2+v^3 h(x,y,z;q)\,.\label{hanoithird}
\end{align}

The functions $f(x,y,z;q)$, $g(x,y,z;q)$ and $h(x,y,z;q)$ on the right hand side of eqs. (\ref{hanoifirst}), (\ref{hanoisecond}) and (\ref{hanoithird}) are the corresponding functions for the Sierpinski system, defined in Eqs. (\ref{f1}), (\ref{f2}) and (\ref{f3}). Also we have introduced the shorthand notation $k = q^3x+3q^2 y +q z$, $w_1= x+2y$ and $w_2=qy+z$.

The Hanoi system can be deduced using what we know already from the Sierpinski system by noting the following:
\begin{itemize}
 \item Besides the terms which are similar to the corresponding terms of the Sierpinski gasket multiplied by $v^3$, in eq. (\ref{hanoifirst}) for $x(n+1)$ there are three new terms. The first term corresponds to the case in which all the three edges of $T_h(n+1)$ have been deleted. The second term is proportional to $v$ and corresponds to the case in which two edges of $T_h(n+1)$ have been deleted while one edge has been contracted. The third term is proportional to $v^2$ and corresponds to the case in which two edges of $T_h(n+1)$ have been contracted while one has been deleted.
 \item In Eq. (\ref{hanoisecond}) for $y(n+1)$ there are two new terms. The first term is proportional to $v$ and corresponds to the case in which two edges of $T_h(n+1)$ have been deleted while one edge has been contracted. The second new term is proportional to $v^2$ and corresponds to the case in which two edges of $T_h(n+1)$ have been contracted while one has been deleted.
 \item In Eq. (\ref{hanoithird}) for $z(n+1)$ there is only one new term: it is proportional to $v^2$ and corresponds to the case in which two edges of $T_h(n+1)$ have been contracted while one has been deleted.
\end{itemize}

Otherwise we can also deduce the dynamical system for the Hanoi lattice directly, by using the algebraic structure provided by the Hanoi composition rule, see Fig. \ref{compositionhanoi}, and a table with the composition products, see Table \ref{connpattablehanoi}.

\subsection{Thermodynamic functions}\label{sectionthermo}

The finite iteration of the system (\ref{dyn-sier1})-(\ref{dyn-sier3}) provide us with the dichromatic polynomial $K(n)$ of the Sierpinski triangle which is the partition function of the Potts model in such a lattice, $Z_n = K(n)$. Similarly, the system (\ref{dyn-hanoi1})-(\ref{dyn-hanoi3}) allows us with the partition function of the Potts model in the Hanoi graph.

It will be relevant to consider the thermodynamical functions per site. The free energy is defined by
\begin{equation}
 f_n \equiv -\frac{T}{V_n} \log Z_n\,,
\end{equation}
where $V_n$ is the number of sites of the lattice at hierarchical step $n$. The internal energy is defined by
\begin{equation}
 u_n = f_n-T \frac{\partial f_n}{\partial T}\,,
\end{equation}
the entropy
\begin{equation}
 s_n = \frac{1}{T}(u_n-f_n)\,,
\end{equation}
and the heat capacity
\begin{equation}
 c_n = \frac{\partial u_n}{\partial T}\,.
\end{equation}

\begin{figure}
  \centering
  \begin{subfigure}{\textwidth}
  \begin{overpic}[scale=0.19]{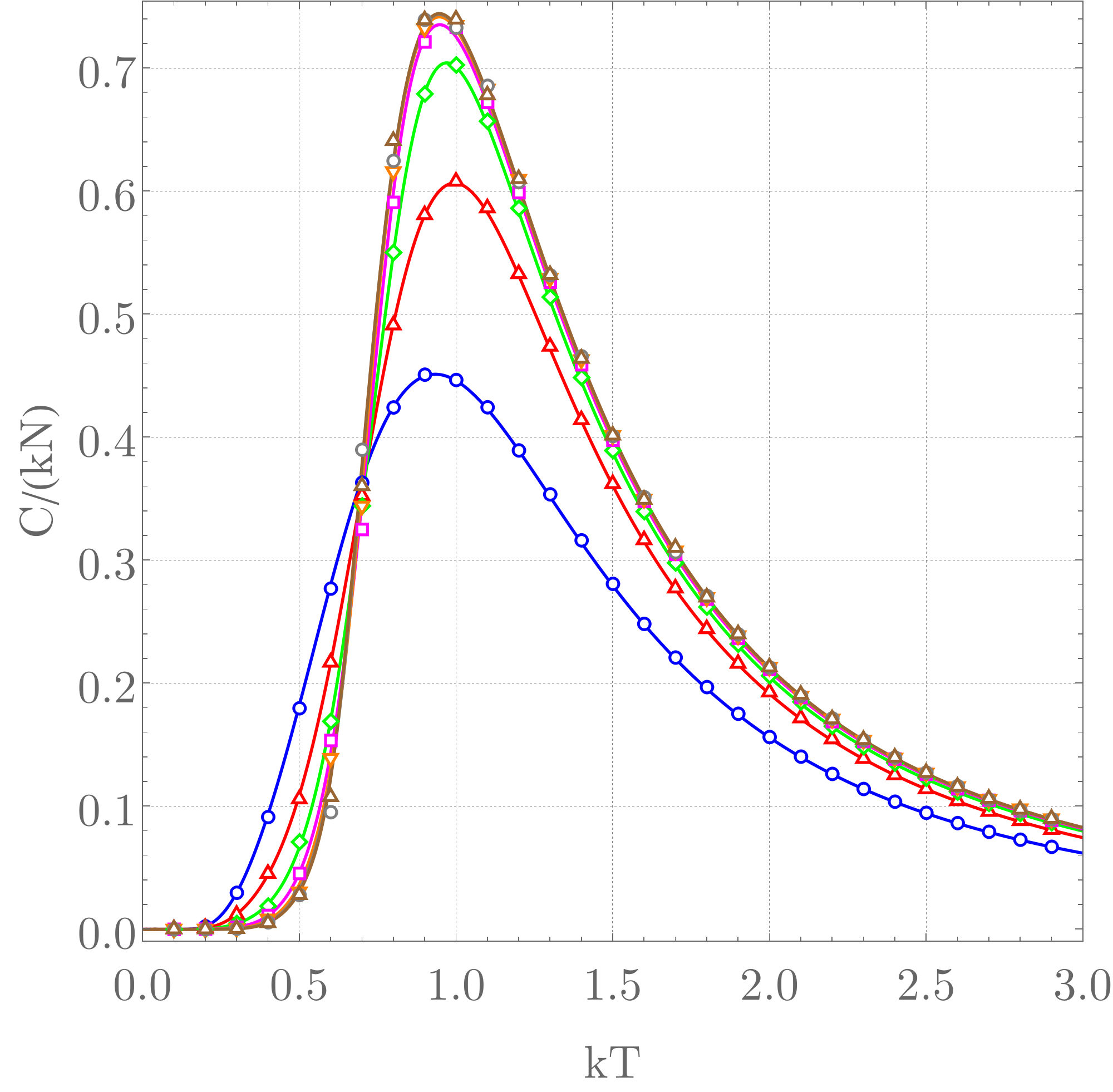}
     \put(75,50){\includegraphics[scale=0.25]{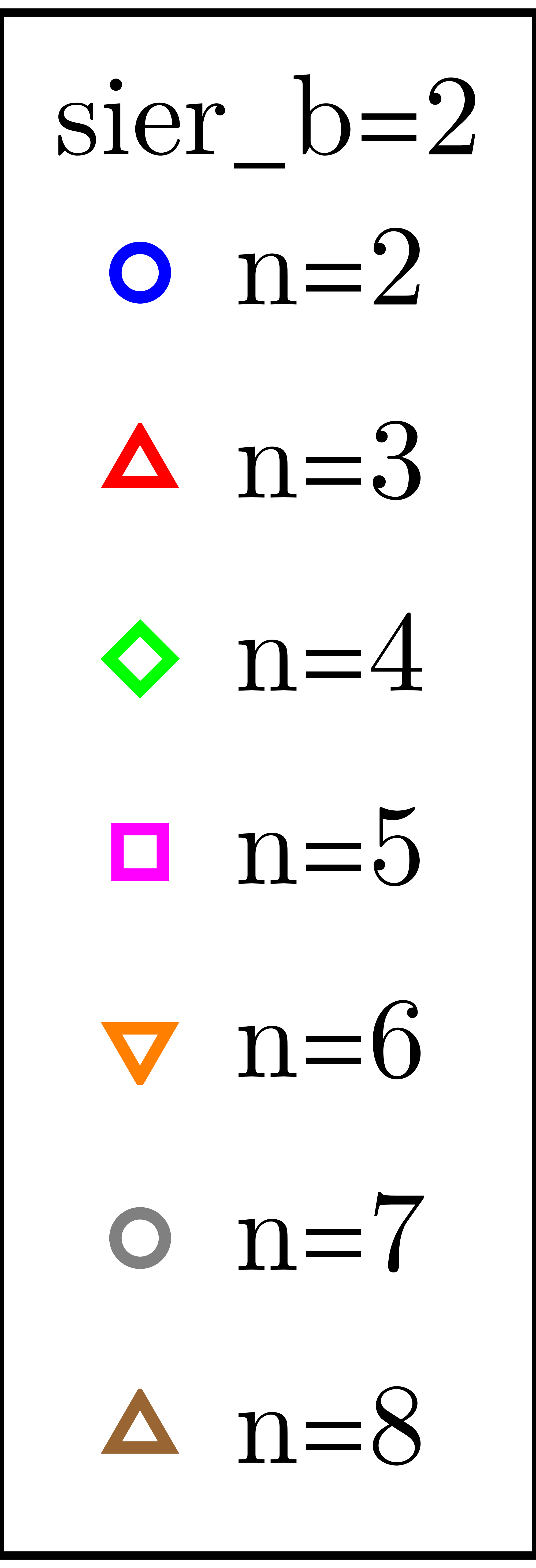}}
  \end{overpic}
  \begin{overpic}[scale=0.19]{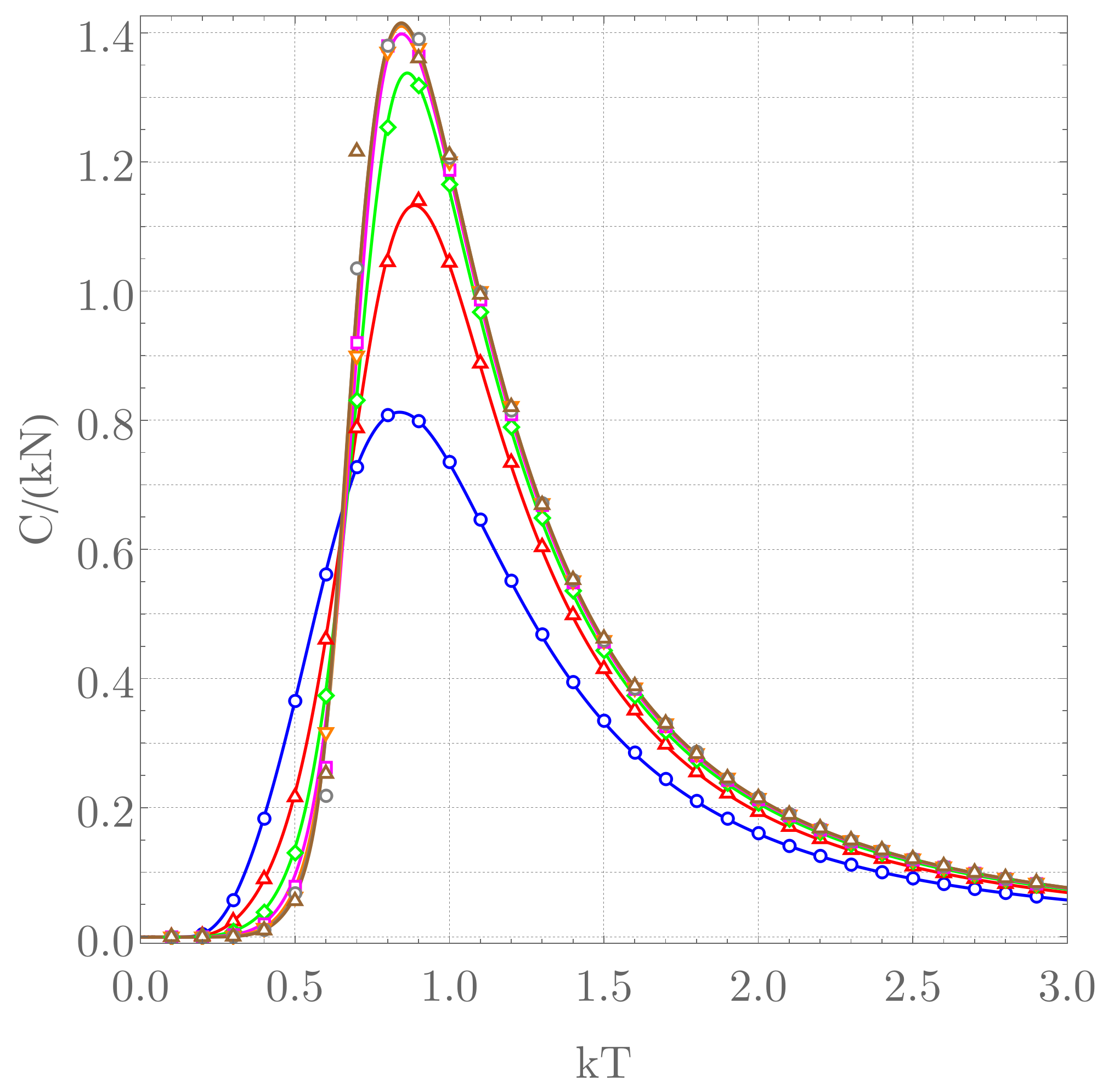}
     \put(75,50){\includegraphics[scale=0.25]{{plot_legend_sier_b=2}.png}}
  \end{overpic}
  \begin{overpic}[scale=0.19]{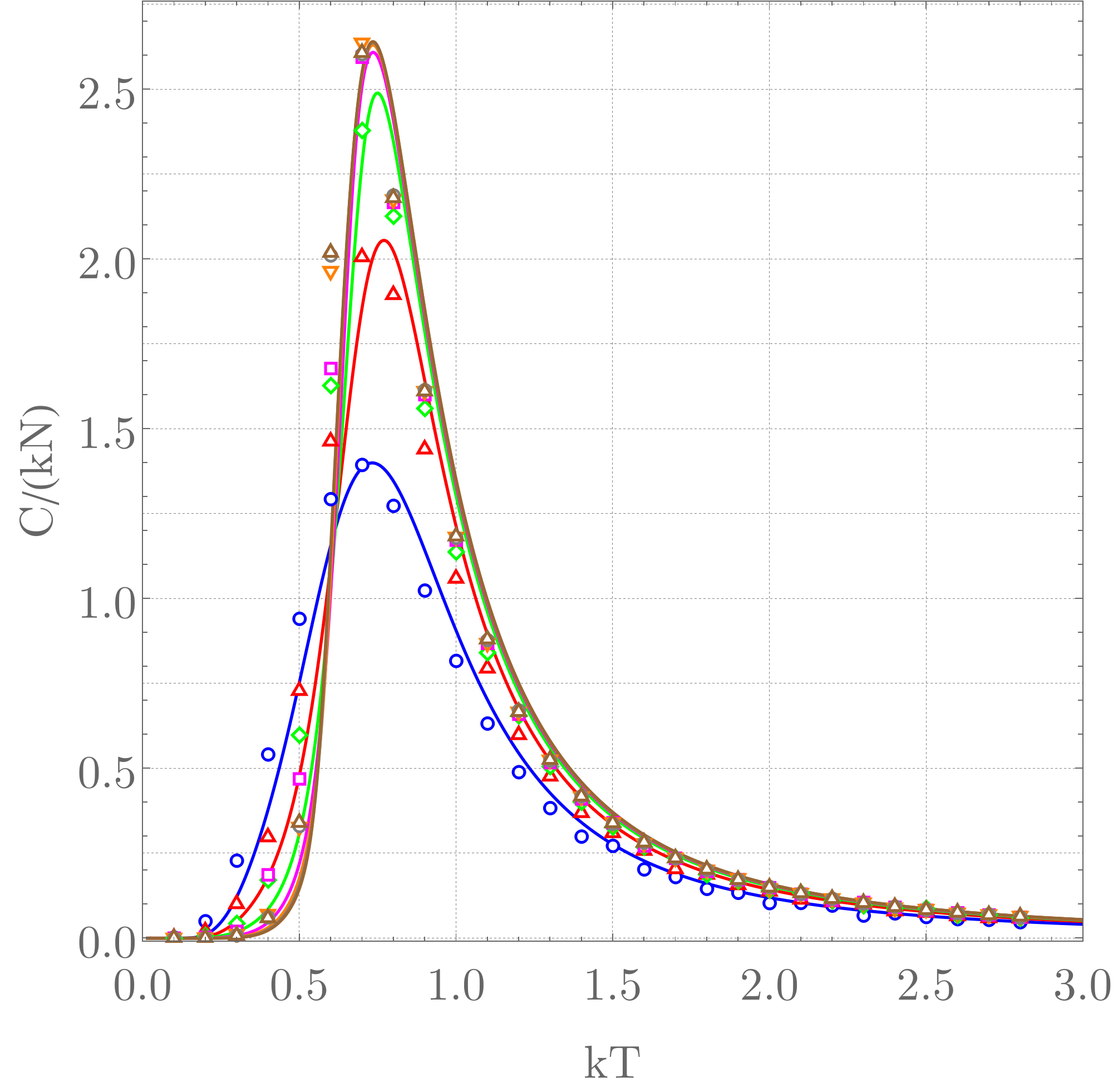}
     \put(75,50){\includegraphics[scale=0.25]{{plot_legend_sier_b=2}.png}}
  \end{overpic}
  \caption{Specific heat in the ferromagnetic regime for $q=2$ (left panel), $q=3$ (middle panel) and $q=5$ (right panel).}
  \label{fig:sfig1shsier}
\end{subfigure}
\begin{subfigure}{\textwidth}
  \begin{overpic}[scale=0.19]{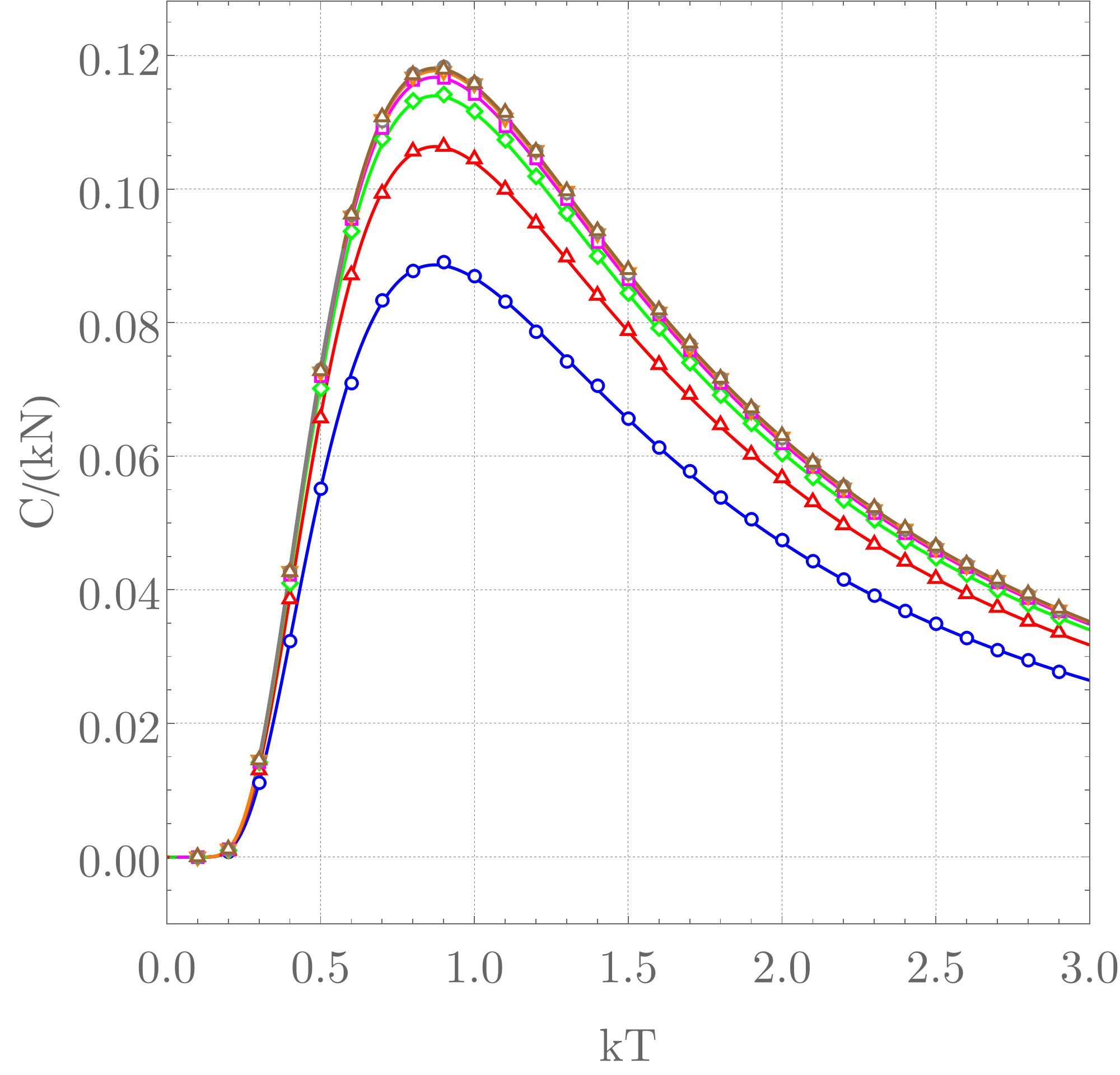}
     \put(75,50){\includegraphics[scale=0.25]{{plot_legend_sier_b=2}.png}}
  \end{overpic}
  \begin{overpic}[scale=0.19]{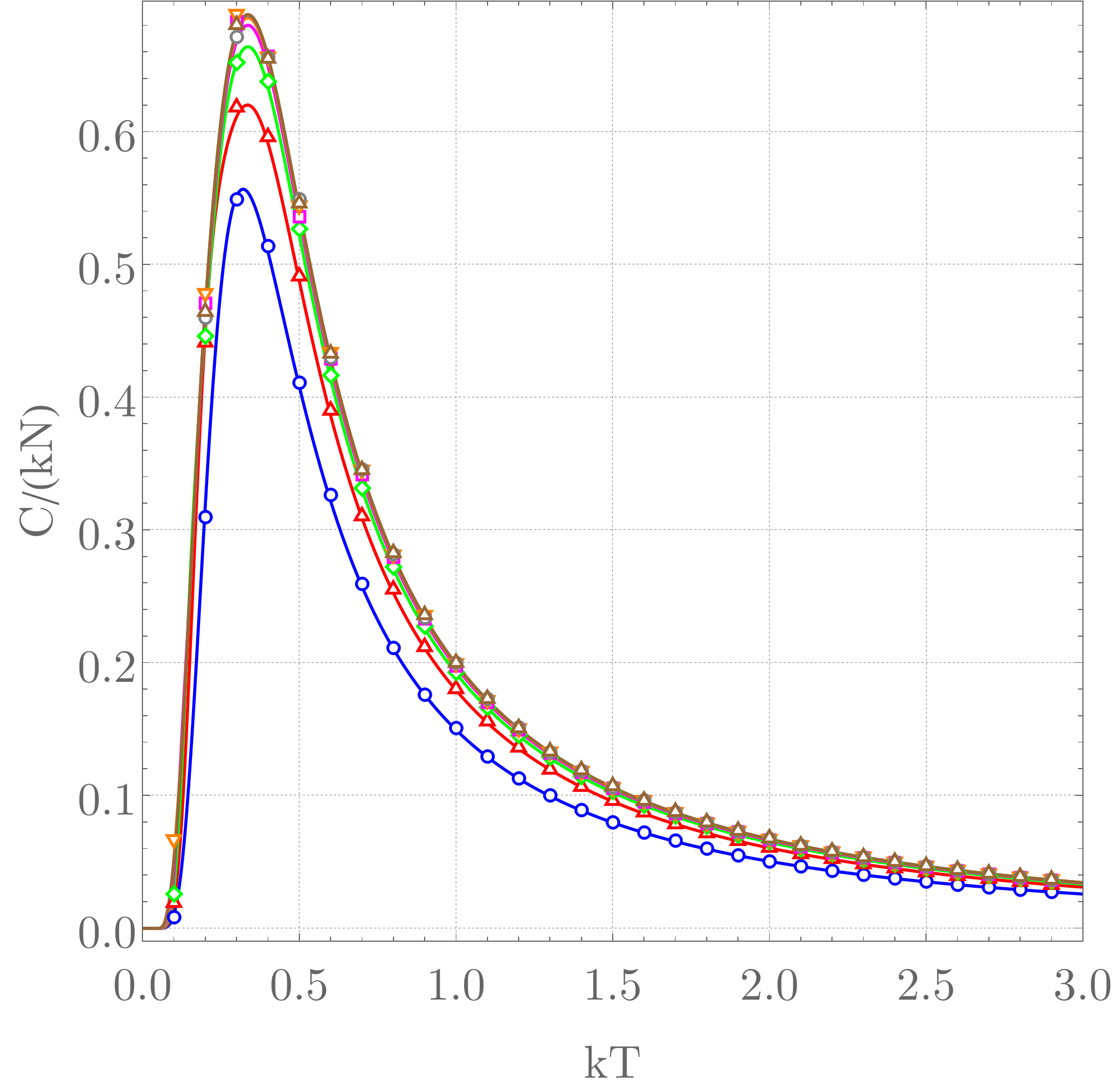}
     \put(75,50){\includegraphics[scale=0.25]{{plot_legend_sier_b=2}.png}}
  \end{overpic}
  \begin{overpic}[scale=0.19]{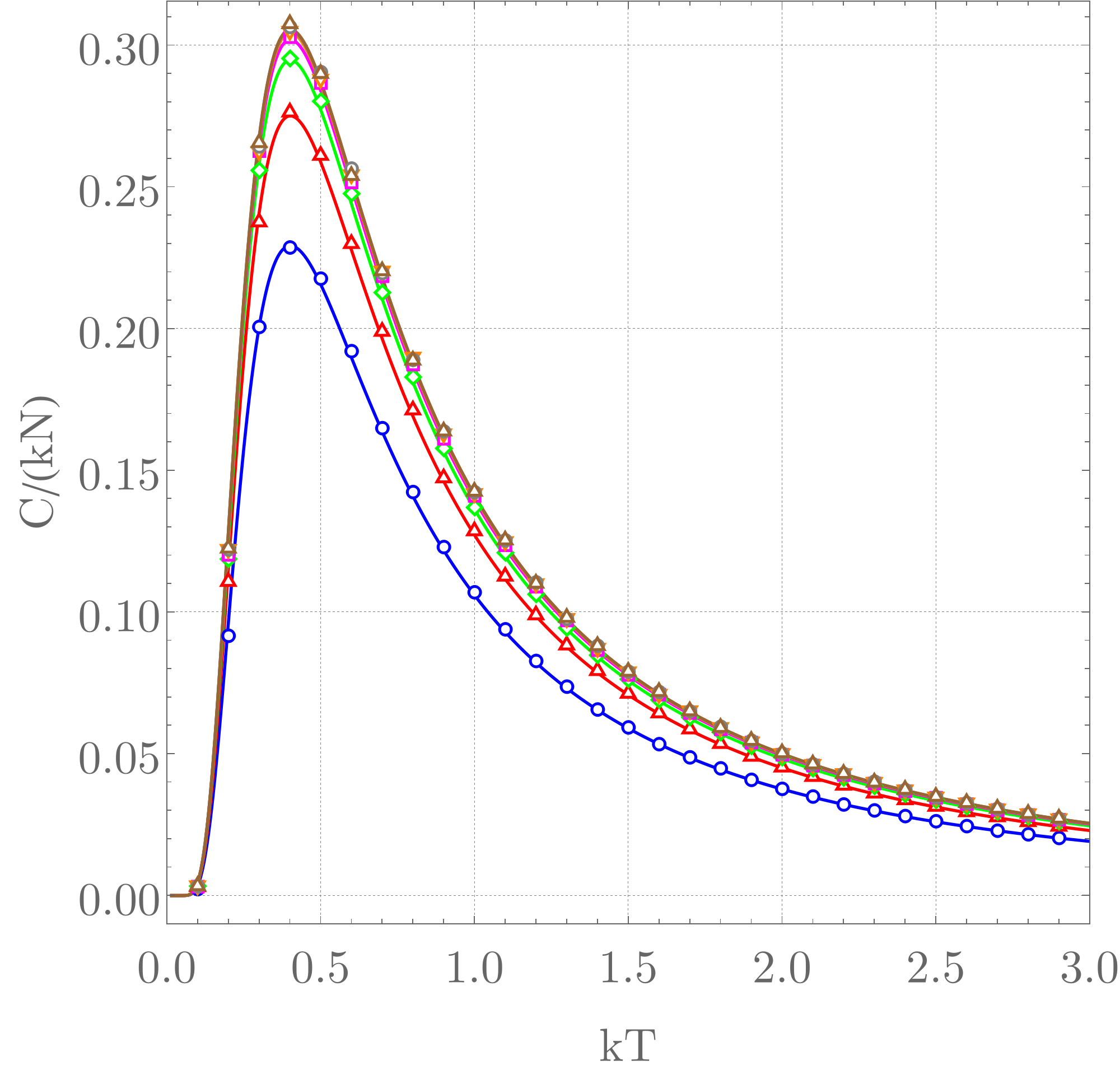}
     \put(75,50){\includegraphics[scale=0.25]{{plot_legend_sier_b=2}.png}}
  \end{overpic}
  \caption{Specific heat in the anti-ferromagnetic regime for $q=2$ (left panel), $q=3$ (middle panel) and $q=5$ (right panel).}
  \label{fig:sfig2shsier}
\end{subfigure}
\caption{Comparison with Metropolis Monte Carlo results for the Sierpinski lattice. Continuous line is the exact result from the dynamical system and the points are the results from simulations. In the metropolis algorithm we used (1000, 1000, 1000, 1230, 3660, 10950, 32820) steps for burnout, (400000, 400000, 400000, 400000, 732000, 2190000, 6564000) sampling steps and we swept (40, 60, 100, 100, 100, 100, 100) steps to grant statistical independence, all these for $n=(2,3,4,5,6,7,8)$ respectively.}\label{metropolissier}
\end{figure}

\begin{figure}
  \centering
  \begin{subfigure}{\textwidth}
  \begin{overpic}[scale=0.19]{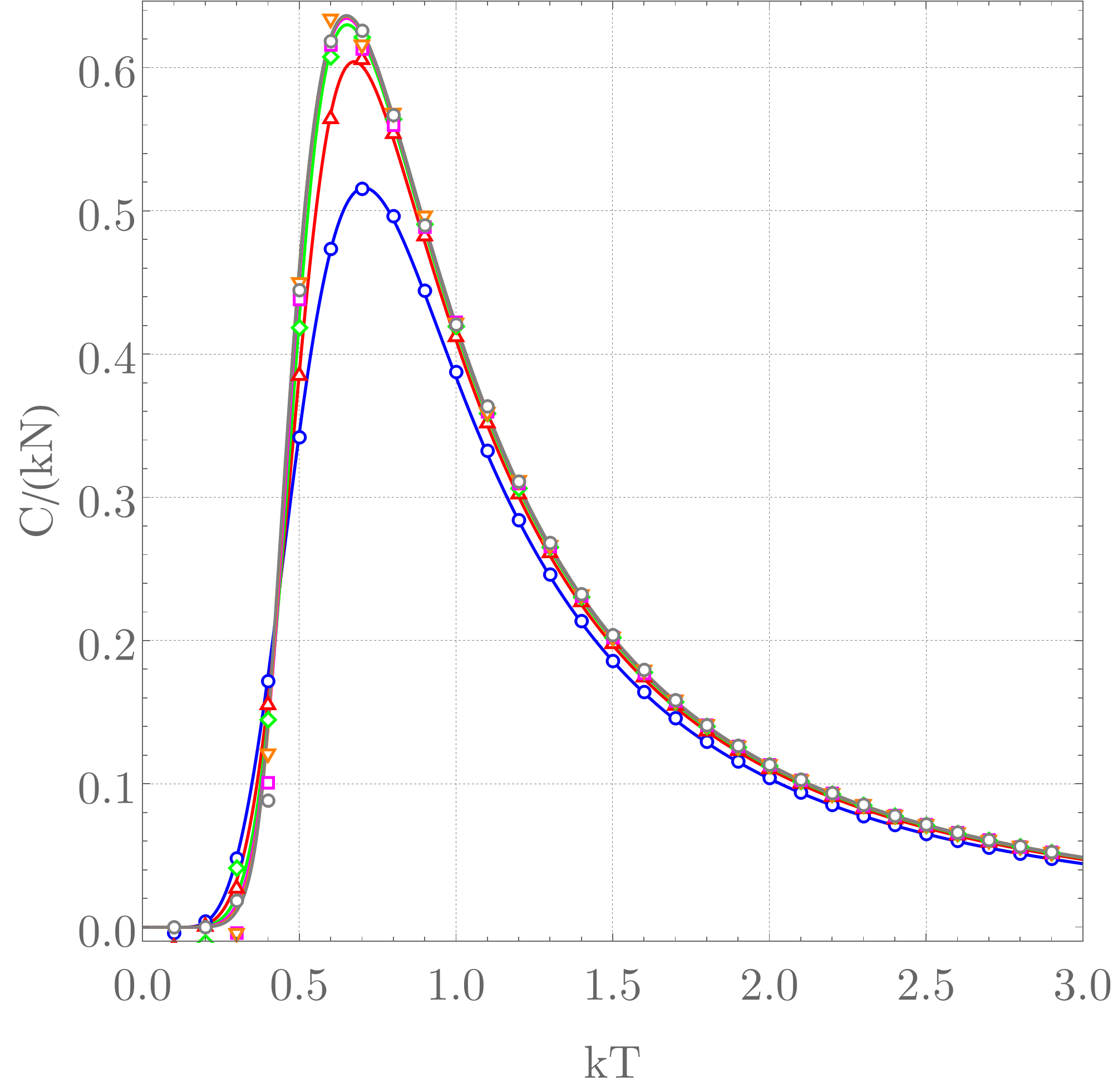}
     \put(75,50){\includegraphics[scale=0.25]{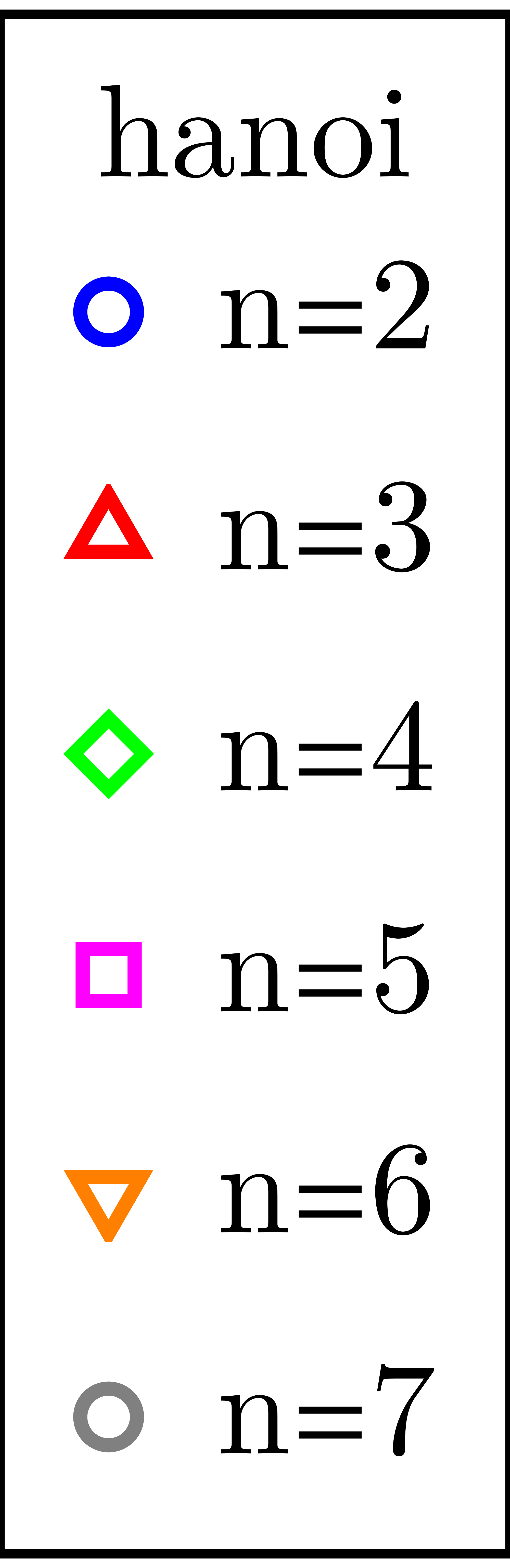}}
  \end{overpic}
  \begin{overpic}[scale=0.19]{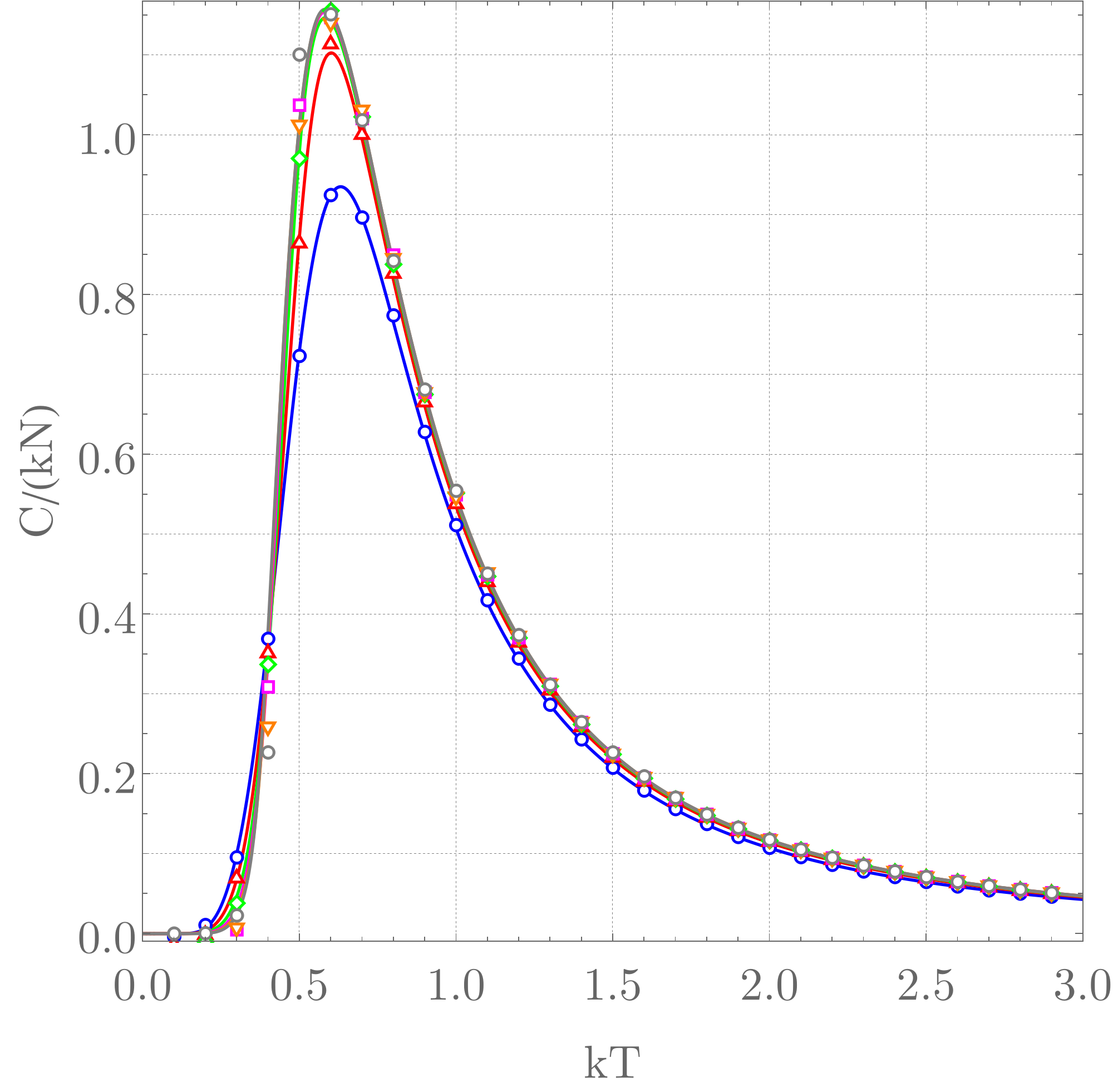}
     \put(75,50){\includegraphics[scale=0.25]{{plot_legend_hanoi}.png}}
  \end{overpic}
  \begin{overpic}[scale=0.19]{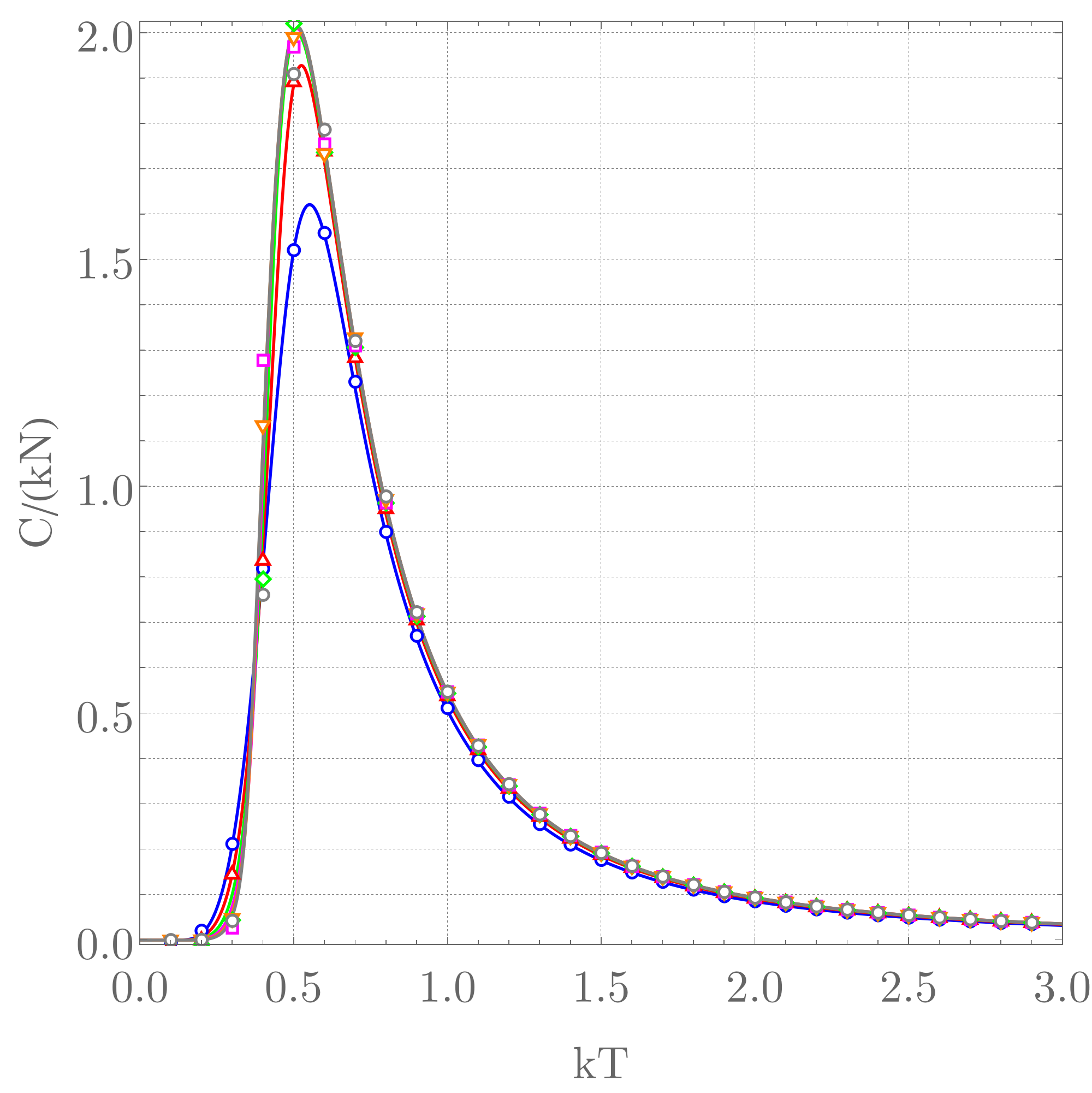}
     \put(75,50){\includegraphics[scale=0.25]{{plot_legend_hanoi}.png}}
  \end{overpic}
  \caption{Specific heat in the ferromagnetic regime for $q=2$ (left panel), $q=3$ (middle panel) and $q=5$ (right panel).}
  \label{fig:sfig1shhan}
\end{subfigure}
\begin{subfigure}{\textwidth}
  \begin{overpic}[scale=0.19]{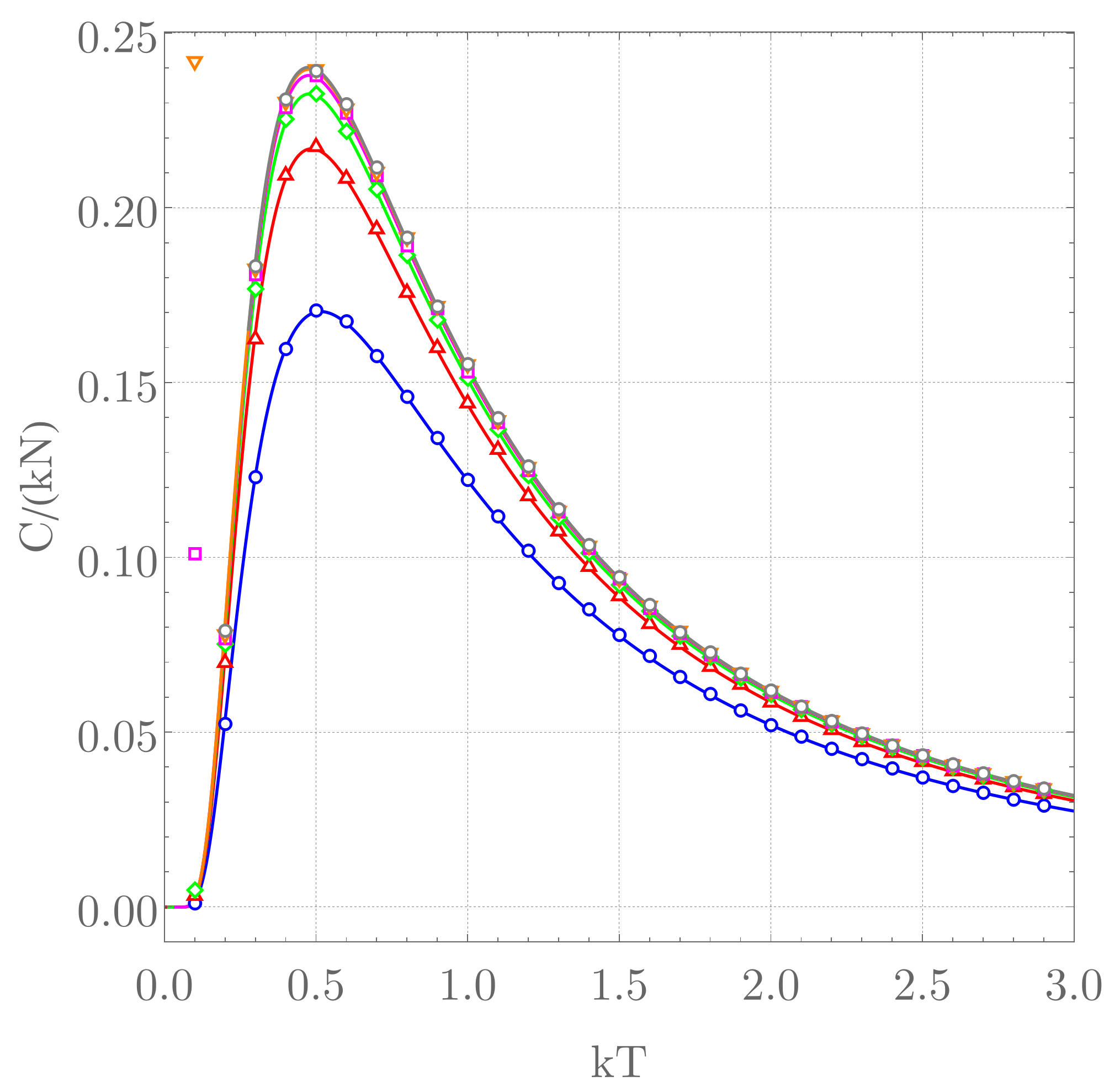}
     \put(75,50){\includegraphics[scale=0.25]{{plot_legend_hanoi}.png}}
  \end{overpic}
  \begin{overpic}[scale=0.19]{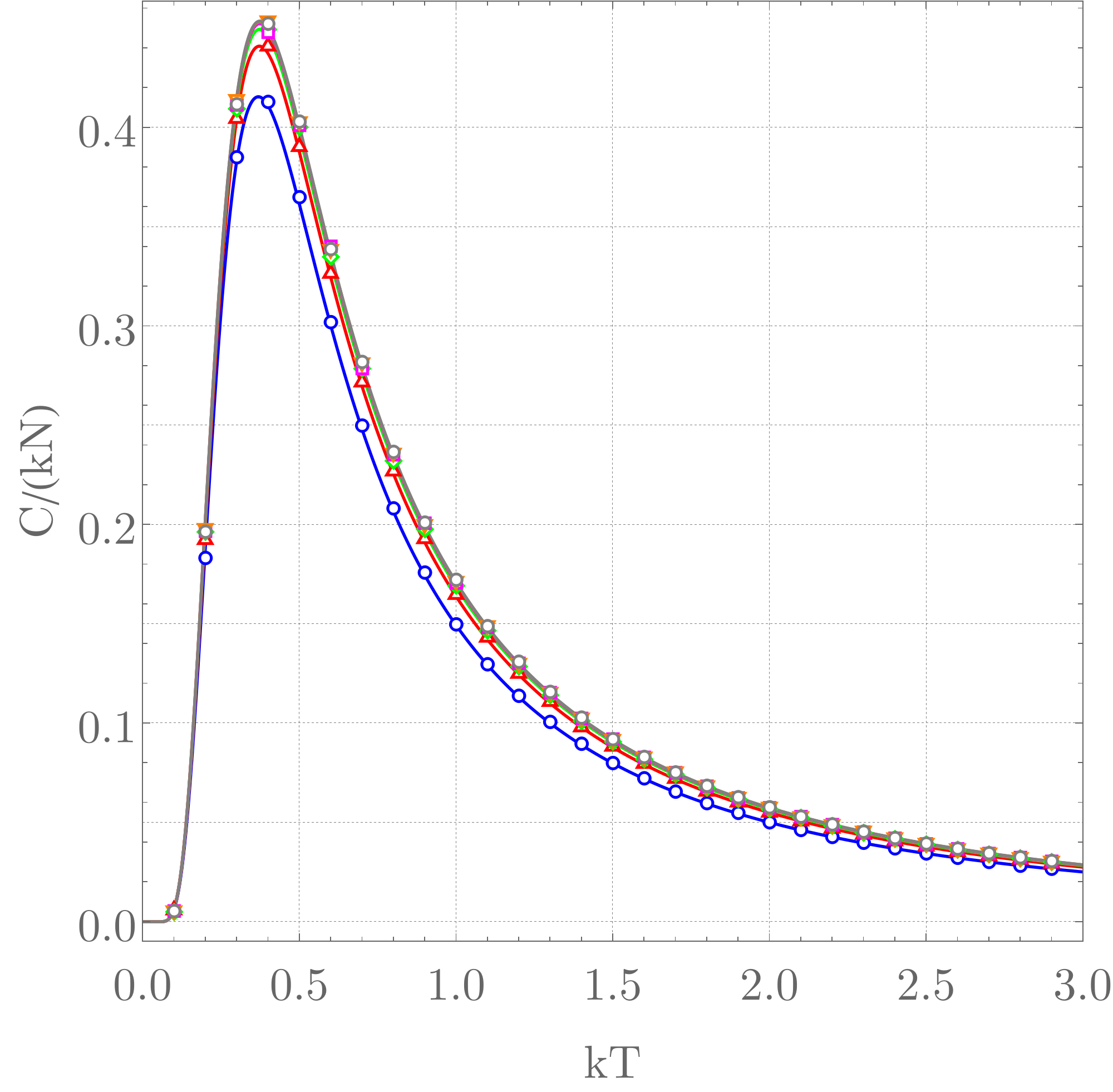}
     \put(75,50){\includegraphics[scale=0.25]{{plot_legend_hanoi}.png}}
  \end{overpic}
  \begin{overpic}[scale=0.19]{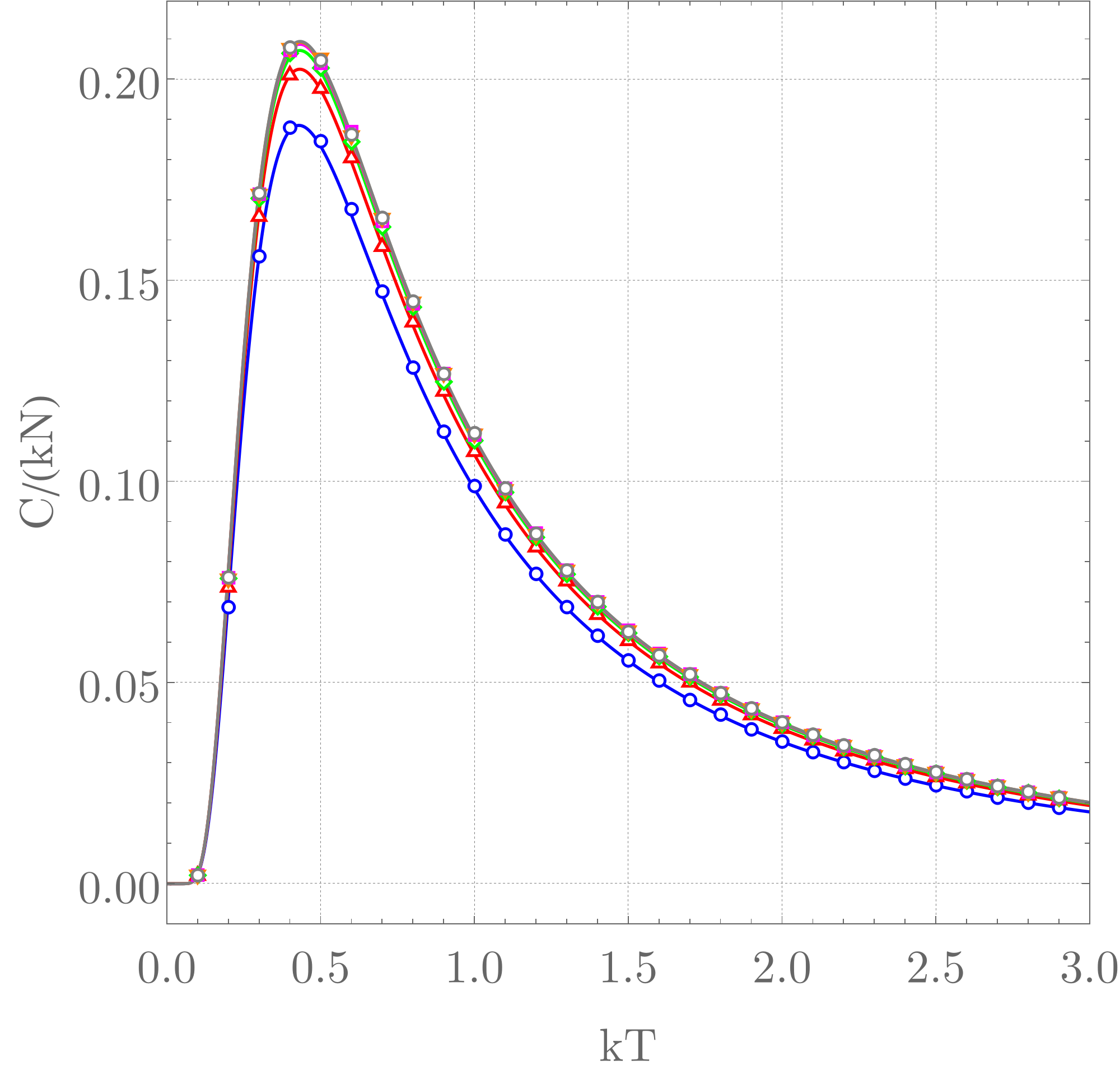}
     \put(75,50){\includegraphics[scale=0.25]{{plot_legend_hanoi}.png}}
  \end{overpic}
  \caption{Specific heat in the anti-ferromagnetic regime for $q=2$ (left panel), $q=3$ (middle panel) and $q=5$ (right panel).}
  \label{fig:sfig2shhan}
\end{subfigure}
\caption{Comparison with Metropolis Monte Carlo results for the Hanoi lattice. Continuous line is the exact result from the dynamical system and the points are the results from simulations. In the metropolis algorithm we used (1000, 1000, 1000, 2430, 7290, 21870) steps for burnout, (400000, 400000, 400000, 486000, 1458000, 4374000) sampling steps and we swept (60, 100, 100, 100, 100, 100) steps to grant statistical independence, all these for $n=(2,3,4,5,6,7)$ respectively.}\label{metropolishan}
\end{figure}

For benchmarks we compared results for finite $n$ with Metropolis Monte Carlo simulations, see Figs. \ref{fig:sfig2shsier} and \ref{fig:sfig2shhan}.

The evaluation for the exact polynomials up to $n=8$ takes a few minutes, while the Monte Carlo simulations take up to few days in a regular pc with an Intel i5-9400 cpu or similar (running on a single core).

\begin{table}[h!]
\centering
        \begin{tabular}{cccc}
         \multicolumn{4}{c}{Sierpinski} \\ \hline
                                              $q$ & $n$ & $T_{max}$ & $c_{max}$ \\ \hline
            \multirow{8}{*}{2} & 1 & 0.924 & 0.055 \\
                               & 2 & 0.870 & 0.089 \\
                               & 3 & 0.870 & 0.106 \\
                               & 4 & 0.870 & 0.114 \\
                               & 5 & 0.870 & 0.117 \\
                               & 6 & 0.870 & 0.118 \\
                               & 7 & 0.870 & 0.118 \\
                               & 8 & 0.870 & 0.118 \\\hline
            \multirow{8}{*}{3} & 1 & 0.351 & 0.342 \\
                               & 2 & 0.321 & 0.557 \\
                               & 3 & 0.333 & 0.620 \\
                               & 4 & 0.336 & 0.664 \\
                               & 5 & 0.336 & 0.680 \\
                               & 6 & 0.336 & 0.686 \\
                               & 7 & 0.336 & 0.688 \\
                               & 8 & 0.336 & 0.688 \\\hline
            \multirow{8}{*}{4} & 1 & 0.396 & 0.204 \\
                               & 2 & 0.375 & 0.324 \\
                               & 3 & 0.375 & 0.389 \\
                               & 4 & 0.375 & 0.416 \\
                               & 5 & 0.375 & 0.427 \\
                               & 6 & 0.375 & 0.430 \\
                               & 7 & 0.375 & 0.431 \\
                               & 8 & 0.375 & 0.432 \\\hline
            \multirow{8}{*}{5} & 1 & 0.420 & 0.147 \\
                               & 2 & 0.402 & 0.230 \\
                               & 3 & 0.402 & 0.276 \\
                               & 4 & 0.402 & 0.296 \\
                               & 5 & 0.402 & 0.303 \\
                               & 6 & 0.402 & 0.305 \\
                               & 7 & 0.402 & 0.306 \\
                               & 8 & 0.402 & 0.306 \\\hline
        \end{tabular}
        \qquad
        \begin{tabular}{cccc}
        \multicolumn{4}{c}{Hanoi} \\ \hline
                                              $q$ & $n$ & $T_{max}$ & $c_{max}$ \\ \hline
            \multirow{8}{*}{2} & 1 & 0.924 & 0.055 \\
                               & 2 & 0.516 & 0.170 \\
                               & 3 & 0.486 & 0.217 \\
                               & 4 & 0.477 & 0.233 \\
                               & 5 & 0.477 & 0.238 \\
                               & 6 & 0.477 & 0.240 \\
                               & 7 & 0.474 & 0.240 \\
                               & 8 & 0.474 & 0.240 \\\hline
            \multirow{8}{*}{3} & 1 & 0.351 & 0.342 \\
                               & 2 & 0.369 & 0.416 \\
                               & 3 & 0.372 & 0.441 \\
                               & 4 & 0.372 & 0.450 \\
                               & 5 & 0.372 & 0.452 \\
                               & 6 & 0.372 & 0.453 \\
                               & 7 & 0.372 & 0.454 \\
                               & 8 & 0.372 & 0.454 \\\hline
            \multirow{8}{*}{4} & 1 & 0.396 & 0.204 \\
                               & 2 & 0.408 & 0.258 \\
                               & 3 & 0.411 & 0.276 \\
                               & 4 & 0.411 & 0.282 \\
                               & 5 & 0.411 & 0.284 \\
                               & 6 & 0.411 & 0.285 \\
                               & 7 & 0.411 & 0.285 \\
                               & 8 & 0.411 & 0.285 \\\hline
            \multirow{8}{*}{5} & 1 & 0.420 & 0.147 \\
                               & 2 & 0.429 & 0.189 \\
                               & 3 & 0.432 & 0.203 \\
                               & 4 & 0.432 & 0.207 \\
                               & 5 & 0.432 & 0.209 \\
                               & 6 & 0.432 & 0.209 \\
                               & 7 & 0.432 & 0.209 \\
                               & 8 & 0.432 & 0.209 \\\hline
        \end{tabular}
    \caption{Values of temperature and peak value of the specific heat for several values of $q$ and $n$ in the antiferromagnetic regime. Models show convergence.}
    \label{tab:Tmaxanti}
\end{table}

\begin{table}[h!]
\centering
        \begin{tabular}{cccc}
         \multicolumn{4}{c}{Sierpinski} \\ \hline
                                              $q$ & $n$ & $T_{max}$ & $c_{max}$ \\ \hline
            \multirow{8}{*}{2} & 1 & 0.699 & 0.340 \\
                               & 2 & 0.933 & 0.451 \\
                               & 3 & 0.993 & 0.607 \\
                               & 4 & 0.969 & 0.704 \\
                               & 5 & 0.948 & 0.735 \\
                               & 6 & 0.945 & 0.742 \\
                               & 7 & 0.945 & 0.744 \\
                               & 8 & 0.945 & 0.745 \\\hline
            \multirow{8}{*}{3} & 1 & 0.630 & 0.605 \\
                               & 2 & 0.837 & 0.813 \\
                               & 3 & 0.885 & 1.133 \\
                               & 4 & 0.864 & 1.338 \\
                               & 5 & 0.843 & 1.399 \\
                               & 6 & 0.843 & 1.411 \\
                               & 7 & 0.843 & 1.415 \\
                               & 8 & 0.843 & 1.416 \\\hline
            \multirow{8}{*}{4} & 1 & 0.585 & 0.829 \\
                               & 2 & 0.777 & 1.124 \\
                               & 3 & 0.819 & 1.612 \\
                               & 4 & 0.798 & 1.930 \\
                               & 5 & 0.780 & 2.020 \\
                               & 6 & 0.780 & 2.037 \\
                               & 7 & 0.780 & 2.043 \\
                               & 8 & 0.780 & 2.045 \\\hline
            \multirow{8}{*}{5} & 1 & 0.552 & 1.025 \\
                               & 2 & 0.735 & 1.402 \\
                               & 3 & 0.771 & 2.059 \\
                               & 4 & 0.750 & 2.493 \\
                               & 5 & 0.735 & 2.613 \\
                               & 6 & 0.735 & 2.635 \\
                               & 7 & 0.735 & 2.642 \\
                               & 8 & 0.735 & 2.645 \\\hline
        \end{tabular}
        \qquad
        \begin{tabular}{cccc}
        \multicolumn{4}{c}{Hanoi} \\ \hline
                                              $q$ & $n$ & $T_{max}$ & $c_{max}$ \\ \hline
            \multirow{8}{*}{2} & 1 & 0.699 & 0.340 \\
                               & 2 & 0.705 & 0.516 \\
                               & 3 & 0.675 & 0.604 \\
                               & 4 & 0.651 & 0.630 \\
                               & 5 & 0.651 & 0.635 \\
                               & 6 & 0.651 & 0.636 \\
                               & 7 & 0.651 & 0.637 \\
                               & 8 & 0.651 & 0.637 \\\hline
            \multirow{8}{*}{3} & 1 & 0.630 & 0.605 \\
                               & 2 & 0.630 & 0.935 \\
                               & 3 & 0.600 & 1.103 \\
                               & 4 & 0.582 & 1.147 \\
                               & 5 & 0.582 & 1.154 \\
                               & 6 & 0.582 & 1.157 \\
                               & 7 & 0.582 & 1.158 \\
                               & 8 & 0.582 & 1.158 \\\hline
            \multirow{8}{*}{4} & 1 & 0.585 & 0.829 \\
                               & 2 & 0.585 & 1.298 \\
                               & 3 & 0.555 & 1.538 \\
                               & 4 & 0.540 & 1.596 \\
                               & 5 & 0.540 & 1.606 \\
                               & 6 & 0.540 & 1.609 \\
                               & 7 & 0.540 & 1.610 \\
                               & 8 & 0.540 & 1.610 \\\hline
            \multirow{8}{*}{5} & 1 & 0.552 & 1.025 \\
                               & 2 & 0.552 & 1.622 \\
                               & 3 & 0.525 & 1.928 \\
                               & 4 & 0.510 & 1.999 \\
                               & 5 & 0.510 & 2.010 \\
                               & 6 & 0.510 & 2.013 \\
                               & 7 & 0.510 & 2.015 \\
                               & 8 & 0.510 & 2.015 \\\hline
        \end{tabular}
    \caption{Values of temperature and peak value of the specific heat for several values of $q$ and $n$ in the ferromagnetic regime. Models show convergence.}
    \label{tab:Tmaxferro}
\end{table}

The partition function contains the chromatic polynomial of the lattice when evaluated at $v=-1$, which is the zero-temperature in the anti-ferromagnetic regime and therefore our method can be used to extract the exact chromatic polynomials.  We can also compute the entropy, which in terms of the $v$-temperature variable takes the form,
\begin{equation}
s = \frac{1}{V}\left(\ln Z -v' \ln v'  \frac{\partial Z/\partial v'}{Z}\right)\,,\label{santiferro}
\end{equation}
where we have omitted the subscript $n$ for the sake of simplicity of notation. The variable $v' = 1+v$ is convenient for working out asymptotic expressions at low temperature in the anti-ferromagnetic regime as it approaches $v'\rightarrow 0$ for $T\rightarrow 0$. In the ferromagnetic regime we can use 
\begin{equation}
s = \frac{1}{V}\left(\ln Z -(1+v) \ln (1+v)  \frac{\partial Z/\partial v}{Z}\right)\,.\label{sferro}
\end{equation}
For finite $n$ we get the exact expressions
\begin{equation}
Z = \begin{cases}
 \sum_{i=0}^{l} c_i v^i \qquad \mathrm{ferromagnetic}\\
 \sum_{i=0}^{l} d_i v'^i \qquad \mathrm{anti-ferromagnetic}
\end{cases}\,,\label{Zexpansion}
\end{equation}
where $l=2V-2$. The term $d_0(q)$ is the chromatic polynomial of the lattice, which can be written in a convenient form for the Sierpinski lattice and the Hanoi lattice
\begin{equation}
 d^{(n)}_0(q) = q(q-1)(q-2)^{x_n}\mathrm{pol}_{y_n}(q)\,,
\end{equation}
where $x_n=1, 3, 6, 9, 12, 15,$... The term $\mathrm{pol}_{y_n}(q)$ is a polynomial of degree $y_n$ on $q$, where the following relation holds $2+x_n+y_n=V_n$. We do not have the general expression for $d^{(n)}_0(q)$ but it can be computed exactly for fixed $n$ iteratively. For $n=1,2,3$ we get
\begin{align}
 d^{(1)}_0(q) =& q (q-1) (q-2)\,,\\
 d^{(2)}_0(q) =& q (q-1) (q-2)^3 (q-2)\,,\\
 d^{(3)}_0(q) =& q (q-1) (q-2)^6 (-272 + 676 q - 831 q^2\notag\\
 &+ 620 q^3 - 292 q^4 + 85 q^5 - 14 q^6 + q^7)\,,
\end{align}
for the Sierpinski lattice and 
\begin{align}
 d^{(1)}_0(q) =& q (q-1) (q-2)\,,\\
 d^{(2)}_0(q) =& q (q-1) (q-2)^3 (5 - 10 q + 10 q^2 - 5 q^3 + q^4)\,,\\
 d^{(3)}_0(q) =& q (q-1) (q-2)^6 (-9500 + 70750 q - 286075 q^2 + 803125 q^3\notag\\
 &- 1709450 q^4 +  2880770 q^5 - 3940970 q^6 + 4444555 q^7 - 4170694 q^8\notag\\
 &+3271840 q^9 - 2147675 q^{10} + 1176423 q^{11} - 534267 q^{12} + 198981 q^{13}\notag\\
 &- 59771 q^{14} + 14125 q^{15} - 2528 q^{16} + 322 q^{17} - 26 q^{18} + q^{19})\,,
\end{align}
for the Hanoi lattice. Higher $n$ polynomials are too long to be presented here. See reference \cite{2010arXiv1006.5333D} for Tutte polynomials using a combinatorial approach.
From (\ref{Zexpansion}) we can obtain exact expressions for the entropy at zero and infinite temperature. We need to treat the cases when $d^{(n)}_0(q)\ne 0$ separately from the cases when $d^{(n)}_0(q)=0$. Firstly, for $q \ge 3$ we have proper colorations of the Sierpinski lattice and therefore $d^{(n)}_0(q)\ne 0$. Replacing (\ref{Zexpansion}) in (\ref{santiferro}) and keeping the leading term we get
\begin{equation}
 s_n = \frac{1}{V_n} \ln (d_0^{(n)})\,.\label{sexactn}
\end{equation}
In the thermodynamic limit we can use
\begin{equation}
 s = \lim_{n \to \infty}\frac{1}{V_n} \ln (d_0^{(n)})\,.\label{sexactnthermolimit}
\end{equation}
For $q=3$ $d_0^{(n)}=6$ and therefore $s=0$, and for $q \ge 4$ (\ref{sexactnthermolimit}) converges to a non-zero finite number, see Tables \ref{svalues} and \ref{svalueshanoi}. On the other hand, the case $q=2$ has a great degeneracy implying $d^{(n)}_j(q)=0$ for $j=0,1,...,k-1$, where $k=3^{n-1} + 1$. The leading term expansion is $Z_{q=2} \sim d^{(n)}_k(q) v'^k$, therefore
\begin{equation}
 s_n = \frac{1}{V_n} \ln (d_k^{(n)})\,.\label{sexactnq=2}
\end{equation}
In the thermodynamic limit
\begin{equation}
 s = \lim_{n \to \infty}\frac{1}{V_n} \ln (d_k^{(n)})\,.\label{sexactnthermolimitq=2}
\end{equation}
converges to a non-zero finite number. Expressions (\ref{sexactn}) and (\ref{sexactnq=2}) are the exact zero temperature entropy for finite Sierpinski and Hanoi lattices for arbitrary $q$ when the exact expresions for $d^{(n)}_0(q)$ and $d^{(n)}_k(q)$ are used. The exact expressions for $d^{(n)}_k(q)$ are the following:
\begin{align}
 d^{(1)}_k(q) =& 3q (q-1)\,,\\
 d^{(2)}_k(q) =& q (q-1) (-15+6q+4q^2)\,,\\
 d^{(3)}_k(q) =& q (q-1) (-442365 + 1042698 q - 868289 q^2 + 225687 q^3 \notag\\
 &+ 62652 q^4- 32718 q^5 - 2990 q^6 + 2038 q^7 + 3 q^8)\,,
\end{align}
for the Sierpinski lattice and 
\begin{align}
 d^{(1)}_k(q) =& 3 q (q-1)\,,\\
 d^{(2)}_k(q) =& q (q-1) (904 - 2009 q + 1790 q^2 - 740 q^3 + 112 q^4 + 3 q^5)\,,\\
 d^{(3)}_k(q) =& q (q-1) (111038005202 - 804578253871 q + 2874454835925 q^2 \notag\\
 &- 6678461432139 q^3 + 11200529335275 q^4 - 14244301463164 q^5 \notag\\
 &+  14066598849017 q^6 - 10873626085915 q^7 + 6542501603993 q^8\notag\\
 &-  2990527322606 q^9 + 973011190480 q^{10} - 179683894208 q^{11}\notag\\
 &-  11089437216 q^{12} + 19894270539 q^{13} - 6859396368 q^{14}\notag\\
 &+ 1289629528 q^{15} - 130897553 q^{16} + 4619718 q^{17} + 171273 q^{18} +  84 q^{19})\,,
\end{align}
for the Hanoi lattice. Again, higher $n$ polynomials are too long to be presented here. The numerical results are summarized in Tables \ref{svalues} and \ref{svalueshanoi}, the results for $q=2$ agree with the ones found in the references \cite{GRILLONBradyMoreira1989,Stinchcombe1990,Nobre-Curado2000}.

The infinite temperature entropy ferromagnetic or antiferromagnetic can be written as
\begin{equation}
 s_n(T\to \infty) = \frac{1}{V_n} \ln (c_0^{(n)})\,.\label{sexactinfty}
\end{equation}
where $c_0^{(n)}=q^{V_n}$ and therefore $s_n(T\to \infty)=\ln q$, which is also valid in the thermodynamic limit. The zero temperature entropy in the ferromagnetic limit is given by
\begin{equation}
 s_n(T\to 0) = \frac{1}{V_n} \ln (c_l^{(n)})\,.\label{sexact0ferro}
\end{equation}
where $c_l^{(n)}=q$, therefore (\ref{sexact0ferro}) converges to zero in the thermodynamic limit.

The analysis presented in this section allows us to conclude that there are many polynomials in the $v$ or $v'$ expansion, beyond the chromatic one, that carry meaningful information of the lattice even at zero temperature.

\begin{table}[h!]
\begin{center}
\tabcolsep=0.11cm
\begin{tabular}{c|cccccccc}
$q$ & $n=$  1 & 2 & 3 & 4 & 5 & 6 & 7 & 8 \\ \hline
 2 & 0.59725 & 0.54302 & 0.51302 & 0.50015 & 0.49545 & 0.49383 & 0.49328 & 0.49310 \\ 
 3 & 0.59725 & 0.29863 & 0.11945 & 0.04266 & 0.01457 & 0.00490 & 0.00164 & 0.00055 \\
 4 & 1.05935 & 0.87625 & 0.77424 & 0.73052 & 0.71453 & 0.70902 & 0.70717 & 0.70654 \\
 5 & 1.36478 & 1.23170 & 1.15615 & 1.12378 & 1.11193 & 1.10785 & 1.10648 & 1.10602 
\end{tabular}
\end{center}
\caption{Exact zero-temperature entropy of the Potts model in the Sierpinski lattice.}\label{svalues}
\end{table}

\begin{table}[h!]
\begin{center}
\tabcolsep=0.11cm
\begin{tabular}{c|cccccccc}
$q$ & $n=$  1 & 2 & 3 & 4 & 5 & 6 & 7 & 8 \\ \hline
 2 & 0.59725 & 0.37024 & 0.29334 & 0.26766 & 0.25911 & 0.25625 & 0.25530 & 0.25499 \\ 
 3 & 0.59725 & 0.46552 & 0.42047 & 0.40546 & 0.40045 & 0.39878 & 0.39822 & 0.39804 \\
 4 & 1.05935 & 0.96391 & 0.93195 & 0.92129 & 0.91774 & 0.91656 & 0.91616 & 0.91603 \\
 5 & 1.36478 & 1.29051 & 1.26572 & 1.25745 & 1.25470 & 1.25378 & 1.25347 & 1.25337 
\end{tabular}
\end{center}
\caption{Exact zero-temperature entropy of the Potts model in the Hanoi lattice.}\label{svalueshanoi}
\end{table}

\section{Fixed points}\label{sectionfixed}

As it is well known in the theory of dynamical systems, the asymptotic behaviour is dominated by the attractive fixed points. In order to analyze the different phases of the Potts model on the Sierpinski triangle, we have to consider all the physically interesting ranges of $v$ and then we have to study the corresponding evolution of the dynamical system in eqs. (\ref{dyn-sier1})-(\ref{dyn-sier3}) for very large $n$.

It is convenient to exploit the homogeneity of the polynomials in Eqs. (\ref{f1}), (\ref{f2}) and (\ref{f3}) in order to reduce the above dynamical system to a system of two equations in two variables, $\eta (n)$ and $\gamma(n)$, as follows: 
\begin{equation}
\eta (n)=\frac{y\left( n\right) }{x\left( n\right) },\qquad \gamma (n)=\frac{%
z\left( n\right) }{x\left( n\right) },  \label{changevariable}
\end{equation}
so that we get a closed system for $\eta $\ and $\gamma $: 
\begin{eqnarray}
\eta (n+1)=\frac{g(1,\eta (n),\gamma (n);q)}{f(1,\eta (n),\gamma (n);q)}\,, \label{reduced1}\\
\gamma (n+1)=\frac{h(1,\eta (n),\gamma (n);q)}{f(1,\eta (n),\gamma(n);q)}\,, \label{reduced2}
\end{eqnarray}
with initial data 
\begin{equation}  \label{icdiscrete}
\eta (1)=v,\qquad \gamma (1)=v^{2}\left( 3+v\right) .
\end{equation}
The formulation of the dynamical system in eqs. (\ref{reduced1})-(\ref{reduced2}) is also convenient as far as the physical interpretation is concerned since the high temperature phase corresponds to the range of temperatures for which, in the large $n$ limit, 
\begin{equation*}
\left\vert \eta (n)\right\vert \underset{n\rightarrow \infty }{\ll }1,\qquad
\left\vert \gamma (n)\right\vert \underset{n\rightarrow \infty }{\ll }1,
\end{equation*}
while, in the antiferromagnetic phase, frustration at macroscopic scales appears if 
\begin{equation*}
\left\vert \eta (n)\right\vert \underset{n\rightarrow \infty }{\gtrapprox } 1\,, \qquad \left\vert \gamma (n)\right\vert \underset{n\rightarrow \infty }{\gtrapprox } 1 \,.
\end{equation*}
(see the discussion below Eqs.(\ref{hight}) and (\ref{modt})). Indeed, from the physical point of view, we are interested in the thermodynamical large $n $ limit of the Sierpinski gasket partition function $K(n)$. Within the present framework, this corresponds to analyse the asymptotic behaviour of the dynamical system (\ref{reduced1})-(\ref{reduced2}) in the large $n$ limit for initial data of the form (\ref{icdiscrete}).

\begin{figure}[h!]
\centering
\begin{subfigure}[b]{0.4\textwidth}
  \begin{overpic}[scale=0.29]{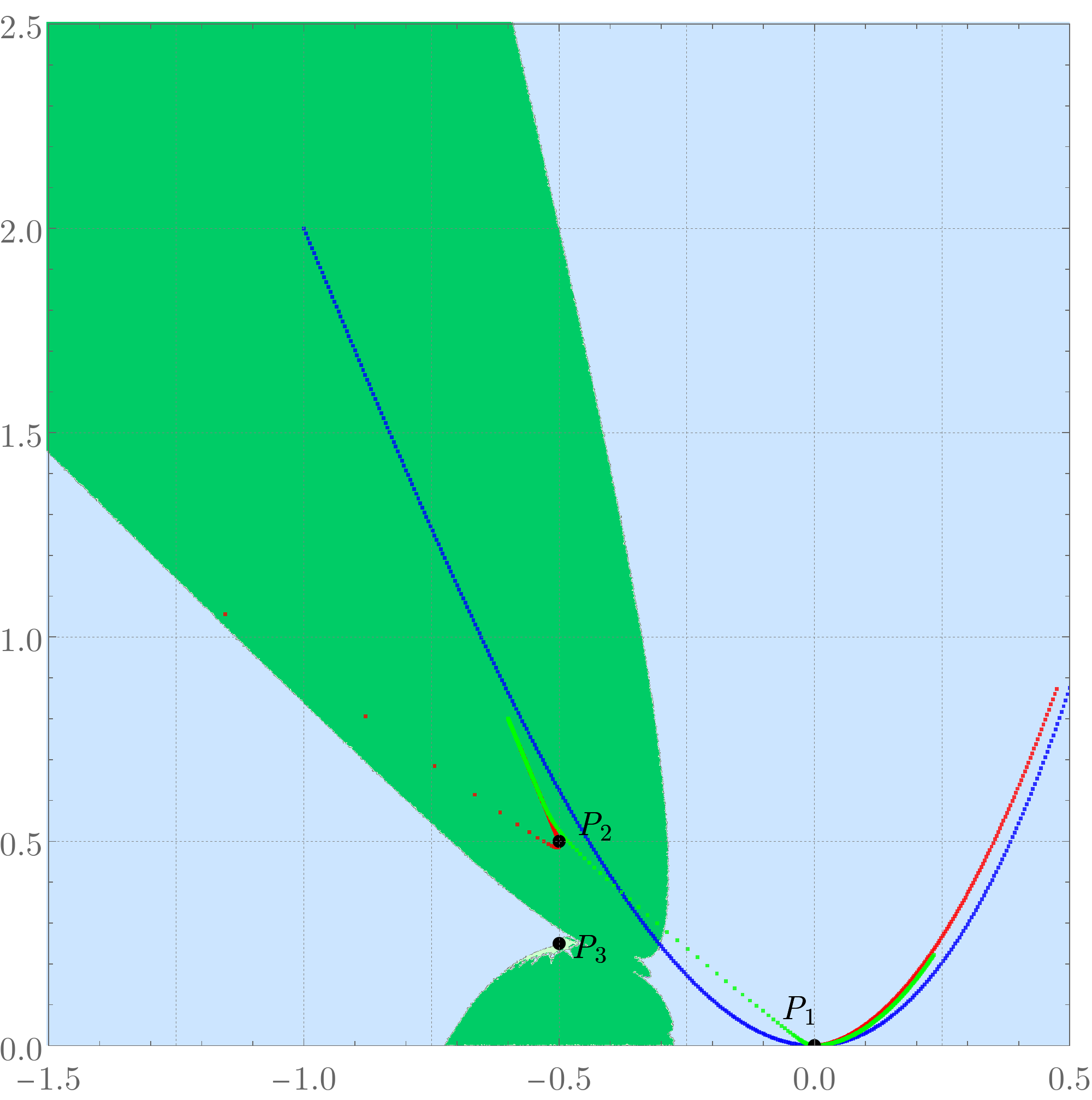}
     \put(54,67){\includegraphics[scale=0.4]{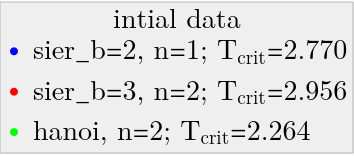}}
     \put(45,87){\includegraphics[scale=0.15]{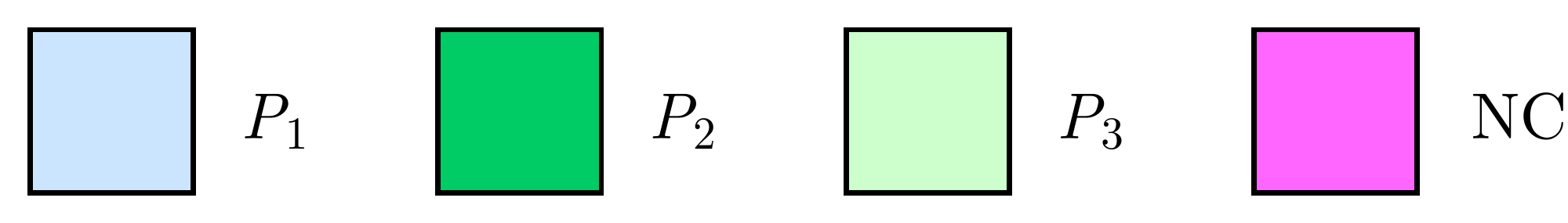}}
  \end{overpic}
  \caption{Basins map for $q=1$.}
  \label{fig:conv_q1}
\end{subfigure}%
\hspace{1.1cm}
\begin{subfigure}[b]{0.4\textwidth}
  \begin{overpic}[scale=0.29]{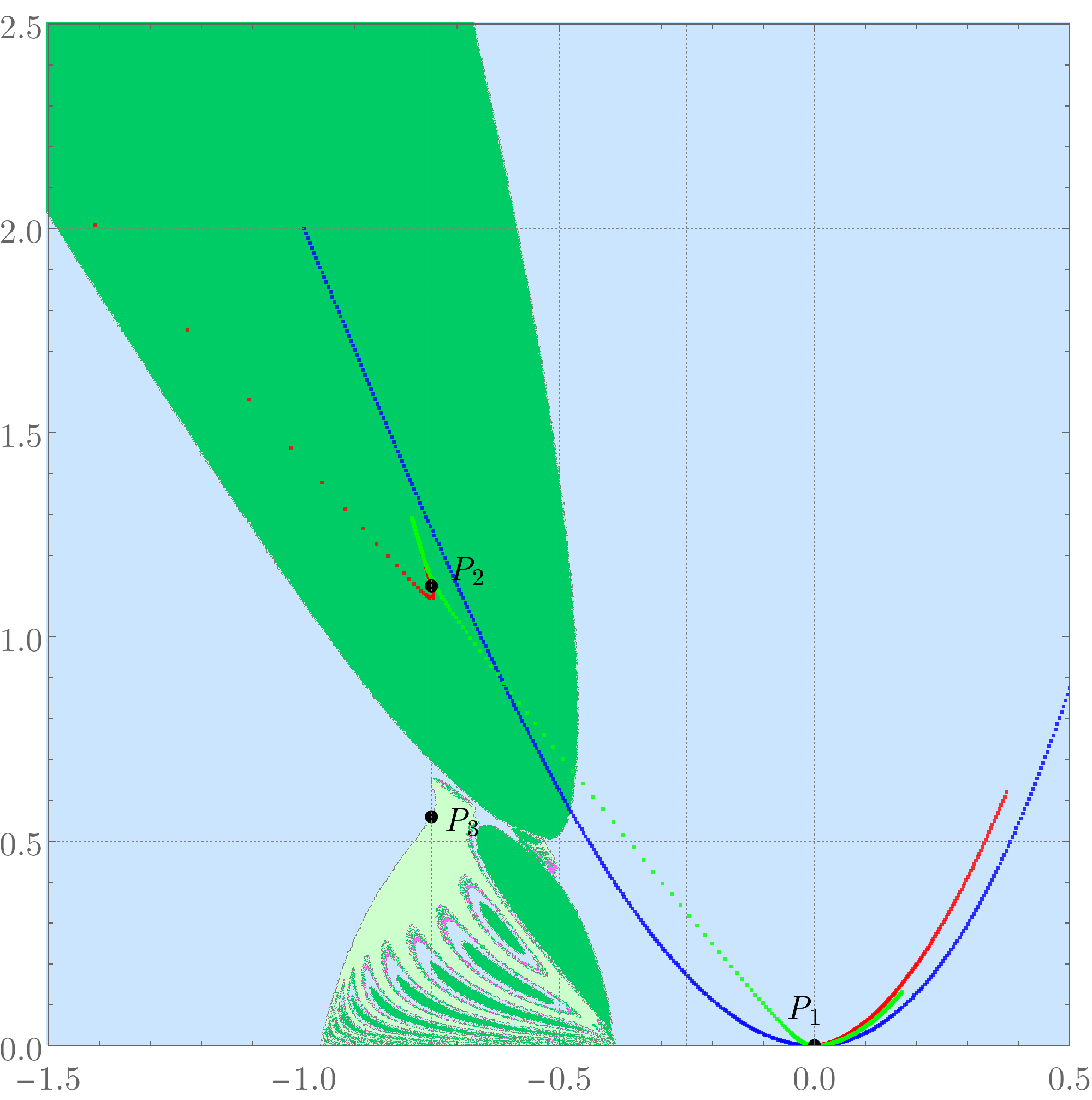}
     \put(54,67){\includegraphics[scale=0.4]{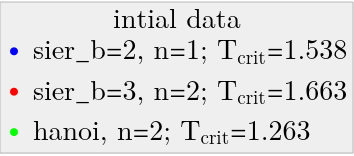}}
     \put(45,87){\includegraphics[scale=0.15]{{sier_b=2_legend_basin-map_complex-v_legend}.png}}
  \end{overpic}
  \caption{Basins map for $q=1.5$.}
  \label{fig:conv_q1.5}
\end{subfigure}\\
  \begin{subfigure}[b]{0.4\textwidth}
  \begin{overpic}[scale=0.29]{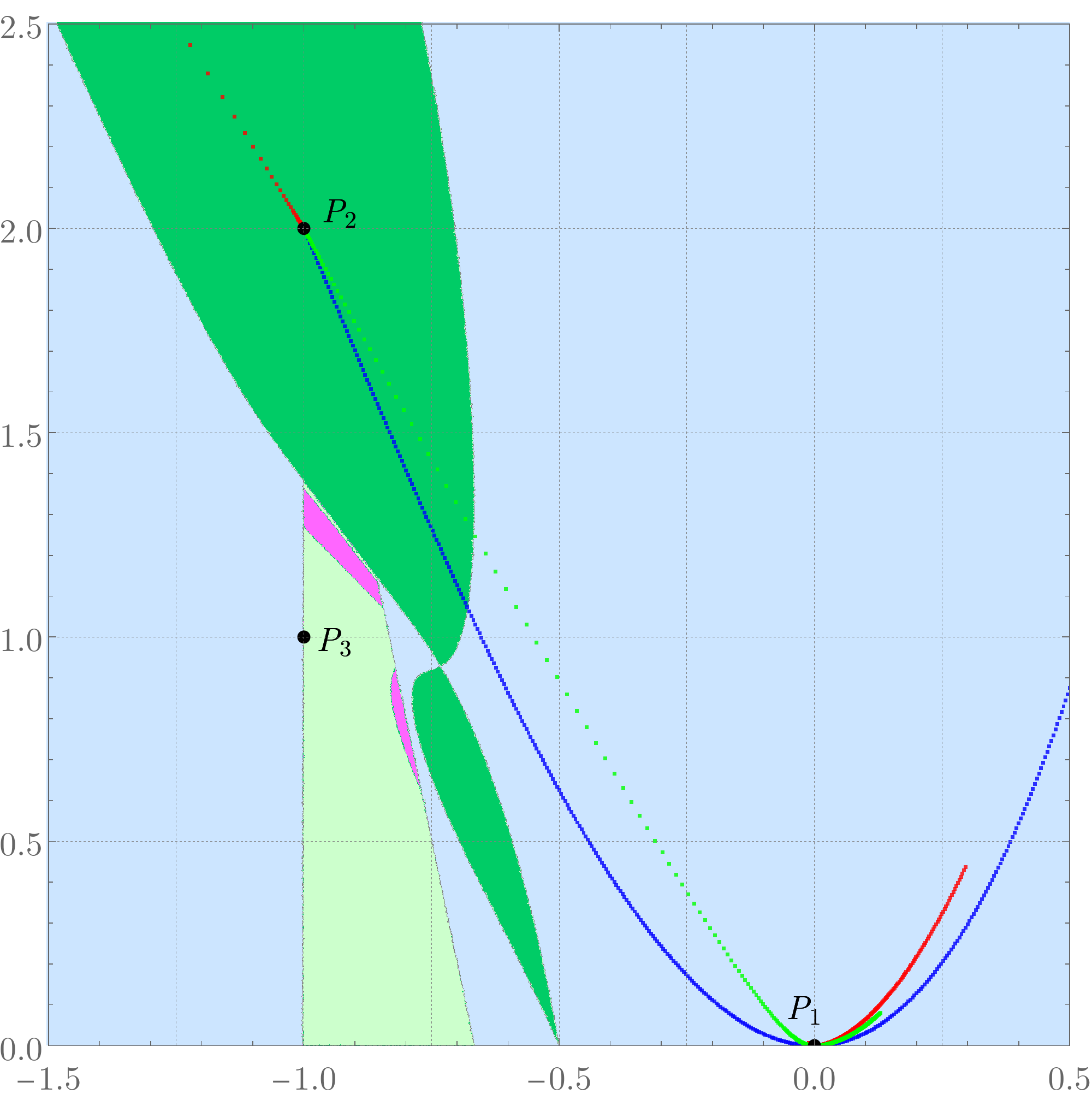}
     \put(54,67){\includegraphics[scale=0.4]{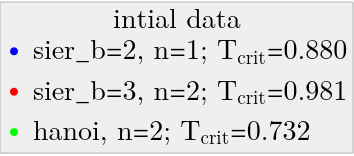}}
     \put(45,87){\includegraphics[scale=0.15]{{sier_b=2_legend_basin-map_complex-v_legend}.png}}
  \end{overpic}
  \caption{Basins map for $q=2$.}
  \label{fig:conv_q2}
\end{subfigure}%
\caption{Basins of convergence of the Sierpinski map in the antiferromagnetic regime for some values of $q$. Horizontal axis corresponds to $\eta$ and vertical axis corresponds to $\gamma$. Two different initial data is overlaid: standard triangle in yellow line and the Hanoi of second step. The sky color indicates the basin of $P_1$, the dark green color indicates the basin of $P_2$, and light green indicates the basin of $P_3$. The temperature at which the jump of basins occur is indicated in the legend.}
\label{fig:conv}
\end{figure}

The system (\ref{reduced1})-(\ref{reduced1}) has ten fixed points which, in general, depend on the parameter $q$ only. The fixed points have the follwing coordinates $(\eta,\gamma)$ and domain
\begin{align}
 P_1 =& (0,0)\,, \quad 0 < q < \infty\,, \quad \mathrm{superstable \ in \ whole \ domain}\\
 P_2 =& (-\tfrac{1}{2} q,\tfrac{1}{2}q^{2})\,, \quad 0 < q < \infty\,, \quad \mathrm{stable \ for \ }0<q<3/2\\
 P_3 =& (-\tfrac{1}{2} q, \tfrac{1}{4} q^{2})\,, \quad 0 < q < \infty\,,\\
 P_4 =& (1 - \sqrt{1 + q}, -q(2 - \sqrt{1 + q}) - 2(1 - \sqrt{1 + q})))\,, \quad 0 < q < \infty\,,\\
 P_5 =& (1 + \sqrt{1 + q}, -q(2 + \sqrt{1 + q}) - 2( 1 + \sqrt{1 + q})))\,, \quad 0 < q < \infty\,,\\
 P_6 =& (q\frac{2 - \sqrt{q-3}}{q-7}, -q^2\frac{2 - \sqrt{q-3}}{q-7})\,, \quad 3 < q < 7 \cup 7 < q< \infty\,,\\
P_7 =& (q\frac{2 + \sqrt{q-3}}{q-7}, -q^2\frac{2 + \sqrt{q-3}}{q-7})\,, \quad 3 < q < 7 \cup 7 < q< \infty\,,\\
P_8 =& (\frac{q}{X}\frac{-40 X+128 (3q-8)+2^{2/3} X^2 }{6(q+14)},0)\,, \quad 22/27 < q < \infty\,,\\
P_9 =& (\frac{-40q^2 X + 128 e^{i\pi/3}(3q-8) -   2^{-4/3}e^{-i\pi/3} q^4 X^2 }{6q X (q+14)}, 0)\,, \quad 0 < q < 2\,,\\
P_{10} =& (\frac{-40q^2 X + 128 e^{-i\pi/6}(3q-8) -  2^{-4/3}e^{i\pi/6} q^4 X^2 }{6q X (q+14)}, 0)\,, \quad 22/27 < q < 2\,,
\end{align}
where
\begin{align}
 X=&\sqrt[3]{-27 q^2+684 q+3 \sqrt{3} (q+14) \sqrt{(q-2) (27 q-22)}-1132}\,.
\end{align}
The stable points are $P_1$ and $P_2$. $P_3$ is marginally stable and the rest are unstable. We constructed the map of basins of attraction in Fig.\ref{fig:conv}. It is worth noting a very peculiar feature of the present dynamical system. Even if the fixed point $P_3$ is unstable, it is relevant as far as the dynamics of the Sierpinski dynamical system is concerned: the technical reason is that $P_3$ lies on the boundary of its own basin of attraction. This is a quite interesting feature as far as the theory of dynamical system is concerned. In the map of basins for $q=2$ we can observe the appearance of lobes that are typical of dynamical systems with vanishing denominators \cite{10.3389/fams.2016.00011}.

Again, the discrete evolution parameter $n$ of the dynamical system corresponds to the hierarchical size of the Hanoi graph itself. Hence, the thermodynamical limit corresponds to the asymptotic analysis of the dynamical system in eqs. (\ref{hanoifirst})-(\ref{hanoithird}) together with the initial data in eqs. (\ref{firsttriangl2}) when $n\rightarrow \infty $.

Let us briefly comment on phase transitions. Indeed, the most obvious way to search for phase transitions would be the analysis of the discontinuities of the free energy density. However, as it has been shown, for instance, in \cite{GASM83} \cite{GASM} the free energy density is smooth in the Sierpinski gasket case. We confirm this observation for $q=2,3$ by finding that the distribution of zeros does not cross the physical temperature axis, see figure \ref{complexplanemapSier} (we also noted this for bigger values of $q$ but for the lack of space we do not present those maps here).

On the other hand, the dynamical system the anti-ferromagnetic Potts model on the Sierpinski gasket displays a quite intricate map of basins characterized by a sudden jump in the amount of macroscopic frustration when one crosses from one basin to another, see Figure \ref{fig:conv}. In \cite{Sim09first,Sim09second}, the authors studied the Potts model on hierarchical lattices using real space RG. The variables defined in \cite{Sim09first,Sim09second} refer to the degeneracy of the energy states of the system rather than to the connectivity patterns as in our approach. We used the dynamical system presented in \cite{Sim09first,Sim09second} to observe that the results are consistent with (\ref{reduced1}) and (\ref{reduced2}). In particular, the transitions, corresponding to the crossing of the boundaries between different basins, occur exactly at the same critical temperatures in both approaches.

\subsection{Fixed points on the Hanoi lattice}

Similarly to what has been done in the study of the Sierpinski dynamical system, it is useful to exploit the fact that the right hand side of eqs. (\ref{hanoifirst})-(\ref{hanoithird}) are homogeneous functions of degree $3$ in order to reduce the dynamical system to a system of two equations in two variables. Thus, we take as basic variables $\eta=y/x$\ and $\gamma=z/x$: 
\begin{eqnarray}
\eta(n+1)=\frac{g_h(1,\eta(n),\gamma(n);q,v)}{f_h(1,\eta(n),\gamma(n);q,v)}\,,\label{hanoireduced1}\\
\gamma(n+1)=\frac{h_h(1,\eta(n),\gamma(n);q,v)}{f_h(1,\eta(n), \gamma(n);q,v)}\,.\label{hanoireduced2}
\end{eqnarray}
The initial data could be again be taken as a simple triangle (\ref{icdiscrete}) for the Hanoi lattice.

From the physical point of view, we are interested in the thermodynamical large $n$ limit of the Potts model partition function on the Hanoi graph $K_{h}(n)$. Within the present framework, this corresponds to analyse the asymptotic behaviour of the dynamical system in eqs. (\ref{hanoifirst})-(\ref{hanoithird}).

The system (\ref{hanoireduced1})-(\ref{hanoireduced2}) has 11 fixed points: $P'_1, \cdots P'_{11}$. The stable points are $P'_1$ and $P'_2$,
\begin{align}
 P'_1 =& (0,0)\,, \quad 0 < q < \infty \quad \mathrm{superstable \ in \ whole \ domain}\,,\\
 P'_2 =& (-\tfrac{1}{2} q,\tfrac{1}{2}q^{2})\,, \quad 0 < q < \infty\,, \quad \mathrm{stable \ for \ }0<q<3/2\,,
\end{align}
where $w_3 = \sqrt{v(q+v)^3}$. $P'_3$ and $P'_4$ are marginally stable,
\begin{align}
 P'_3 =& ( -\frac{q(2q + 3v) + v^2 + w_3}{4q + 3v}, \frac{2(q + v)^3 + (3q + 2v)w_3}{4q + 3v})\,, \quad 0 < q < \infty\,,\\
 P'_4 =& ( -\frac{q(2q + 3v) + v^2 - w_3}{4q + 3v}, \frac{2(q + v)^3 - (3q + 2v)w_3}{4q + 3v})\,, \quad 0 < q < \infty\,,
\end{align}
The fixed points $P'_3$ and $P'_4$ are marginally stable and form a complex conjugate pair when $v(q+v)^3 < 0$. They are real and different for $v(q+v)^3 > 0$. We will skip the expressions for $P'_5$--$P'_{11}$ since they are long and cumbersome and also because $P'_5$--$P'_{11}$ are unstable fixed points. In Fig. \ref{fig:convHanoi} we have constructed the maps of basins of attraction for the Hanoi system (\ref{hanoireduced1})-(\ref{hanoireduced2}). 

It is interesting to note the invariance of $T_\mathrm{crit}$ under the change in initial conditions $T(1) \rightarrow H(2)$ is present as it should be since both lattices converge to the same lattice in the thermodynamical limit. In fact, we checked that the invariance of $T_\mathrm{crit}$ under the change in initial conditions $T(1) \rightarrow T(n)$.

\subsection{Changing the lattice by changing the initial data}\label{sec-change}

\begin{figure}
\centering
\begin{tikzpicture}[scale=0.7]
\tikzmath{\shiftx = 1;\shifty = 3.464/2; \rad=0.15;}

\draw [line width=1pt] (1,3.464/2) -- (2,3.464) -- (3,3.464/2) -- cycle;
\draw [line width=1pt] (0-\shiftx,0-\shifty) -- (1-\shiftx,3.464/2-\shifty) -- (2-\shiftx,0-\shifty) -- cycle;
\draw [line width=1pt] (0+\shiftx,0-\shifty) -- (1+\shiftx,3.464/2-\shifty) -- (2+\shiftx,0-\shifty) -- cycle;
\draw [line width=1pt] (0,0) -- (2,0) -- (1,0+\shifty) -- cycle;
\draw [line width=1pt] (0+\shiftx+\shiftx,0) -- (2+\shiftx+\shiftx,0) -- (1+\shiftx+\shiftx,0+\shifty) -- cycle;
\draw [line width=1pt] (2+\shiftx,0-\shifty) -- (3+\shiftx,3.464/2-\shifty) -- (4+\shiftx,0-\shifty) -- cycle;

\coordinate[label=above:$e_1$] (e1) at (2,3.464);
\coordinate[label=left:$e_2$]  (e2) at (0-\shiftx,0-\shifty);
\coordinate[label=right:$e_3$] (e3) at (4+\shiftx,0-\shifty);

\fill   (2,3.464) circle (\rad)
        (0-\shiftx,0-\shifty) circle (\rad)
        (4+\shiftx,0-\shifty) circle (\rad)
        
        (1,3.464/2) circle (\rad)
        (0,0) circle (\rad)
        (0+\shiftx+\shiftx,0) circle (\rad)
        (2-\shiftx,0-\shifty) circle (\rad)
        (2+\shiftx,0-\shifty) circle (\rad)
        (4,0) circle (\rad)
        (3,3.464/2) circle (\rad);
        
\tikzmath{\shiftx = 1; \shiftX = 8; \shifty = 3.464/2; \rad=0.15;}

\draw [line width=1pt] (1+\shiftX,3.464/2) -- (2+\shiftX,3.464) -- (3+\shiftX,3.464/2) -- cycle;
\draw [line width=1pt] (0-\shiftx+\shiftX,0-\shifty) -- (1-\shiftx+\shiftX,3.464/2-\shifty) -- (2-\shiftx+\shiftX,0-\shifty) -- cycle;
\draw [line width=1pt] (2+\shiftx+\shiftX,0-\shifty) -- (3+\shiftx+\shiftX,3.464/2-\shifty) -- (4+\shiftx+\shiftX,0-\shifty) -- cycle;

\draw [line width=1pt, fill=gray!50] (1+\shiftX,3.464/2) -- (0+\shiftX,0);
\draw [line width=1pt, fill=gray!50] (1+\shiftX,-3.464/2) -- (3+\shiftX,-3.464/2);
\draw [line width=1pt, fill=gray!50] (3+\shiftX,3.464/2) -- (4+\shiftX,0);

\coordinate[label=above:$e_1$] (e1) at (2+\shiftX,3.464);
\coordinate[label=left:$e_2$]  (e2) at (0-\shiftx+\shiftX,0-\shifty);
\coordinate[label=right:$e_3$] (e3) at (4+\shiftx+\shiftX,0-\shifty);

\fill   (2+\shiftX,3.464) circle (\rad)
        (0-\shiftx+\shiftX,0-\shifty) circle (\rad)
        (4+\shiftx+\shiftX,0-\shifty) circle (\rad)
        
        (1+\shiftX,3.464/2) circle (\rad)
        (0+\shiftX,0) circle (\rad)
        (2-\shiftx+\shiftX,0-\shifty) circle (\rad)
        (2+\shiftx+\shiftX,0-\shifty) circle (\rad)
        (4+\shiftX,0) circle (\rad)
        (3+\shiftX,3.464/2) circle (\rad);

\end{tikzpicture}
\caption{$S(3,2)$ on the left and $H(2)$ on the right. }\label{idatalattices}
\end{figure}
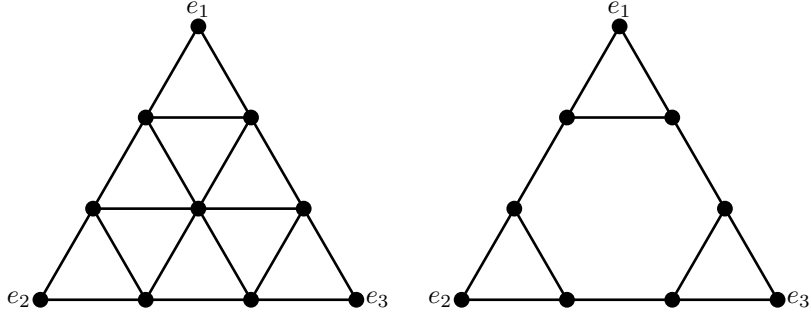

We can change the lattice by changing the initial data. So far we have consider the 3-vertex cycle as the initial data, see eq.  (\ref{firsttriangl2}). Such lattice has three external vertices as requires and can be seen as simple triangle. However we can construct new lattices, either under the Sierpinski or the Hanoi iteration, by simply changing the initial lattice from the simple triangle to any other lattice with three external vertices and invariance under $(2\pi/3)$-rotations. For illustration purposes we will consider the Sierpinski with three triangles along the side and $n=2$ ($S(3,2)$) and the Hanoi graph with $n=2$ ($H(2)$)as examples of alternatives to the standard 3-vertex cycle, see fig. \ref{idatalattices} 

For $S(3,2)$ we get
\begin{align}
  x(1)=& (128, 1152, 4896, 13152, 25128, 36144, 40036, 34218, 22461, 11344, \notag\\
       & 4422, 1320, 290, 42, 3)\cdot (1,v,\cdots,v^{14})\,,\label{firsttrianglS32-1}\\
  y(1) = & (0, 0, 0, 32, 336, 1640, 4840, 9470, 12824, 12384, 8810, 4720, 1912, \notag\\
         & 574, 121, 16, 1)\cdot (1,v,\cdots,v^{16})\,,\\
  z(1) =& (0, 0, 0, 0, 0, 0, 232, 1908, 6876, 14346, 19485, 18624, 13092, 6888, \notag\\
        &2700, 768, 150, 18, 1)\cdot (1,v,\cdots,v^{18})\label{firsttrianglS32-3}\,.
\end{align}
where
\begin{equation}
 (a_0,a_1,\cdots,a_n)\cdot (1,v,\cdots,v^{n})=a_0+a_1 v+\cdots a_n v^n\,.
\end{equation}

For $H(2)$ we get
\begin{align}
  x(1)=& (64, 384, 1056, 1760, 1956, 1476, 724, 204, 27, 1)\cdot (1,v,\cdots,v^9)\,,\label{firsttrianglH2-1}\\
  y(1) = & (0, 0, 0, 16, 88, 216, 298, 240, 105, 23, 2)\cdot (1,v,\cdots,v^{10})\,,\\
  z(1) =& (0, 0, 0, 0, 0, 0, 36, 138, 213, 156, 60, 12, 1)\cdot (1,v,\cdots,v^{12})\label{firsttrianglH2-3}\,.
\end{align}

\begin{figure}[h!]
\centering
\begin{subfigure}[b]{0.4\textwidth}
  \begin{overpic}[scale=0.29]{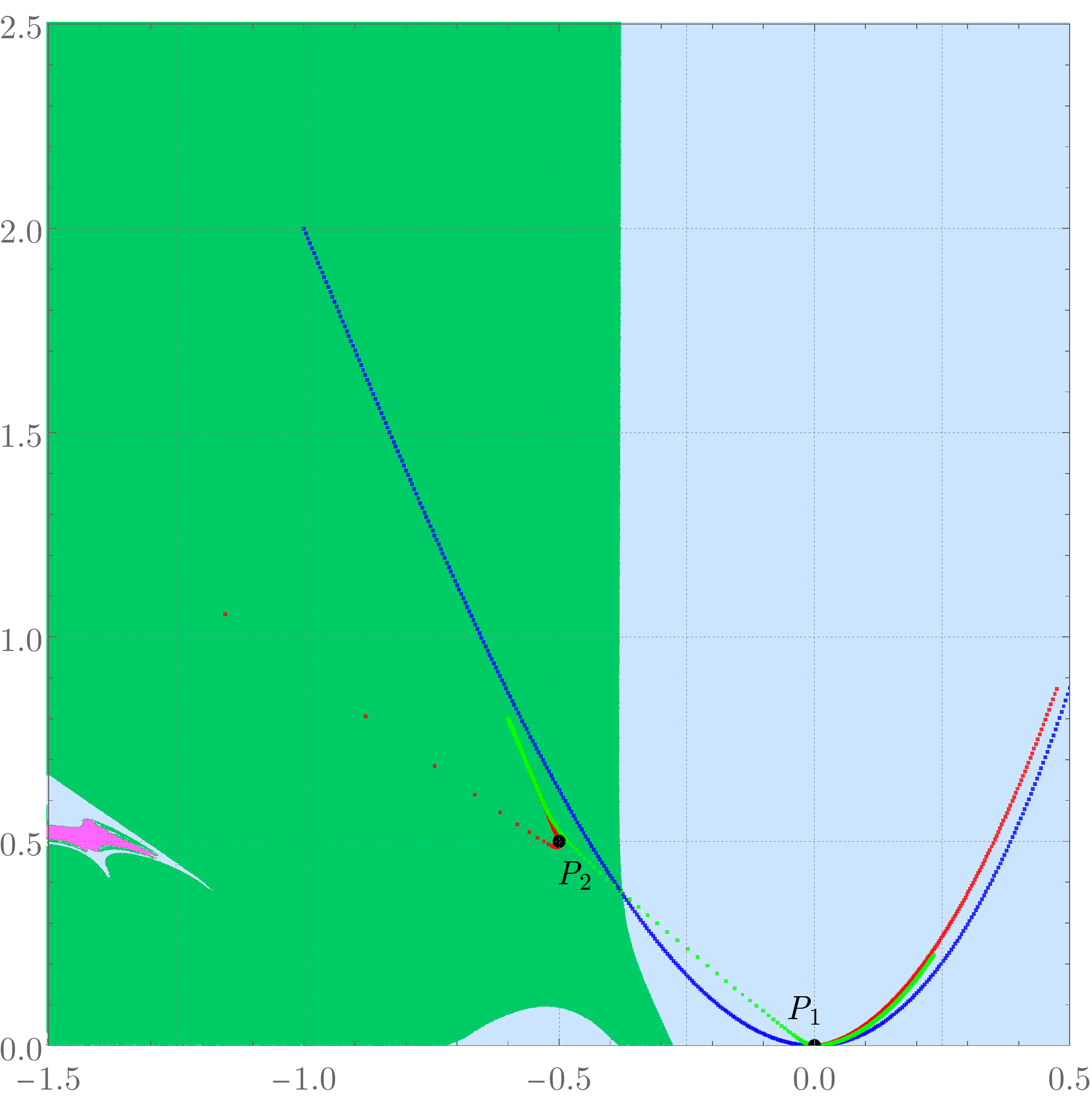}
     \put(54,67){\includegraphics[scale=0.4]{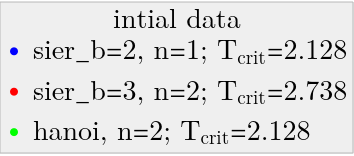}}
     \put(45,87){\includegraphics[scale=0.15]{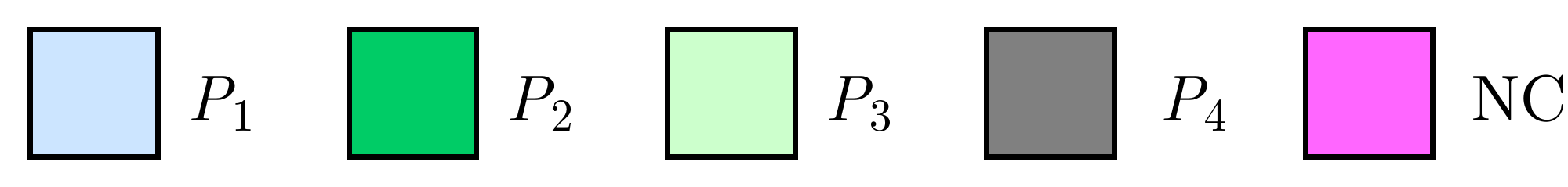}}
  \end{overpic}
  \caption{Basins map for $q=1$.}
  \label{fig:conv_q1hanoi}
\end{subfigure}%
\hspace{1.1cm}
\begin{subfigure}[b]{0.4\textwidth}
  \begin{overpic}[scale=0.29]{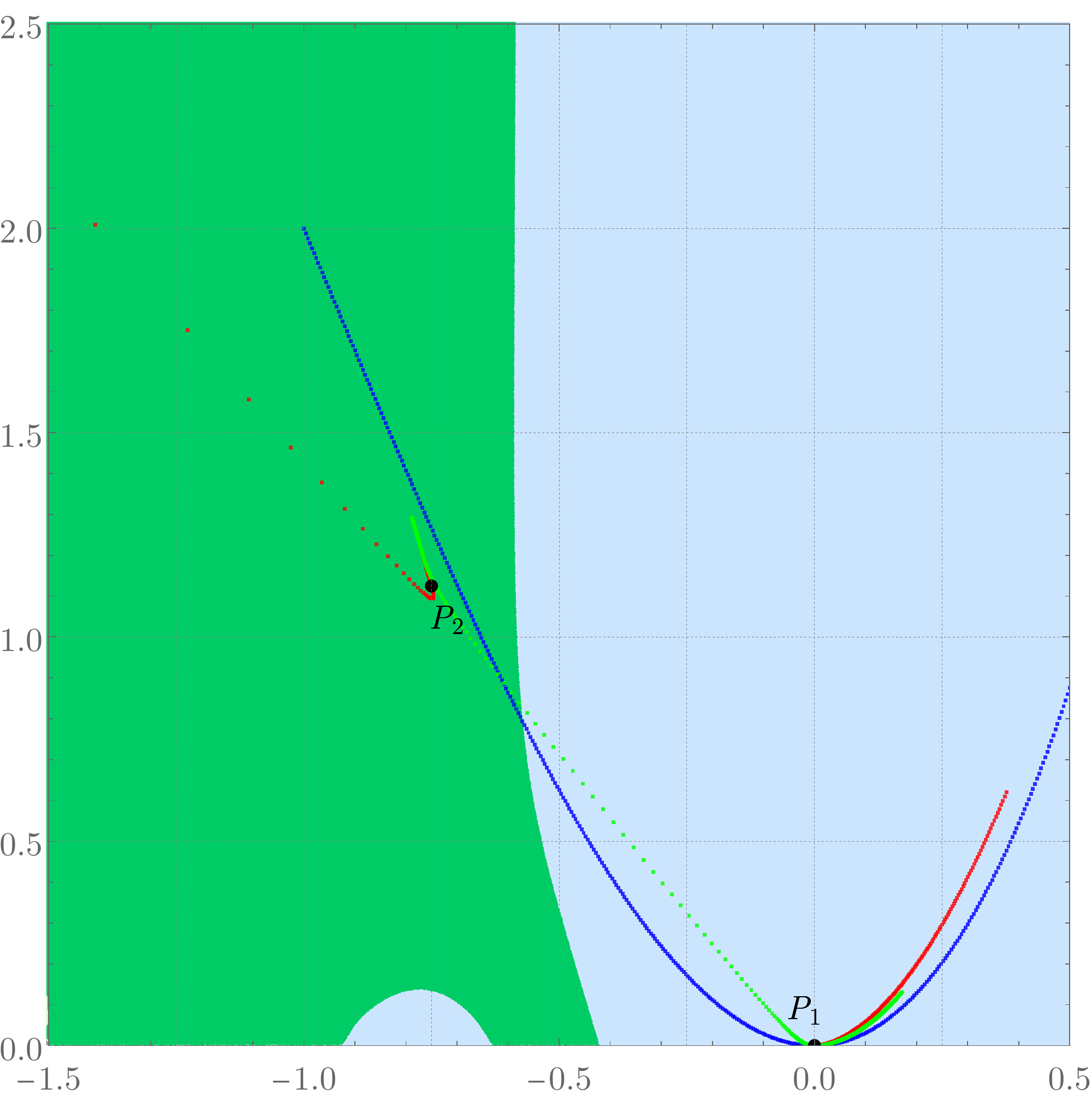}
     \put(54,67){\includegraphics[scale=0.4]{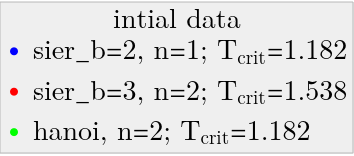}}
     \put(45,87){\includegraphics[scale=0.15]{{hanoi_legend_basin-map_complex-v_legend}.png}}
  \end{overpic}
  \caption{Basins map for $q=1.5$.}
  \label{fig:conv_q1.5hanoi}
\end{subfigure}\\
  \begin{subfigure}[b]{0.4\textwidth}
  \begin{overpic}[scale=0.29]{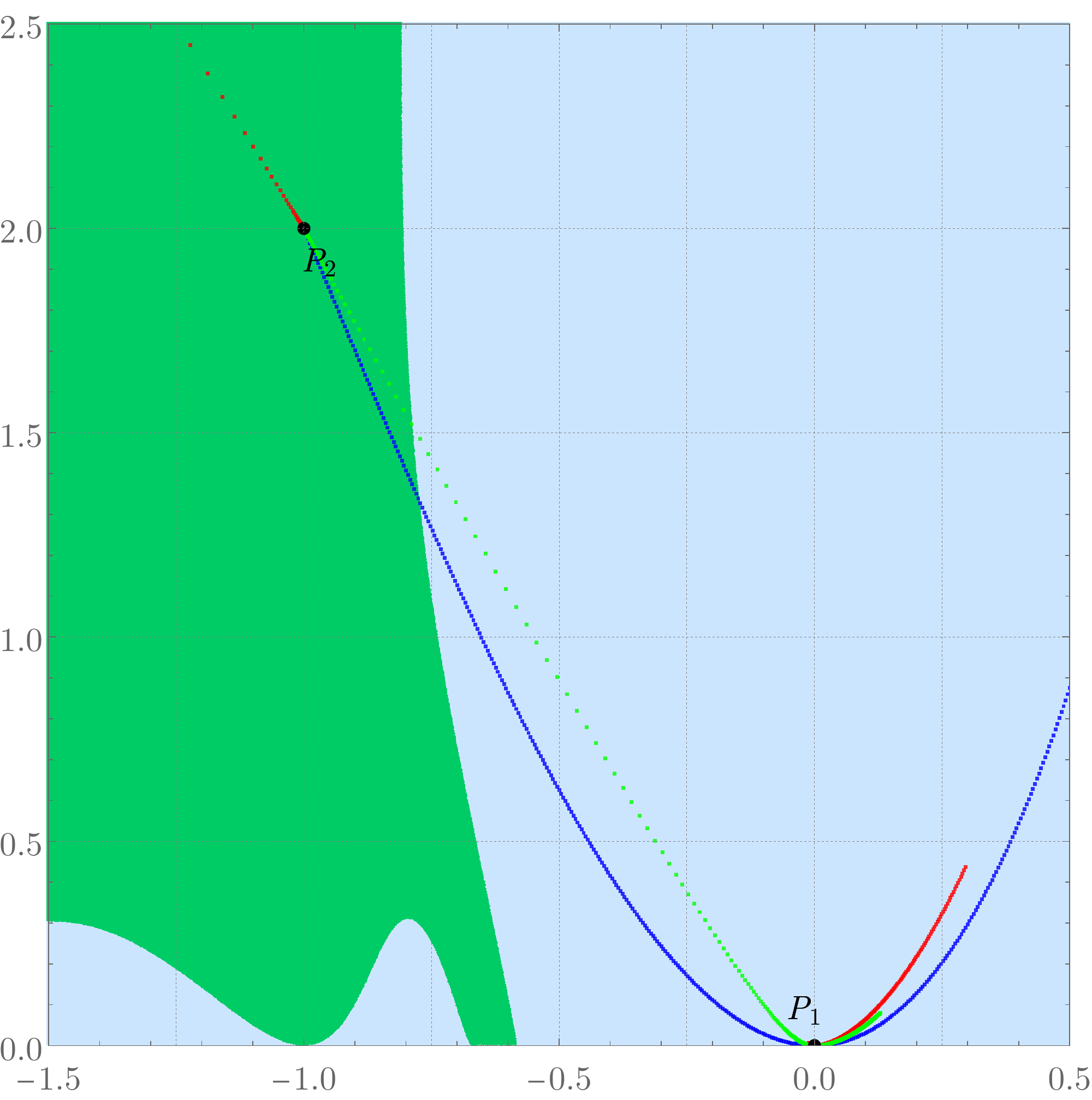}
     \put(41,67){\includegraphics[scale=0.4]{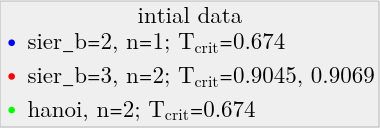}}
     \put(45,87){\includegraphics[scale=0.15]{{hanoi_legend_basin-map_complex-v_legend}.png}}
  \end{overpic}
  \caption{Basins map for $q=2$. For the $S(3,2)$ initial conditions we found jumps at two different temperatures.}
  \label{fig:conv_q2hanoi}
\end{subfigure}%
\caption{Basins of convergence of the Hanoi map in the antiferromagnetic regime for some values of $q$. Horizontal axis corresponds to $\eta$ and vertical axis corresponds to $\gamma$. Three different initial data are shown: standard triangle in red, $S(3,2)$ in purple and $H(2)$ in blue. The temperature at which the jump of basins occur is indicated in the legend.}
\label{fig:convHanoi}
\end{figure}

By considering different seeds for the initial data, such as (\ref{firsttriangl2}); (\ref{firsttrianglS32-1})-(\ref{firsttrianglS32-3}); or (\ref{firsttrianglH2-1})-(\ref{firsttrianglH2-3}), the resulting lattices will be different but they may share the same dynamical system depending on which hierarchical step we are using. In order to determine the thermodynamic limit it will be sufficient to overlap the curves of initial data $(\eta(1),\gamma(1))$, see fig. \ref{fig:conv} for the Sierpinski iteration and fig. \ref{fig:convHanoi} for the Hanoi iteration. The change in the parametric curve may or may not be strong (curves may almost overlap each other), however there is also a change in the parametrization that induces a change of the critical temperature ($T_c$) at which the line crosses from one to another basin of attraction. The results of $T_c$ are reported in Fig. \ref{fig:conv} for the Sierpinski iteration and fig. \ref{fig:convHanoi} for the Hanoi iteration.

\section{Locci of the zeros of the partition function}\label{zerosandbasins}

\begin{figure}
  \centering
  \begin{subfigure}{\textwidth}
  \begin{overpic}[scale=0.6]{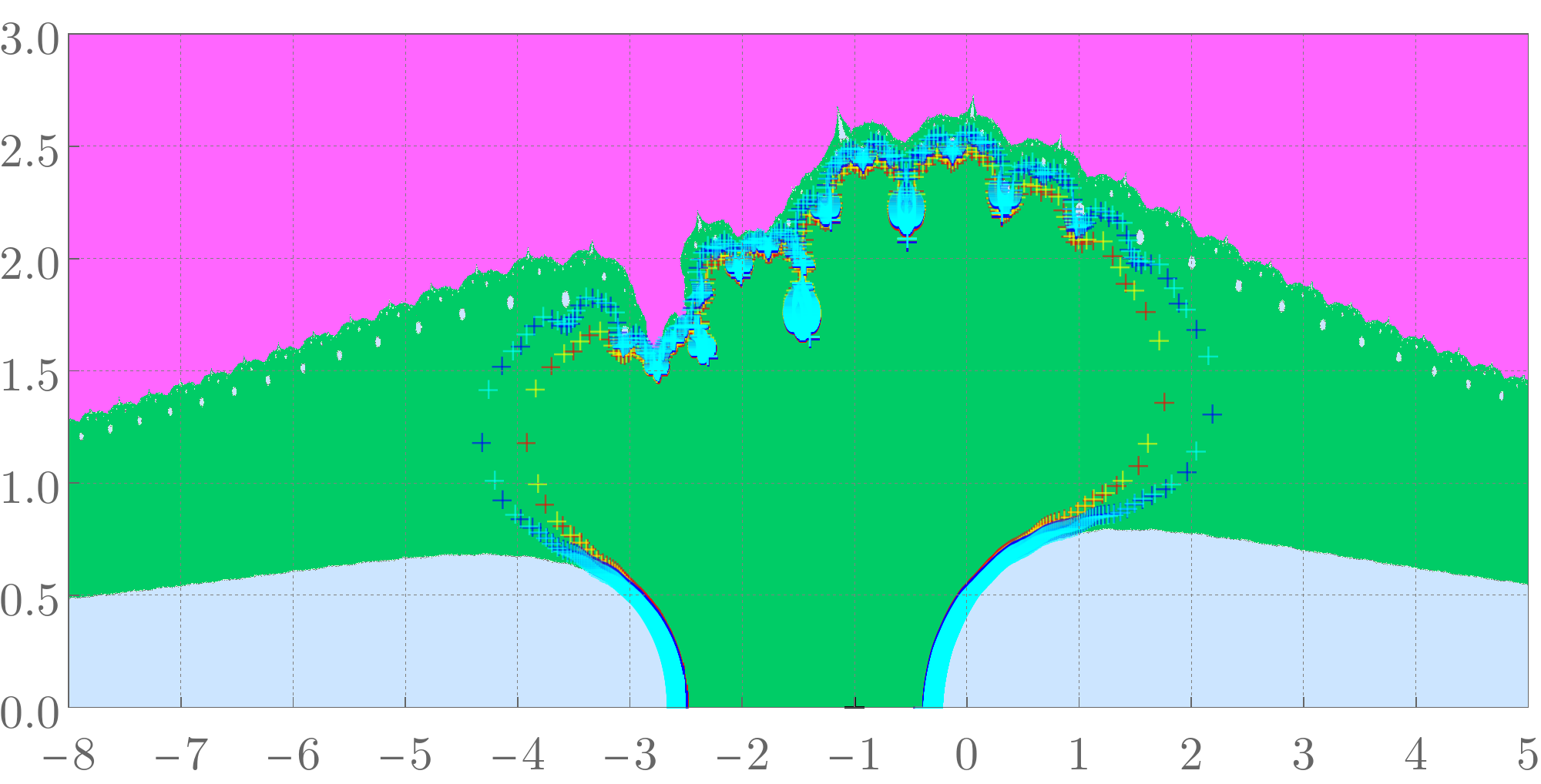}
     \put(6,30){\includegraphics[scale=0.5]{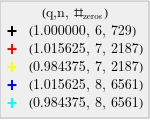}}
     \put(6,6){\includegraphics[scale=0.15]{{sier_b=2_legend_basin-map_complex-v_legend}.png}}
  \end{overpic}
  \caption{Basins map for $q=1$. The model shows degeneracy of zeros at $q=1$ and therefore we included cases $q=1 \pm \epsilon$ that remove the degeneracy.}
  \label{fig:sfig1_sier-complex}
\end{subfigure}
\begin{subfigure}{\textwidth}
  \begin{overpic}[scale=0.6]{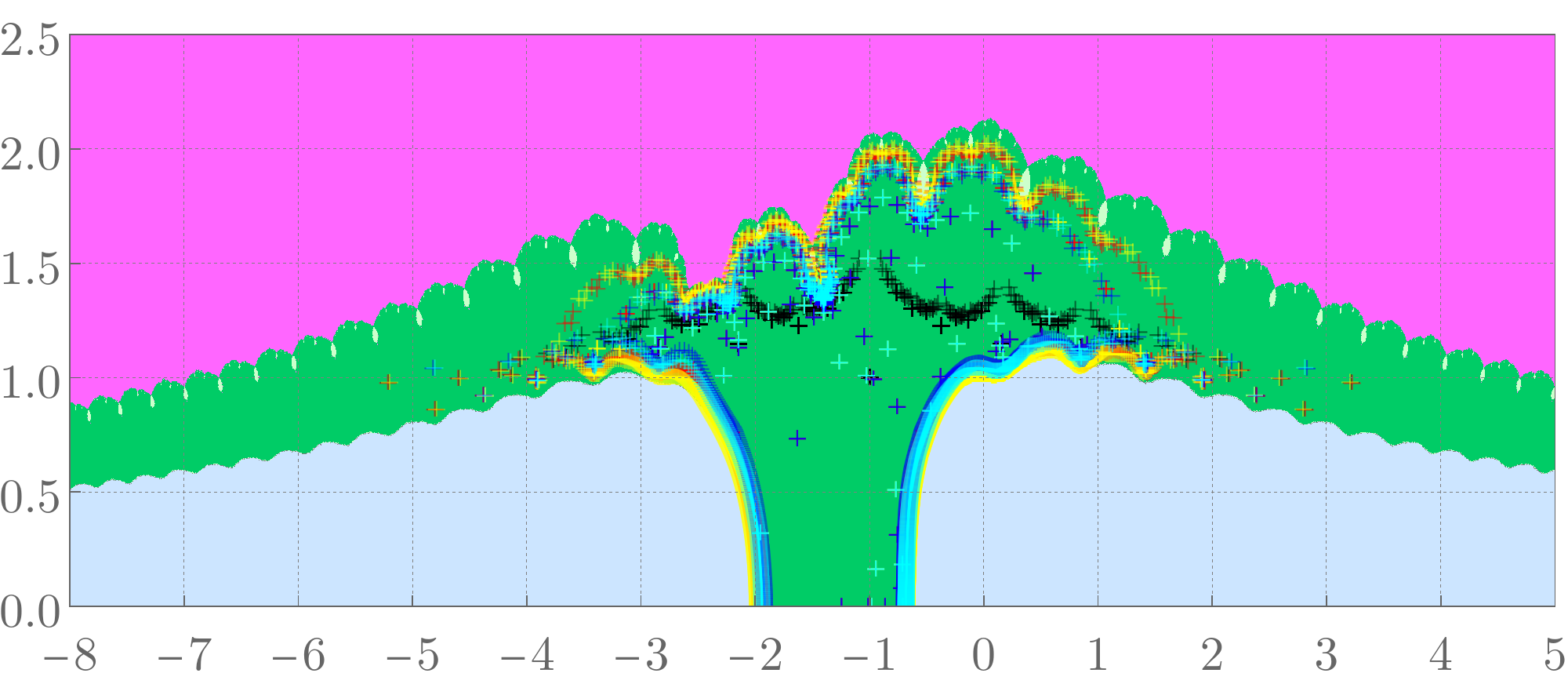}
     \put(6,23){\includegraphics[scale=0.25]{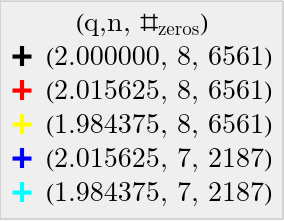}}
     \put(6,6){\includegraphics[scale=0.15]{{sier_b=2_legend_basin-map_complex-v_legend}.png}}
  \end{overpic}
  \caption{Basins map for $q=2$. The model shows degeneracy of zeros at $q=2$ and therefore we included cases $q=2 \pm \epsilon$ that remove the degeneracy.}
  \label{fig:sfig2_sier-complex}
\end{subfigure}
\begin{subfigure}{\textwidth}
  \begin{overpic}[scale=0.6]{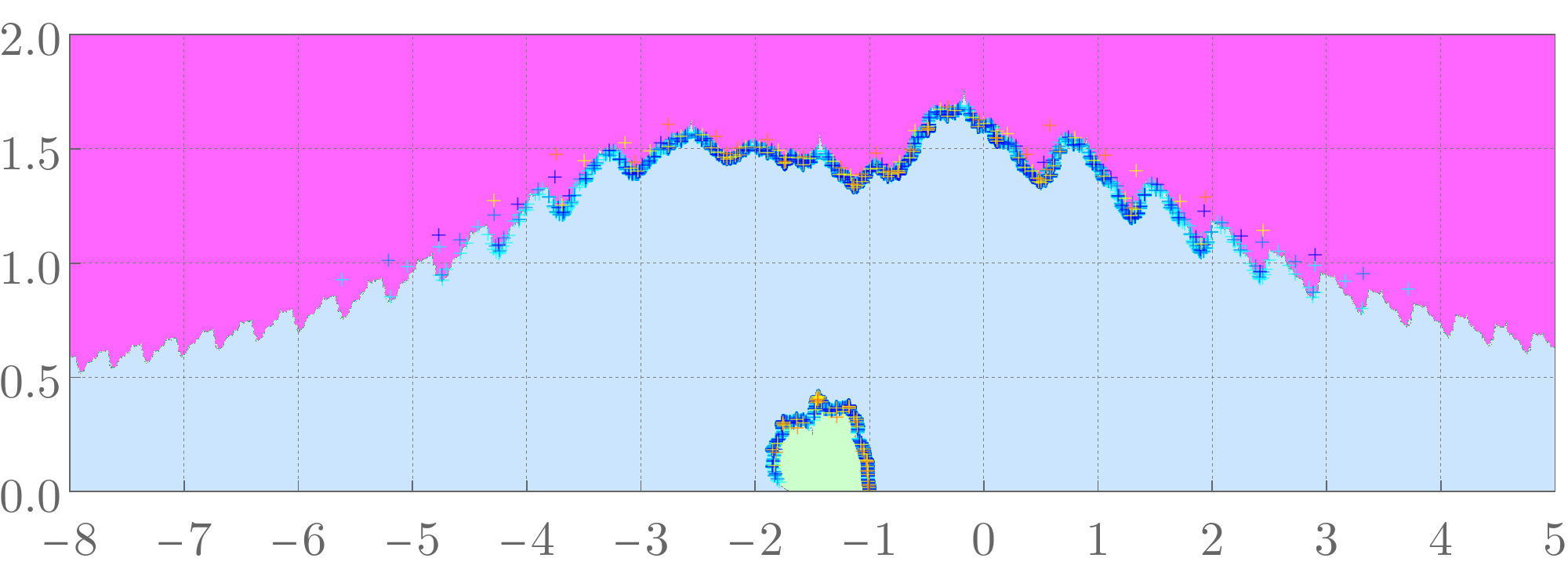}
     \put(6,15){\includegraphics[scale=0.5]{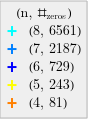}}
     \put(6,6){\includegraphics[scale=0.15]{{sier_b=2_legend_basin-map_complex-v_legend}.png}}
  \end{overpic}
  \caption{Basins map for $q=3$.}
  \label{fig:sfig3_sier-complex}
\end{subfigure}
\caption{Map of basins of attraction on the complex $v$-plane for the Sierpinski lattice and zeros of the partition function for finite $n$ cases. To remove degeneracy we used $\epsilon=1/64$.}\label{complexplanemapSier}
\end{figure}

\begin{figure}
  \centering
  \begin{subfigure}{\textwidth}
  \begin{overpic}[scale=0.15]{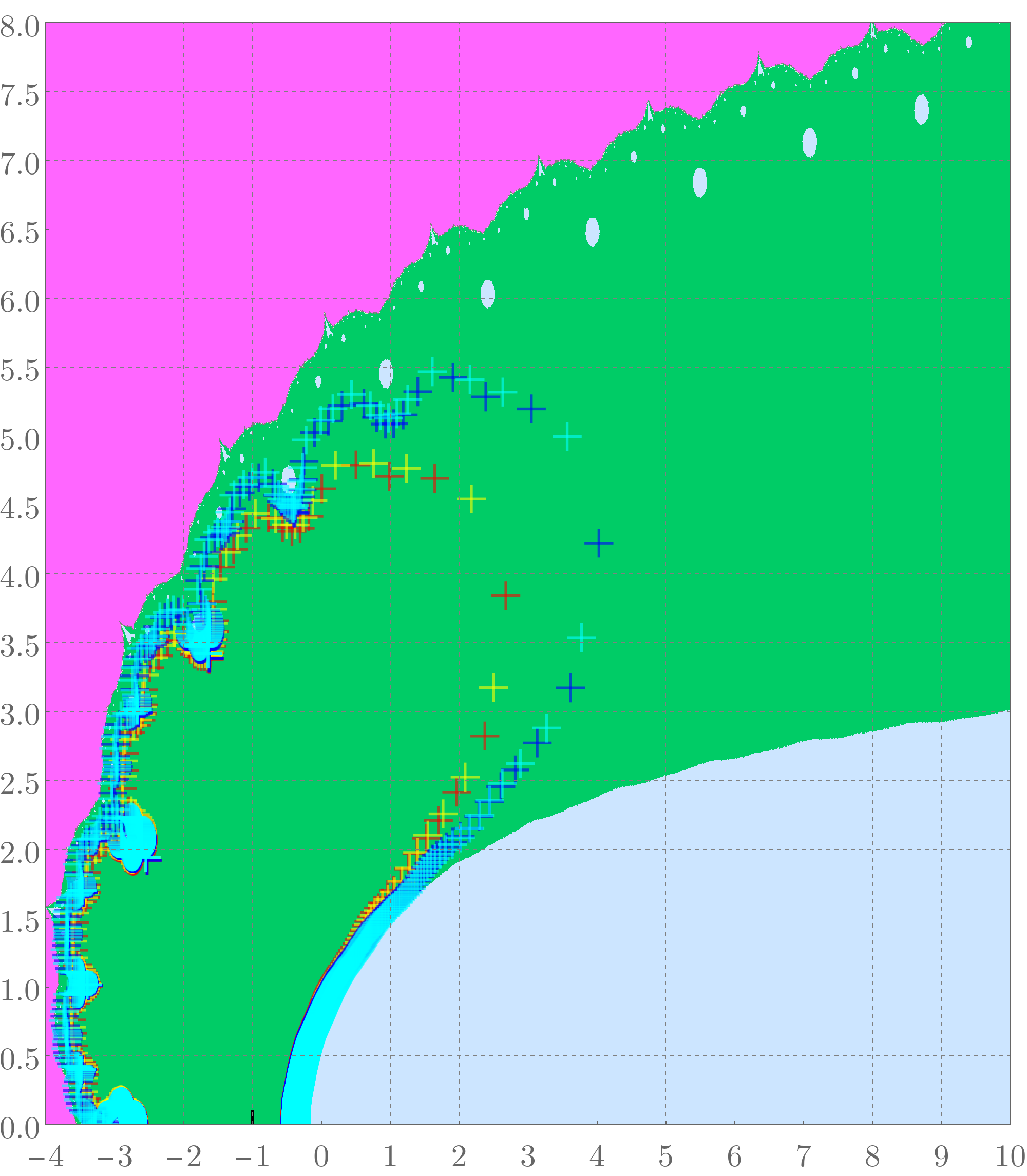}
     \put(6,68){\includegraphics[scale=0.25]{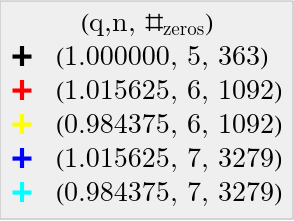}}
     \put(40,6){\includegraphics[scale=0.15]{{hanoi_legend_basin-map_complex-v_legend}.png}}
  \end{overpic}
  \begin{overpic}[scale=0.15]{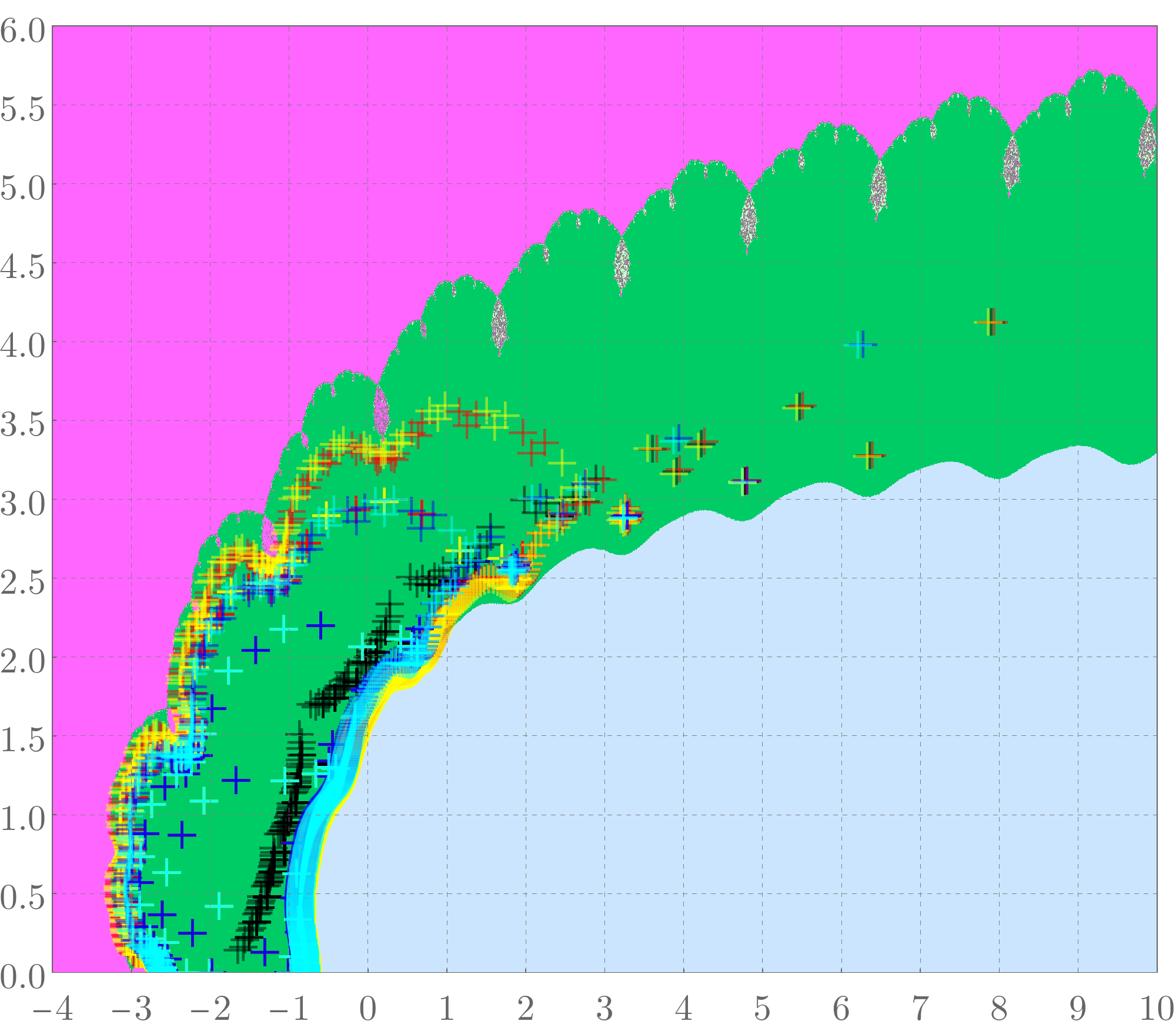}
     \put(6,52){\includegraphics[scale=0.25]{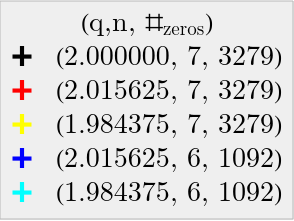}}
     \put(46,6){\includegraphics[scale=0.15]{{hanoi_legend_basin-map_complex-v_legend}.png}}
  \end{overpic}
  \caption{Basins map for $q=1$ (left panel) and $q=2$ (right panel). The model shows degeneracy of zeros at $q=1,2$ and therefore we included cases $q=1\pm\epsilon$ (left panel) and $q=2 \pm \epsilon$ (right panel) that remove the degeneracy.}
  \label{fig:sfig1_hanoi-complex}
\end{subfigure}
\begin{subfigure}{\textwidth}
  \begin{overpic}[scale=0.2]{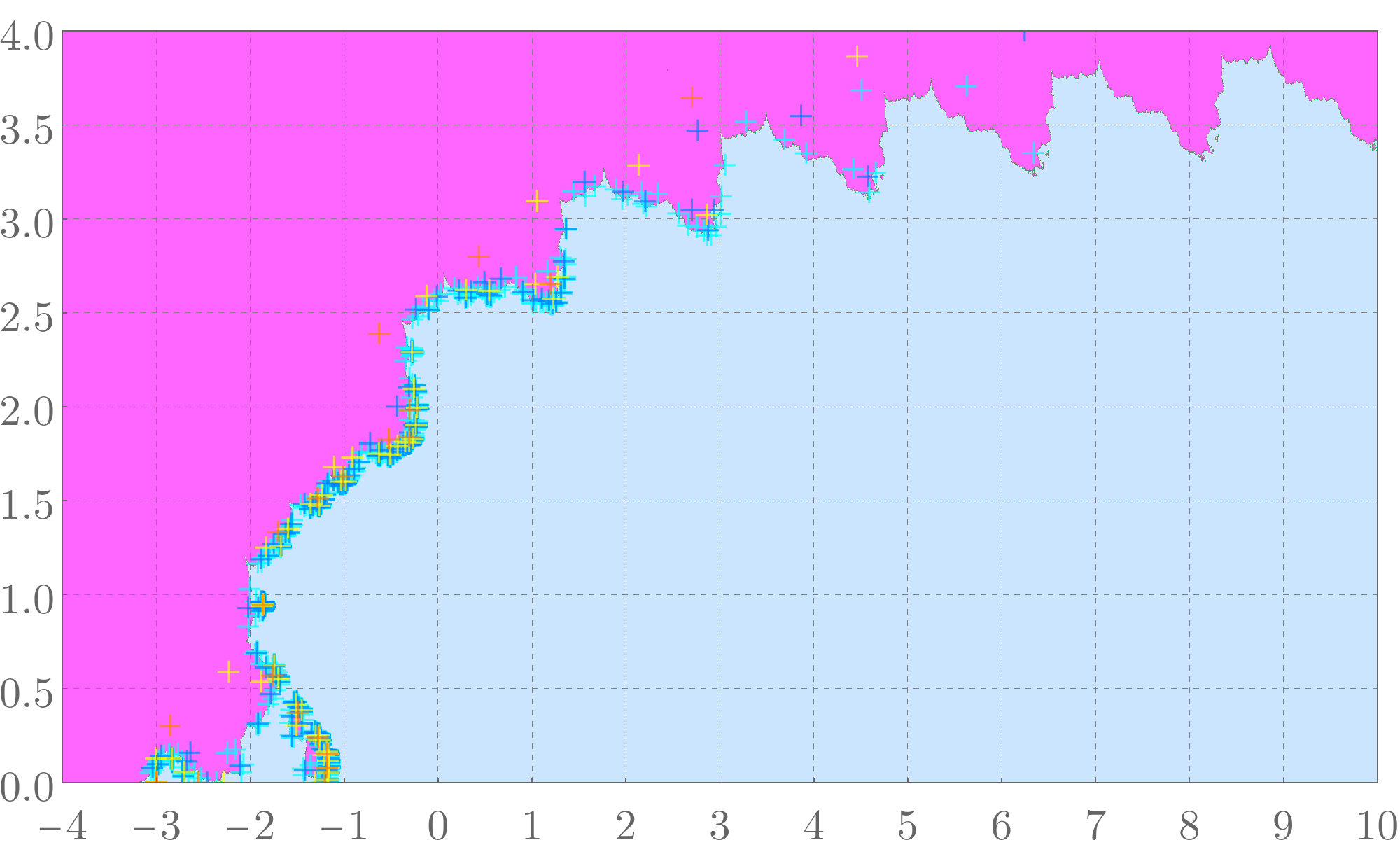}
     \put(6,32){\includegraphics[scale=0.5]{{sier_b=2_legend-zeros_basin-map_complex-v_q=2._n=200_step=multi}.png}}
     \put(60,6){\includegraphics[scale=0.15]{{hanoi_legend_basin-map_complex-v_legend}.png}}
  \end{overpic}
  \caption{Basins map for $q=3$.}
  \label{fig:sfig2_hanoi-complex}
\end{subfigure}
\caption{Map of basins of attraction on the complex $v$-plane for the Hanoi lattice and zeros of the partition function for finite $n$ cases. To remove degeneracy we used $q=\epsilon=1/64$.}\label{complexplanemapHan}
\end{figure}

Here we comment on a very interesting finding related to the location of the zeros of the partition function in the plane of complex temperature. We study the maps of basins of fixed points of the system (\ref{reduced1})-(\ref{reduced2}) (or (\ref{hanoireduced1})-(\ref{hanoireduced2})), with, for the sake of definiteness, initial data (\ref{icdiscrete}). Now we promote the temperature variable to the complex plane $v \in \mathbb{C}$. We constructed the polynomials $K(n)(q,v)$ for fixed values of $q$ for many values of $n=2,3,\cdots$. The maximal value of $n$ that we can achieve depends on the type of programming language and libraries used, but values $n\sim 5,6,7,8,\cdots$ allow to see the pattern of distribution of zeros, see Figs. \ref{complexplanemapSier} and \ref{complexplanemapHan}. This lead us to the following observation:\\
\begin{center}
\begin{minipage}{.8\textwidth}
  \emph{The zeros of $Z_n$ in the complex-temperature plane have a location that asymptotically tends to the boundary of the basin of attraction of superstable fixed points. Exemption to this rule may appear for small enough and discrete $q$-values where a large degeneracy may appear in the polynomials.}
\end{minipage}
\end{center}

Therefore we learned that the study of the basins of attraction of the fixed points (for hierarchical lattices) provides the generalization of the region $|\lambda_1| = |\lambda_2|$ for systems with a transfer matrix, where $|\lambda_1|$ $|\lambda_2|$ are the larger eigenvalues of the transfer matrix (in absolute value). It is important to mention that the present formulation is based on partial sums over the space of subgraphs in the Fortuin-Kasteleyn formulation, as opposed to sums over restricted partition functions in spin space. To the best knowledge of the authors, this is a novel result that corresponds to an extension of the Beraha-Kahane-Weiss theorem \cite{BERAHA198052}.

When the initial data is such that it crosses the boundary of islands of attraction of stable fixed points a jump in the fixed point will be generated, see Figs. \ref{fig:conv} and \ref{fig:convHanoi}. The existence of this jump will depend on the dynamical system and the initial conditions, and the temperature at which this jump occurs will also depend on them. Whether this jump leads to a phase transition or not can be decided by studying the continuity conditions of the thermodynamic functions deduced from the partition function,
\begin{equation}
K\left( n\right) =x\left( n\right) \left( q^{3}+3q^{2}\eta \left( n\right)+q\gamma (n)\right)\,. \label{parfuc}
\end{equation}

\begin{figure}[h!]
\begin{subfigure}[b]{.45\linewidth}
\includegraphics[width=\linewidth]{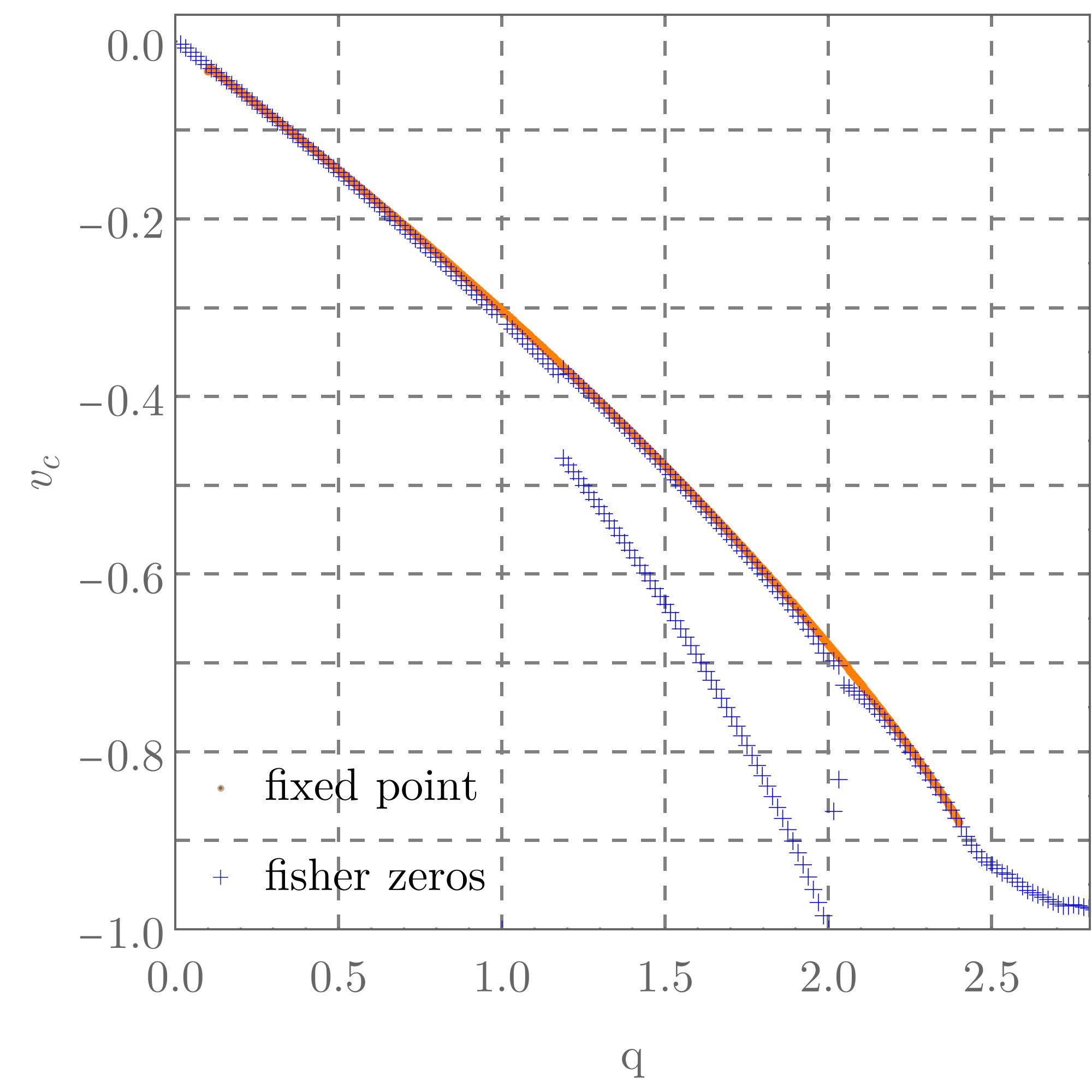}
\caption{Sierpinski case.}\label{fig:zerosvsjump_Sier}
\end{subfigure}
\begin{subfigure}[b]{.45\linewidth}
\includegraphics[width=\linewidth]{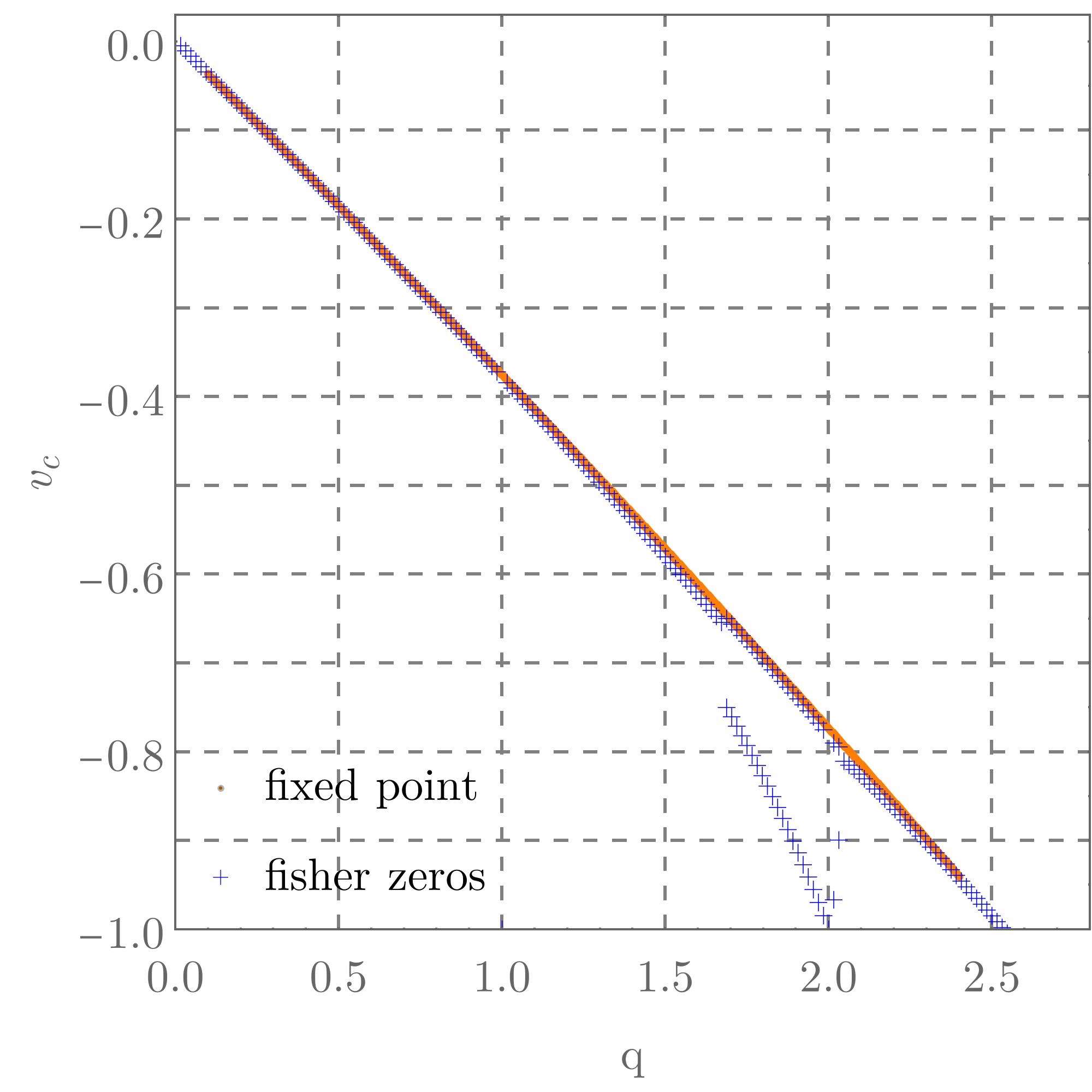}
\caption{Hanoi case.}\label{fig:zerosvsjump_Hanoi}
\end{subfigure}
\caption{Points are the Fisher zeros that lie on the real axis. Continuous line is the $v$-temperature at which the jump of basins occur.}
\label{fig:zerosvsjump}
\end{figure}

As it is well known, the relation between the existence of Fisher zeros pinching the real axis of temperature indicates phase transitions. This suggests to look for a relation between zeros on the real axis of temperature and the temperature at which the jump  occurs. In the antiferromagnetic regime we found one or more than one zeros lying on the real axis (within certain numerical tolerance the zero takes the form $(v_x,0)$) in the range $q \sim 0 \rightarrow 2.5$. The present method, that allows for non-integer $q$, allows to scan a range of the $q$-domain. The results are very impressive and are summarized in Fig. \ref{fig:zerosvsjump}. This lead us to propose the following general result:
\begin{center}
\begin{minipage}{.8\textwidth}
\emph{The Fisher zero with a larger real value of $v_x$ asymptotically (in the sense of $n\rightarrow \infty$) lies at the line at which the jump on basins of attraction occurs.} 
\end{minipage}
\end{center}

Zeros on the real axis cease to exist at a slightly higher value of $q$ than the boundary of the stability region of $P_2$ and $P'_2$ ($q \sim 2.5$).

In Fig. \ref{fig:zerosvsjump} we can also see subtle jumps in the zeros at the integer values $q=1,2$. This is related to the appearance of a large degeneracy of the Fisher zeros at those values. Such degeneracy can also be seen by comparing the Fisher zeros for values $q = q_\mathrm{integer} \pm \epsilon$ where $\epsilon$ is a small value.

We have also repeated the same computation with the dynamical system presented in \cite{Sim09first,Sim09second}. The dynamical variables defined in \cite{Sim09first,Sim09second} refer to the degeneracy of the energy states of the system rather than to the connectivity patterns as in the present case. Nevertheless, the two dynamical systems produce consistent results: the phase transitions described by the dynamical system of \cite{Sim09first,Sim09second} occur exactly at the same critical temperatures found in our approach. It is reasonable to assume that the peculiarity of the free energy density of the system to be a piecewise constant non-vanishing function can explain why the analysis of the Fisher zeros reported in \cite{Sim09first,Sim09second} does not detect any phase transition even though the direct computation of the free energy density with both dynamical systems reveals multiple \textquotedblleft $0^{th}$ transitions\textquotedblright. To the best of authors' knowledge, this is the first example in which the Fisher zeros technique does not detect a phase transition. A similar analysis can be carried out for the Hanoi dynamical system.

Let us comment of other studies of the zeros in the complex plane and the fixed points of the associated dynamical systems in hierarchical lattices \cite{10.1063/1.5127667}.

\section{Conclusions}\label{sectionconclusions}

We presented an analytic exact study of the Potts model partition function on self-similar triangular lattices. Two cases have been analyzed, the Sierpinski gasket and the Hanoi graph, by means of a formalism based on ideas of graph theory. By introducing suitable geometric coefficients related to the connectivity pattern of the vertices of subgraphs on the lattices is possible to reduce the computation of the partition function to a dynamical system. Then, using known results in the theory of dynamical systems, one can study properties of the system by analyzing the map of basins of fixed points. We run benchmarks with Metropolis Monte Carlo simulations. We found temperatures at which jumps in basins occur ($T_\mathrm{crit}$), for which we agree with results of \cite{Sim09first,Sim09second} that are based on real space renormalization group. It has been shown that the formalism can be easily extended to other families of recursive lattices that are obtained by changing initial conditions. We noted an exact invariance of $T_\mathrm{crit}$ under the change in initial conditions $T(1) \to T(n)$. We do not know a demonstration of this fact and it is a very interesting topic to study in a future work.

We commented on the relation of the chromatic polynomial and the zero-temperature entropy of the antiferromagnetic model. However, the special case of $q=2$, that has no proper colorings, revealed that $d_k^{(n)}$ provides the zero-temperature entropy of the antiferromagnetic model. This shows that other terms, besides $d_0^{(n)}(q)$ are invariant quantities of the lattice with a physical meaning. The natural question that arises is the following: What coefficients in the expansion $c_i^{(n)}(q)$ or $d_i^{(n)}(q)$ can have physical or invariant meaning? Are there close exact forms for $c_i^{(n)}(q)$ or $d_i^{(n)}(q)$ and their asymptotic expressions? These are very important questions that can be addressed with the formalism presented here and we left for a future study.

In section \ref{zerosandbasins}, we studied the loci of zeros and their relation with the map of basins of attraction. Based on our findings, we provided a conjecture of the locci in terms of the boundary region of different basins of stable points. We determined the physical temperature at which jumps in fixed points occur and we provided a conjecture about its relation to the zero with a larger value of $v_x$, where $v=(v_x,v_y)$ is the complex plane temperature.

\section*{Acknowledgements}

The author thanks Luca Parisi for useful discussions and to Fabrizio Canfora for powerful insights without which the present study could not have been carried out.

\bibliography{paper}
\bibliographystyle{ieeetr}

\end{document}